\documentclass[]{mn2e}

\usepackage[dvips]{graphicx,color}
\usepackage{times}
\usepackage{multirow}

\usepackage{amsmath}
\usepackage{pifont}

\usepackage{lscape}
\usepackage{graphicx,rotating}
\usepackage{txfonts,amsfonts}
\usepackage{pifont}

\newcommand{\tickYes}{\checkmark}
\newcommand{\tickNo}{\hspace{1pt}\ding{55}}

\def\nms{\mathsurround=0pt}
\def\oversim#1#2{\lower 2pt\vbox{\baselineskip 
0pt \lineskip 1pt
    \ialign{$\nms#1\hfil##\hfil$\crcr#2\crcr\sim\crcr}}}

   \title[On core-collapse SNe and H\,{\sc ii} regions]
{On the association between core-collapse supernovae and 
H\,{\sc ii} regions}

   \author[Paul A. Crowther]{Paul A. Crowther
\thanks{Paul.Crowther@sheffield.ac.uk}
\vspace{3mm} \\
Dept of Physics \& Astronomy, University of Sheffield, Hicks Building,
             Hounsfield Rd, Sheffield, S3 7RH, United Kingdom}

\date{\today}

\pagerange{\pageref{firstpage}--\pageref{lastpage}} \pubyear{2012}

\begin{document}

\maketitle

\label{firstpage}

\begin{abstract} 
Previous studies of the location of core-collapse supernovae (ccSNe) in 
their host galaxies have variously claimed an association with H\,{\sc ii} 
regions; no association; or an association only with hydrogen-deficient 
ccSNe. Here, we examine the immediate environments of 39 ccSNe whose 
positions are well known in nearby ($\leq$15 Mpc), low inclination 
($\leq$65$^{\circ}$) hosts using mostly archival, continuum-subtracted 
H$\alpha$ ground-based imaging. We find that 11 out of 29 hydrogen-rich 
ccSNe are spatially associated with H\,{\sc ii} regions (38 $\pm$ 11\%), 
versus 7 out of 10 hydrogen-poor ccSNe (70 $\pm$ 26\%). Similar 
results from 
Anderson et al. led to an interpretation that the progenitors of type Ib/c 
ccSNe are more massive than those of type II ccSNe. Here, we quantify the 
luminosities of H\,{\sc ii} region either coincident with, or nearby to 
the ccSNe. Characteristic nebulae are long-lived ($\sim$20 Myr) giant H\,{\sc ii} 
regions rather than short-lived ($\sim$4 Myr) isolated, compact H\,{\sc 
ii} regions. Therefore, the absence of a H\,{\sc ii} region from most type II 
ccSNe merely reflects the longer lifetime of stars with $\lessapprox$12 
$M_{\odot}$ 
than giant H\,{\sc ii} regions. Conversely, the association of a H\,{\sc 
ii} region with most type Ib/c ccSNe is due to the shorter lifetime of stars with
$>$12 $M_{\odot}$ stars than the duty cycle of giant H\,{\sc ii} regions. 
Therefore, we conclude that the observed association between certain ccSNe 
and H\,{\sc  ii}  provides only weak constraints upon their progenitor 
masses. Nevertheless, we do favour lower mass progenitors for two type 
Ib/c ccSNe  that lack associated nebular emission, a host cluster or a 
nearby giant H\,{\sc ii} region. 
Finally, we also reconsider the association between long Gamma Ray Bursts 
and the  peak  continuum light from their (mostly)  dwarf hosts, and 
conclude that  this is suggestive of very high mass progenitors, in common 
with previous studies.
\end{abstract}

\begin{keywords}
stars: early-type -- stars: supernovae: general -- ISM: HII 
regions -- galaxies: star clusters, galaxies: ISM
\end{keywords}

\section{Introduction}

The past decade has seen major advances in establishing the progenitors of
core-collapse supernovae (ccSNe, Smartt 2009). Three discrete 
sub-populations of hydrogen-rich ccSNe are known, exhibiting plateau's 
(II-P), slow declines (II-L) and rapid declines (IIb) in their light 
curves (Arcavi et al. 2012), representing progressively lower hydrogen 
envelope masses. It has been empirically established that the most common 
of these (II-P) are the direct progeny of red supergiants (Smartt et al. 
2009). Some of the rarer subtypes (II-L and IIb) have been proposed to 
originate from yellow supergiants, hydrogen-rich Wolf-Rayet stars or 
interacting binaries, while hydrogen-rich ccSNe with narrow components in 
their spectra (IIn) seem to involve interactions with dense circumstellar 
material (Kiewe et al. 2012). 

No progenitors of  hydrogen-deficient (Ib/c) ccSNe have yet been 
detected  (Crockett et al.  2007; Yoon et al. 2012) which are
believed to either arise from massive Wolf-Rayet stars that have stripped 
their  hydrogen from powerful stellar  winds (e.g. Conti 1976;
Crowther 2007), or lower mass stars in  close binary systems 
(Podsiadlowski et al. 1992;
Nomoto et al. 1995; Fryer et al. 2007), or some  combination thereof 
(Smith et al. 2011; Langer 2012). It is likely that helium-strong IIb and Ib  ccSNe 
possess similar  progenitor channels (Arcavi et al. 2012), while helium-weak 
Ic ccSNe may arise from disparate progenitors (e.g. Dessart et al. 2012).
In particular, broad-lined type Ic ccSNe are notable in several ways; they
represent the majority of hydrogen-deficient ccSNe in  
dwarf hosts (Arcavi et al. 2010), there is a broad-lined Ic-Gamma Ray 
Burst (GRB) connection (Woosley \& Bloom 2006), while long
GRBs prefer metal-poor hosts (Levesque et al. 2010).


In view of the scarcity of nearby ccSNe amenable to the direct detection 
of the progenitor star, studies have turned to the host 
environment. For example, the population of hydrogen-deficient 
ccSNe are dominated by Ic ccSNe in large galaxies, versus Ib and 
broad-lined Ic ccSNe in dwarf galaxies (Arcavi et al. 2010), although
the overall statistics of Ib/c versus II ccSNe are relatively insensitive 
to host galaxy. 

\begin{table*}
  \begin{center}
  \caption{Examples of nearby H\,{\sc ii} regions, spanning a range of luminosities, for an assumed O7V Lyman continuum ionizing flux of 10$^{49}$ 
(ph\,s$^{-1}$), adapted from Kennicutt (1984, 1998).}
  \label{table2}
  \begin{tabular}{lccrrrr}\hline 
  Region & Type & galaxy & Distance & Diameter & L(H$\alpha)$ & N(O7V) \\
         &      &        &  (kpc)    & (pc)     & (erg\,s$^{-1}$) &      
\\
  \hline
Orion (M 42) &  Classical & Milky Way & 0.5 & 5\phantom{:}    &  1$\times 10^{37}$ & $<$1 \\ 
Rosette (NGC 2244)& Classical & Milky Way & 1.5 & 50\phantom{:} & 9$\times 10^{37}$ & 7 \\ 
%
%
N66 & Giant & SMC & 60\phantom{.0} & 220: & 6$\times 10^{38}$ & 50 \\
Carina (NGC 3372) & Giant    & Milky Way & 2.3 & 300:  &   1.5$\times 10^{39}$ & 120 \\ 
NGC 604 & Giant    & M 33 & 800\phantom{.0} & 400\phantom{:}   & 4.5$\times 10^{39}$ & 320  \\ 
30 Doradus  & Supergiant& LMC & 50\phantom{.0} & 370\phantom{:}   & 1.5$\times 10^{40}$ & 1100  \\
NGC 5461 & Supergiant &M 101 & 6400\phantom{.0} & 1000: &  7$\times 10^{40}$ & 5000  \\ 
  \hline
  \end{tabular}   
  \end{center}
\end{table*}

In addition, the immediate environment from which the ccSNe 
originated has also been examined. The first serious attempt to assess 
their association with H\,{\sc ii} regions was by van Dyk (1992). From a 
sample of 38 core-collapse SNe of all subtypes, he concluded that 
approximately  50\% were  associated with a H\,{\sc ii} region, with no 
statistically significant  difference between type II and Ib/c SNe, albeit 
hampered by poor positional accuracy (up to $\pm$10$''$). Bartunov et al. 
(1994), also  concluded that  both H-rich and H-deficient core-collapse 
SNe were  concentrated towards  H\,{\sc ii} regions, implying similar 
ages/masses. Improved statistics (49 ccSNe) enabled van Dyk et al (1996) 
to confirm earlier results, concluding that the (massive) Wolf-Rayet 
scenario could be excluded for most type Ib/c ccSNe, albeit once 
again subject to poor positional accuracy for many targets.

More recently, James \& Anderson (2006), Anderson \& James 
(2008) and Anderson et al. (2012) have taken a statistical approach
to the environment of ccSNe, involving its position with respect to 
the  cumulative distribution of H$\alpha$ emission in the host galaxy, 
whose  recession velocities extended up to $cz$ = 10,000 km\,s$^{-1}$. 
Anderson \& James (2008) found a low fraction of type II 
SNe to be associated with H\,{\sc ii} regions, concluding that the 
{\it ``type II 
progenitor population does not trace the underlying star formation.''}
In  contrast, they noted that  type Ib, and especially Ic 
ccSNe {\it are} spatially coincident with  H\,{\sc ii} regions, suggesting 
a progenitor mass sequence from 
II  $\rightarrow$ Ib $\rightarrow$ Ic.
Anderson et al. (2012) include additional statistics for hydrogen-rich 
ccSNe from which they claim a mass sequence 
IIn $\rightarrow$ II-P $\rightarrow$ II-L $\rightarrow$ IIb. 
The latter naturally connects IIb
and Ib ccSNe, but the low progenitor masses inferred for IIn ccSNe does
not readily match expectations that these arise from massive Luminous
Blue Variables (Smith 2008).

To add to the puzzle, Smartt (2009) argued against a monotonic
mass sequence for progenitors of II  $\rightarrow$ Ib $\rightarrow$ Ic
on the basis of the rate of Ib/c ccSNe, the lack of direct detections
of Ib/c progenitors (e.g. Crockett 2009) and inferred low Ib/c ccSNe 
ejecta masses.
From a qualitative study of the environment of the volume- and epoch- 
limited sample of 
ccSNe
of Smartt et al. (2009), 0 from 17 type II SN observed at high spatial
resolution are located in bright 
H\,{\sc ii} regions (Smartt, priv. comm.). Meanwhile, only 1 case 
from 9 Ib/c ccSNe from  Crockett (2009) for which high spatial resolution 
imaging was available,  is located in a  large star forming  region, 
albeit spatially offset from  H\,{\sc ii} emission (Smartt, priv. 
comm.). Therefore, high resolution imaging does not appear to 
support any significant association between ccSNe and H\,{\sc ii} 
regions.

In addition to the association of ccSNe 
with H\,{\sc ii} regions, or lack thereof, studies of the location of 
ccSNe with respect to the host galaxy light have also been performed. 
Kelly et al. (2008) found that type Ic ccSNe are located in the brightest 
regions of their host galaxies, type II ccSNe are randomly distributed, 
with intermediate properties for type Ib ccSNe. Long duration GRBs, 
in common with type Ic ccSNe, are also strongly biased 
towards the brightest regions of their hosts (Fruchter et al. 2006), 
adding to the GRB-Ic SN link.

In the present study we re-assess the the degree of association of nearby 
ccSNe with H\,{\sc ii} regions in their hosts, in an attempt to reconcile 
the recent Smartt (2009) and Anderson et al. (2012) studies. 
Section~\ref{clusters} provides a background to H\,{\sc ii} regions, star 
clusters and massive stars. Section~\ref{sn_section} assesses
the association of nearby core-collapse SN with H$\alpha$ emission, 
while Section~\ref{implications} looks into how/whether these 
results contribute to the question of progenitor masses for different 
flavours of ccSNe. In Section~\ref{discussion} we briefly reassess the 
significance that certain ccSNe and long-duration GRBs are 
located in the brightest regions of their host galaxies, while
brief conclusions are drawn in Section~\ref{conclusions}.

\section{H\,{\sc ii} regions and massive stars}\label{clusters}

In this section we provide a brief background to the expected connections 
between massive stars and H\,{\sc ii} regions,  of relevance to our empirical 
study set out in Section~\ref{sn_section}.

\subsection{Clusters and massive stars}

It is widely accepted that the majority of stars form within star 
clusters (Lada \& Lada 2003), although recent evidence suggests star 
formation occurs across a broad continuum of stellar densities (e.g. Evans 
et al. 2009) from dense star clusters to diffuse OB associations (Gieles 
\& Portegies Zwart 2011).  Nevertheless, given their short-lifetimes 
(2.5--50 Myr) only a few percent of massive stars ($\geq 8 M_{\odot}$) 
appear genuinely `isolated' (de Wit et al. 2005) such that they either 
tend to be associated with their natal cluster or are potential runaways 
from it\footnote{Runaways may be ejected from their cluster either 
dynamically during the formation process or at a later stage after 
receiving a kick following a supernova explosion in a close binary 
system.}.

According to Weidner \& Kroupa (2006), there is a tight relation between 
cluster mass, and the most massive star formed within the cluster, 
although this remains controversial (Calzetti et al. 2010, Eldridge 2012). 
If this is so, the galaxy-wide stellar initial mass function (IMF) will 
also depend upon the cluster mass function and the mass range spanned by 
star clusters (Pflamm-Altenburg et al. 2007). By way of 
example, a cluster with a mass of $\sim\,10^{2}$ $M_{\odot}$, similar to 
the $\rho$ Oph star  forming region (Wilking et al. 1989), will barely 
produce any massive stars. In  contrast, a cluster with a  stellar mass of 
$10^{4}$ $M_{\odot}$ such as NGC~3603 (Harayama et al. 2008), would be 
expected to form $\sim$100 massive stars, with the most massive examples
exceeding 100 $M_{\odot}$ (Schnurr et al. 2008; Crowther et al. 2010).
Massive stars, therefore, tend to be intimately connected with the 
youngest, brightest star clusters. 

\subsection{H\,{\sc ii} regions: from classical to supergiant}

In view of the Salpeter IMF slope for high mass
stars (Bastian et  al. 2010), 8--20 $M_{\odot}$ early B-type stars
form $\sim$75\% of their overall statistics. However, 
the most  frequently used indicator of active star formation is nebular 
hydrogen  emission (e.g. H$\alpha$) from gas ionized by young, massive 
stars. The Lyman continuum ionizing output from such stars is a very sensitive 
function of temperature (stellar mass), such that one O3 dwarf ($\sim$75 
$M_{\odot}$) will emit more ionizing photons than 25,000 B2 dwarfs 
($\sim$9 $M_{\odot}$, Conti et al. 2008). Therefore, H\,{\sc ii} regions 
are biased towards the $\sim$25\% of high mass stars exceeding
20 $M_{\odot}$, namely O-type stars (B stars will produce extremely 
faint H\,{\sc ii} regions).

Beyond several Mpc, current sensitivities limit the detection of H\,{\sc 
ii} regions to relatively bright examples, involving several ionizing 
early O-type stars (Pflamm-Altenburg et al. 2007). Still, the H$\alpha$ 
luminosity of bright H\,{\sc ii} regions can be converted into the 
corresponding number of Lyman continuum ionizing photons, for which the 
number of equivalent O7 dwarf stars, N(O7V), serves as a useful reference 
(Vacca \& Conti 1992). Table~\ref{table2} lists examples of nearby
H\,{\sc 
ii} regions (adapted from Kennicutt 1984, 1998) which range from classical 
H\,{\sc ii} regions powered typically by one or a few  stars (e.g. M~42), 
through giant, extended H\,{\sc ii} regions powered by 
tens of O stars (Carina Nebula) to exceptionally bright `supergiant' 
regions powered by hundreds of O stars (30 Doradus). We will follow 
these template H\,{\sc ii} regions when we investigate the nebular 
environment of ccSNe in Section~\ref{sn_section}. Although there is a 
spread in H\,{\sc ii} region size at a particular H$\alpha$ luminosity 
(e.g. Lopez et al. 2011, their figure 1), faint regions are typically 
small ($\leq$10 pc), giant regions are extended ($\sim$100 pc) and 
supergiant regions tend to be very extended  (several hundred pc).

\begin{figure}
\begin{center}
 \includegraphics[width=6cm]{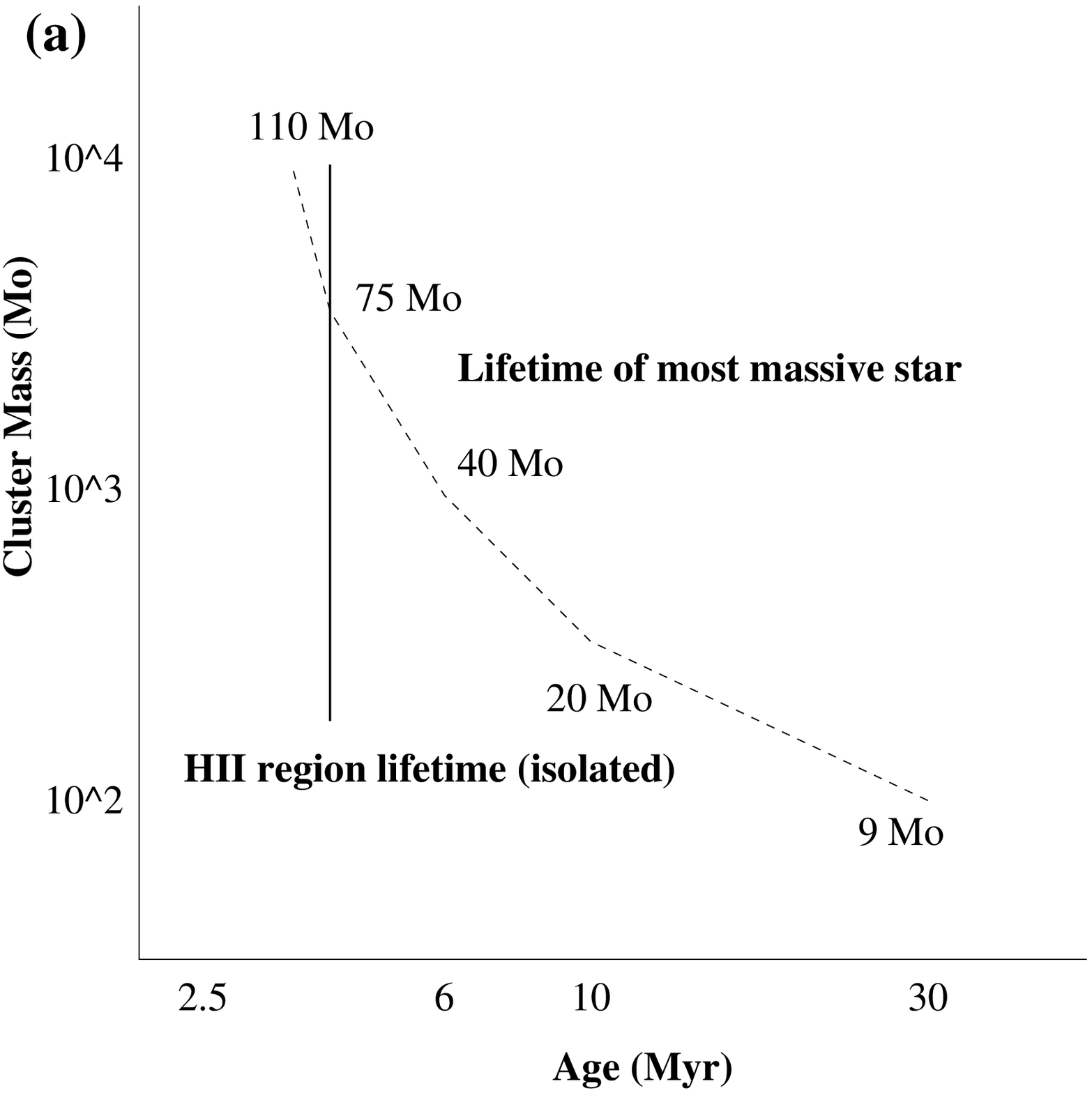} 
 \includegraphics[width=6cm]{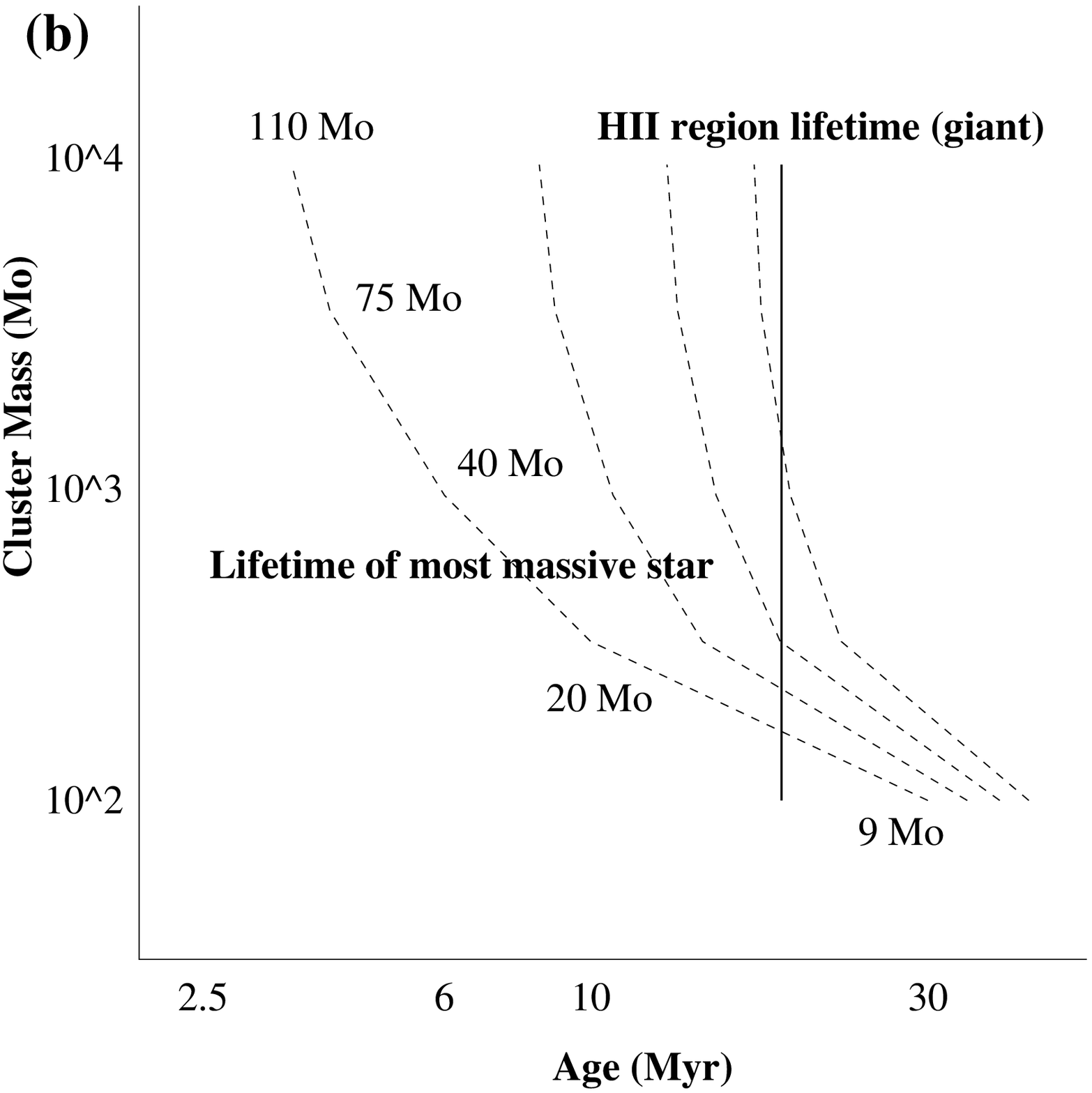} 
 \caption{
(a) Schematic comparing the lifetime of the most massive star in a cluster 
(dotted line, according to Pflamm-Eltenberg et al. 2007) and isolated 
H\,{\sc ii}  regions (vertical solid line, adapted from Walborn 2010). 
Core-collapse SNe should only be 
associated with isolated H\,{\sc ii} regions for very massive progenitors; 
(b) as (a) except for (super)giant H\,{\sc ii} regions, comprising
4 distinct star forming episodes separated by 5 Myr, with the age
referring to the first stellar generation. For a total duty cycle
of 20 Myr, an association between ccSNe and giant H\,{\sc ii} 
regions would be expected except for progenitors with masses below 
$\sim$12$M_{\odot}$ or more massive progenitors from subsequent
stellar generations as indicated.}
\label{fig1}
\end{center}
\end{figure}

Kennicutt et al. (1989) have studied the behaviour of the H\,{\sc ii} 
region luminosity function in nearby spirals and irregular galaxies. 
Early-type (Sa-Sb) spirals possess a steep luminosity function, with the 
bulk of massive star formation occuring in small regions ionized by one of 
a few O stars (M~42-like), plus a low cut-off to the luminosity function. 
Late-type spirals and irregulars possess a shallower luminosity function, 
in which most of the massive stars form within (30 Dor-like) large H\,{\sc 
ii} regions/OB complexes. For example, although the LMC contains 
considerably fewer H\,{\sc ii} regions than M~31 (SAb), it contains ten 
H\,{\sc ii} regions more luminous than any counterpart in M~31 (Kennicutt 
et al. 1989).

\subsection{Lifetime of H\,{\sc ii} regions}\label{lifetimes}

Before turning to our survey of nearby ccSNe, let us first assess the
empirically determined duration of the H\,{\sc ii} region phase, for which
both a plentiful supply of ionizing photons (from O stars) and neutral gas 
(left  over from  the star  formation process) are required. The former is 
limited to $\sim$10 Myr  according to the latest evolutionary model 
predictions for 20 $M_{\odot}$  stars (Ekstr\"{o}m et al. 2012), and is often
merely adopted as the H\,{\sc ii} lifetime, while the latter depends 
sensitively upon its environment.  Walborn (2010) studied the
properties of young, intermediate mass star clusters within the Local 
Group which indicate that the H\,{\sc ii} region phase is present for 
only the first $\sim$3--4 Myr. The gas is swept up and expelled via 
radiative and mechanical feedback from stars, and subsequently supernovae
(e.g. Dale et al. 2012). 

Gas has already been removed from relatively high mass, isolated clusters 
such as Westerlund 1 after 5 Myr ($\sim 10^{5} M_{\odot}$, Clark et al. 
2005). Therefore, one would {\it not} expect ccSNe to be spatially 
coincident with {\it isolated} H\,{\sc ii} regions unless the mass of the 
progenitor was sufficiently high ($>$75 $M_{\odot}$) for its 
lifetime to be comparable to the gas dispersion timescale.

This is illustrated in Figure~\ref{fig1}(a) where we compare the lifetime 
of the most massive stars in clusters\footnote{Stellar lifetimes are
adopted from rotating, solar metallicity models of Ekstr\"{o}m et al. (2012)}
(according to Eqn.~10 from Pflamm-Altenberg et al. 2007), with an estimate of 
the duration of isolated H\,{\sc ii} regions (adapted from Walborn 2010). 
Solely very massive ($> 75 M_{\odot}$) stars would end their life 
before the gas in the associated H\,{\sc ii} region had dispersed.


Of course, not all massive star formation occurs within isolated, compact 
star clusters. Giant and supergiant H\,{\sc ii} regions extend to several 
hundred parsec in size (Table~\ref{table2}), are ionized by successive 
generations of star clusters, with a total duty cycle of $\sim$20 Myr. 
Therefore, older (lower mass) populations will appear co-located with 
younger (higher mass) stars in external (super)giant H\,{\sc ii} regions. 
Indeed, 30 Doradus would only subtend $\sim$1 arcsec at a distance of 50 
Mpc.

For giant H\,{\sc ii} regions such as the Carina Nebula, several clusters 
exist with distinct ages, ranging from 1--2 Myr (Trumpler 14, e.g. Vazquez 
et al. 1996) to 5--10 Myr (Trumpler 15, e.g. Wang et al. 2011). Indeed, 
the proximity of the supergiant H\,{\sc ii} region 30 Doradus 
enables individual stars to be studied in detail (e.g. Evans et al. 2011). 
Walborn \& Blades (1997) identified five distinct spatial structures 
within 30 Dor, (i) the central 1--2 Myr cluster R136;  (ii) a surrounding 
triggered generation embedded in dense knots ($<$ 1 Myr); (iii) a OB 
supergiants spread throughout the region (4--6 Myr); (iv) an OB 
association to the southeast surrounding R143 ($\sim$5 Myr); (v) an older 
(20--25 Myr) cluster to the northwest (Hodge 301, Grebel \& Chu 2000). 

Therefore, 
a massive star from the first stellar generation exploding within such an 
environment as a SN after 5--10+ Myr would still be associated with a 
bright H\,{\sc ii} region, even if  its natal star cluster had cleared the 
gas from its immediate vicinity, as  illustrated in Fig.~\ref{fig1}(b). A 
total duty cycle of 20 Myr for giant  H\,{\sc ii} regions sets a 
lower threshold of 12 $M_{\odot}$ to the progenitor mass (from the initial 
stellar generation) for possible association with the giant H\,{\sc ii} 
region.  Progressively  higher mass  progenitors from subsequent generations 
(for illustration, four  generations separated by 5 Myr are shown in 
Fig.~\ref{fig1}b) would be  required. For example, a star formed 10 Myr
after the initial burst would only be associated with the giant H\,{\sc ii} 
region if its initial mass exceeded 20 $M_{\odot}$.

Finally, (nuclear) starburst regions  of galaxies, in which gas is 
continuously accreted, may have still longer  duty cycles of $\sim 100$ 
Myr, preventing any constraints upon progenitor masses.

In summary we set out five scenarios, depending on whether or not 
the ccSNe are spatially coincident with H\,{\sc ii} regions, as follows:
\begin{itemize}

\item \underline{Class 1:} The ccSNe progenitor is coincident with an 
isolated, young, bright star cluster from which the gas has 
been expelled, so a H\,{\sc ii} region is {\it absent}. At optical 
wavelengths, a star 
cluster rapidly fades (by 1 mag) at early times between 5--10 Myr, with a 
slower decline of 0.5 mag between 10--30  Myr (e.g. Bik et al. 2003), so a 
high mass progenitor ($\geq$20 $M_{\odot}$) might be anticipated;

\item \underline{Class 2:} Either the ccSNe progenitor is coincident 
with an isolated, faint star cluster, or is not coincident with any
detectable star cluster. In this case the cluster may have already 
dissolved and an H\,{\sc ii} region again {\it absent}. This scenario 
would favour an  older ($\gg$10 Myr) cluster, and correspondingly lower 
mass progenitor ($<$ 20 $M_{\odot}$);

\item \underline{Class 3:} The ccSNe progenitor was formed in a star 
cluster, but was subsequently ejected via either dynamical interactions or 
after receiving a kick from the supernova of a close companion, so it is 
not directly associated with a H\,{\sc ii} region, although a bright H\,{\sc ii}
region is {\it nearby}. 
For a dynamical interaction origin, if a massive star has a space velocity
of 50 km\,s$^{-1}$ with respect to its natal cluster, typical of a runaway, its 
projected distance would be up to 150 pc after  3 Myr,  or 1.5 kpc after 30 Myr. 
According to Fujii \& Portegies Zwart  (2011) the fraction of dynamical runaways 
is low ($<$0.1\%) for high mass (10$^{5}  M_{\odot}$) star clusters,  but 
increasing to 1--10\% for high mass stars  in less massive (10$^{4} M_{\odot}$) 
clusters.  Therefore, one would anticipate a nearby  (few hundred pc),  
high mass, dense star cluster, given the  short lifetime of 
high mass ($\geq 25 M_{\odot}$) stars, which would most likely lie within
a giant H\,{\sc ii} region (e.g. Carina Nebula, 30 Doradus).

\item \underline{Class 4:} A H\,{\sc ii} region is {\it present} at the 
ccSNe position, albeit with a low luminosity.  This favours a very high mass 
progenitor ($>$75 $M_{\odot}$) if the star forming region is 
compact/isolated, or a significantly lower mass ($\leq$20 $M_{\odot}$?) if 
it is merely an older, extended star forming region;

\item \underline{Class 5:} A H\,{\sc ii} region is again {\it present} at 
the ccSNe position, albeit spatially extended, with a high luminosity. 
The progenitor was formed in a  cluster within a 
large star forming complex in which ongoing 
star  formation is  maintaining the Lyman continuum radiation. The natal 
cluster may be detected, it could have 
already  dissolved, or the star could be a runaway from a nearby high mass 
cluster.  For a giant H\,{\sc ii} region duty cycle of 20 Myr, 
a lower  limit of 12 $M_{\odot}$ can be assigned for a progenitor
formed in the first generation of stars, with an increasing mass limit for 
subsequent generations (recall Fig.~\ref{fig1}b). 
If the ccSNe is associated with a very long-lived starburst 
region, no robust limit to the progenitor mass can be assigned.
\end{itemize}

\section{Are ccSNe associated with H\,{\sc ii} 
regions?}\label{sn_section}

In this section we discuss the supernova sample investigated, together 
with methods used and re-assess whether nearby ccSNe are associated with 
H\,{\sc ii} regions in their host galaxy.


\begin{table*}
  \begin{center}
  \caption{Properties of type Ib/c ccSNe used in the present study, 
including spectral  types (col 2), host galaxies (col 3), distance (col 4),
positions (col 5--6, from Asiago Catalogue
except where noted), deprojected galactocentric distances (col 7, $R_{\rm SN}$),
source of H$\alpha$ imaging (col 8, key in Table~5), whether the SN is
associated with a cluster (col 9), or nebular emission (col 10), 
including results from HST imaging in parentheses. Col 11 provides
information on the closest H\,{\sc ii} region, including its offset and
deprojected distance from the SN and whether it is spatially extended (ext.)
or compact (comp.). Cols 12--14 present the flux ($F$, $\leq \pm$0.1 dex), 
radius ($r$, $\pm$0$''$.5) and H$\alpha$ luminosity ($L$, factor of two)
of the H\,{\sc ii} region, while Col 15 shows the class of environment.}
  \label{type1_sn}
\begin{small}
\begin{tabular}{l@{\hspace{2mm}}l@{\hspace{2mm}}l@{\hspace{2mm}}r
@{\hspace{2mm}}l@{\hspace{2mm}}r@{\hspace{2mm}}l@{\hspace{2mm}}
l@{\hspace{2mm}}c@{\hspace{1mm}}c@{\hspace{2mm}}l@{\hspace{-3mm}}r@{\hspace{2mm}}c
@{\hspace{1mm}}r@{\hspace{4mm}}r}
\hline 
 SN & SN   &Host & $d$   & \multicolumn{2}{c}{SN (J2000)} & $R_{\rm SN}$ 
&Tel& Cl? & 
H$\alpha$? & Comment & 
$F$(H$\alpha$+[N\,{\sc ii}])& $r$(H\,{\sc ii}) & $L$(H$\alpha$) & Class\\
    & type &     & Mpc & \multicolumn{2}{c}{ $\alpha$, $\delta$} & $R_{\rm 
25}$& ID &\multicolumn{2}{c}{(HST)}& 
&erg\,s$^{-1}$\,cm$^{-2}$&arcsec&\multicolumn{2}{l}{$10^{37}$ 
erg\,s$^{-1}$}\\
  \hline
1983N$^{1}$ & Ib          & M 83     & 4.92 & 13 36 51.28 & --29 54 02.8 
& 0.48 & j1 & \tickNo & \tickYes & Ext. H\,{\sc ii} reg. 1$''$/25 pc E & $6 
\times 10^{-15}$ & 1.5 & 3 & 4\\ 
      &           &           &                  & & & & & (\tickNo)  
& (\tickNo?) & Twin GH\,{\sc ii} reg. 8$''$/180 pc SE & $1.4 \times 
10^{-13}$ & 5 & 80 \\
1985F & Ib/c        & NGC 4618  & 9.2\phantom{0} & 12 41 33.01 & +41 
09 
05.9 & 0.10 & i & \tickYes &
\tickYes  & Coincident w GH\,{\sc ii} reg. & $5 \times 
10^{-14}$ & 
2.5 & 46 &  5\\ 
      &           &           &                  & & & & &(\tickYes) &
 (\tickYes?) & & & & \\
1994I & Ic        & M~51a & 8.39 & 13 29 54.07 & +47 11 30.5 & 0.06 & h & \tickNo
& \tickYes & Ext. GH\,{\sc ii} reg. 2$''$/80 pc W & $1.0 \times 
10^{-14}$ & 2 & 13 &  5 \\  
      &           &           &                  &            &
&                      &    & (\tickNo)  & (\tickYes)
 & Ext. GH\,{\sc ii} reg. 9$''$/0.38 kpc SW & $2 \times 
10^{-14}$ & 2 & 30 \\
1997X  & Ib       & NGC 4691 & 12.0\phantom{0} & 12 48 14.28 &--03 
19 
58.5 & 0.14 & l1 
& \tickNo & \tickNo &H\,{\sc ii} reg. 1$''$.5/100 pc SW & -- & 
1 & -- & 3\\
      &           &           &                  &            
&                      &    &   & &
 & (Giant) H\,{\sc ii} reg. 6$''$/0.4 kpc SSW & -- & 3 & --\\
2002ap & Ic        & M 74 & 9.0\phantom{0} & 01 36 23.85 & +15 45 13.2 & 
0.89 & e1,n & \tickNo  & \tickNo 
& Ext. H\,{\sc ii} 10$''$/400 pc SE & {\it $2 \times 10^{-15}$} & 4 & 2 & 2\\ 
      &           &           &                  & & & & &(\tickNo) &
 (\tickNo) & & & & \\
2003jg & Ib/c      & NGC 2997 &11.7\phantom{0}& 09 45 37.91 & --31 11 21.0 
& 0.07 & f & \tickNo & \tickNo & 
Ext H\,{\sc ii} reg. 2$''$/140 pc E & $1.2 \times 10^{-15}$ & 
1.5 & 4 & 2\\ 
      &  &         &           &                  &            &    
&&(\tickNo)& & Nuclear starburst 12$''$/650 pc E & $7 \times 10^{-13}$ & 10 & 
2200 \\
2005at & Ic        & NGC 6744 & 11.6\phantom{0} & 19 09 53.57 & --63 49 
22.8 & 0.30 & j1 & \tickYes & \tickYes & Coincident with ext. H\,{\sc ii} reg. & $2 \times 10^{-15}$ & 1.5 & 
7 &  4\\ 
2005kl & Ic        & NGC 4369  & 11.2  & 12 24 35.68 & +39 23 03.5 & 0.17 & m
& \tickNo & \tickYes & (Giant?) H\,{\sc ii} reg. 1$''$.5/80 pc SE & -- & 2 
& -- &  5\\
2007gr & Ic        & NGC 1058  &  9.86 & 02 43 27.98 & +37 20 44.7 & 0.45 & i & 
\tickYes & \tickYes & 
Ext. H\,{\sc ii} reg. 1$''$.5/75 pc NE & $1.1 \times 10^{-14}$ & 2 & 
16 & 5 \\ 
      &           &           &                  &            & & & 
&(\tickYes) & (\tickYes?)  & Ext. GH\,{\sc ii} reg. 2$''$/100 pc W & $4 
\times 10^{-14}$ & 2.5 & 60 \\
2008eh & Ib/c?      & NGC 2997  & 11.3\phantom{0}  & 09 45 48.16 & --31 10 
44.9 & 0.50 & f & \tickNo & \tickYes & 
Edge of GH\,{\sc ii} reg. & $1.2 \times 
10^{-14}$  
& 2 & 40 &  5 \\ 
      &           &           &                  &            & & & & &
 & Ext. GH\,{\sc ii} reg. 4$''$.5/250 pc S & $2 \times 
10^{-14}$ & 2.5 & 70 \\
 \hline
  \end{tabular}   
\end{small}
  \end{center} 
\begin{small}
\begin{flushleft}
1: SN 1983N coordinates from Sramek et al. (1984)
\end{flushleft}
\end{small}
\end{table*}

\begin{table*}
  \begin{center}
  \caption{Properties of type II ccSNe used in the present study. 
Column headings are as in Table~2.}
  \label{type2_sn}
\begin{small}
\begin{tabular}{l@{\hspace{2mm}}l@{\hspace{2mm}}l@{\hspace{2mm}}r
@{\hspace{2mm}}l@{\hspace{2mm}}r@{\hspace{2mm}}l@{\hspace{2mm}}
l@{\hspace{2mm}}c@{\hspace{1mm}}c@{\hspace{2mm}}l@{\hspace{-3mm}}r@{\hspace{2mm}}c
@{\hspace{1mm}}r@{\hspace{4mm}}r
}
\hline 
 SN & SN   &Host & $d$   & \multicolumn{2}{c}{SN (J2000)} & $R_{\rm SN}$ & 
Tel& Cl? & 
H$\alpha$? & Comment & 
$F$(H$\alpha$+[N\,{\sc ii}])& $r$(H\,{\sc ii}) & $L$(H$\alpha$) & Class\\
    & type &     & Mpc & \multicolumn{2}{c}{ $\alpha$, $\delta$} & $R_{\rm 
25}$ & ID 
&\multicolumn{2}{c}{(HST)}& 
&erg\,s$^{-1}$\,cm$^{-2}$&arcsec&\multicolumn{2}{l}{$10^{37}$erg\,s$^{-1}$}\\
  \hline
%
1923A & II-P:      & M 83      & 4.92  & 13 37 09.2 &-29 51 04\phantom{.0} 
& 0.33 & j1 & \tickNo & \tickYes?&
Ext. GH\,{\sc ii} reg. $\sim 4''$/100 pc N & 
$3 \times 10^{-14}$ & 4 & 78  & 5\\ 
      &           &           &                  &            &                      
& &    & (\tickNo)  &
 (\tickYes) & &  &  &  &\\
%
%
%
1964H$^{5}$ & II        & NGC 7292  & 12.9\phantom{0} & 22 28 24.06 & +30 
17 23.3 & 0.51 & d2,l2 & 
\tickNo  & \tickNo & H\,{\sc ii} reg. 5$''$/0.35 kpc NW & $1.4 \times 
10^{-15}$ & 2 & 4 & 2\\ 
      &           &           &                  &            &                      
& &    & (\tickNo)  &
  &GH\,{\sc ii} reg 7$''$/0.5 kpc W  & $4 \times 10^{-15}$ & 2 & 13 &\\
1968D & II        & NGC 6946  & 7.0\phantom{0} & 20 34 58.32 & +60 09 34.5 
& 0.14 & k & \tickNo &
\tickNo & H\,{\sc ii} reg. 4$''$.3/150 pc SSE & $5 \times 10^{-15}$ & 2 
& 3 &  2\\ 
      &           &           &                  &            &                      
& &   & (\tickNo)  &
(\tickNo)  & Ext. H\,{\sc ii} reg. 12$''$/400 pc NNE & $1.1 \times 
10^{-14}$ & 3 & 
7 \\
1968L & II-P       & M 83      &  4.92 & 13 37 00.51 & --29 51 59.0& 0.02 & j1 & 
\tickNo & \tickYes & 
GH\,{\sc ii} reg. 1$''$/25 pc E & $3 \times 10^{-13}$ & 1 & 150 & 5\\ 
      &           &           &                  & & & & &(\tickNo) &
 (\tickYes) & Double GH\,{\sc ii} reg. 3$''$/75 pc N & $2 \times 10^{-12}$ 
& 2 
& 1200 \\
1970G$^{5}$ & II-L       & M 101     & 6.96 & 14 03 00.76 & +54 14 33.2 & 0.46 &
b,l2 
& 
\tickYes 
& \tickYes  & Edge of GH\,{\sc ii} reg. & $2 \times 
10^{-12}$ & 20 & 2070 & 5\\ 
      &           &           &                  & & & & &(\tickNo) &
 (\tickNo?) & &  & & & \\
1980K & II-L       & NGC 6946  & 7.0\phantom{0} & 20 35 30.07 & +60 06 
23.7 & 0.99& h & \tickNo  & \tickNo 
& H\,{\sc ii} reg. 60$''$/2.1 kpc WNW  & $6 \times 10^{-15}$ & 2 & 4 &  2\\ 
      &           &           &                  & & & & &(\tickNo) &
 (\tickNo?) & & & & \\
1986L$^{5}$  & II-L       & NGC 1559 & 12.6\phantom{0} & 04 17 29.4 & --62 
47 
01\phantom{.0} & 0.55 & j2 & \tickNo &
\tickNo & H\,{\sc ii} reg. 2$''$/150 pc N & $2 \times 10^{-15}$ & 1 
& 8 & 3 \\ 
      &           &           &                  &            &                      &    &   &
 & & GH\,{\sc ii} reg. 3$''$/230 pc SW & $3 \times 10^{-14}$ & 2 & 
95 \\
1987A & IIpec    & LMC & 0.05 & 05 35 28.01 & --69 16 11.6 & -- & a1,a2& 
\tickNo$^{\rm 1}$ & \tickYes 
& H\,{\sc ii} complex 2.75$'$/40 pc NW & {\it $1.3 \times 10^{-10}$} & 
2$'$ 
& 7 & 4\\ 
      &           &           &                  &            &                      &    &   &
 & & GH\,{\sc ii} reg. 20.5$'$/300 pc NE & {\it $1.4 \times 
10^{-8}$\phantom{0}} & 10.5$'$ & 735 \\
1992ba & II       & NGC 2082& 13.1\phantom{0} & 05 41 47.1 & --64 18 
00.9 & 0.55 &   &   & $^{\rm 2}$ \\
1993J & IIb       & M 81 & 3.6\phantom{0}  & 09 55 24.95 & +69 01 13.4 & 0.32 & h 
& \tickNo & SNR? & Coincident with faint emission & $1.1
\times 10^{-15}$ & 2 & 0.1 & 2\\ 
      &           &           &                  &            &                 &     
&    & (\tickNo)  & (SNR) 
 & Ext. H\,{\sc ii} reg. 21$''$/580 pc NE & $1.1 \times 10^{-14}$ & 3 
& 1.3 \\
1995V &  II-P       & NGC 1087  & 14.4 & 02 46 26.77 & --00 29 55.6 & 0.36 &
d2 & \tickNo? & \tickNo? & GH\,{\sc ii} reg. 5$''$/0.6 kpc SW & $5 
\times 10^{-15}$ & 3.5 & 18  & 2\\
1995X  & II        & UGC 12160 & 14.4 & 22 40 51.30 & +75 10 11.5 & 0.42 & &  
& $^{\rm 3}$ \\
1996cr & IIn:      & Circinus & 4.21 & 14 13 10.01 & --65 20 44.4 & 0.18 & c &
\tickNo & \tickYes & GH\,{\sc ii} reg. 3$''$/100 pc SE & $3 \times 
10^{-14}$ & 
4 & 31 &  5\\  
      &           &           &                  &            &                 &     
&    &(\tickNo)   &
(\tickYes) & Ext. GH\,{\sc ii} reg. 15$''$/0.5 kpc NNW & $7 
\times 10^{-14}$ & 5 & 67 \\
1998dn & II        & NGC 337A & 11.4\phantom{0} & 01 01 27.08 & --07 36 36.7 
& 1.21 & d1 &  \tickNo & \tickNo  & Bright H\,{\sc ii} reg. 5$''$.5/500 pc NW & 
$5 \times 
10^{-14}$ & 2.5 & 14 & 3\\ 
1999em$^{5}$ & II-P       & NGC 1637 & 9.77 & 04 41 27.05 & --02 51 45.8 & 0.22 &
e1 & 
\tickNo &
\tickNo & H\,{\sc ii} reg. 6$''$/300 pc SE & {\it $7 \times 10^{-16}$} & 2 
& 1.6 & 2
\\  
      &           &           &                  &            &                  &    
&    &(\tickNo)   &
 & H\,{\sc ii} reg. 9$''$.5/0.5 kpc SE & {\it $7 
\times 10^{-15}$} & 3 & 15 \\
1999eu & II-P       & NGC 1097 & 14.2\phantom{0} & 02 46 20.79 & --30 19 
06.1& 0.83 & e2 & \tickNo &
\tickNo & H\,{\sc ii} reg. 3$''$.75/375 pc W & {\it $8 \times 10^{-16}$} & 
2 & 8 &  2\\ 
1999gi & II-P       & NGC 3184 & 13.0\phantom{0} & 10 18 16.66 & +41 26 
28.2 & 0.28 & h & \tickNo? &
\tickYes & Ext. H\,{\sc ii} reg. 2$''$/125 pc SW & $3 \times 
10^{-15}$ 
& 2 & 8 & 5\\ 
      &       &    &           &                  &            & & 
&(\tickNo) & (\tickNo?) & Ext. H\,{\sc ii} reg. 2$''$/125 pc NE & $9 
\times 
10^{-15}$ & 3 & 22 \\
2001X$^{5}$  & II-P       & NGC 5921  & 14.0 & 15 21 55.46 &+05 03 43.1 & 0.34
&d2 
& 
\tickNo 
& \tickNo & Diffuse H\,{\sc ii} reg. 3$''$/0.3 kpc SE & $3 \times 
10^{-15}$ & 2 & 12  & 3\\
      &           &           &                  &            & & & 
&&  & Ext. H\,{\sc ii} reg. 4$''$/0.4 kpc N & $9 \times 
10^{-15}$ & 2 & 36 \\
2001ig & IIb       & NGC 7424  & 7.94 & 22 57 30.69 & --41 02 25.9 & 1.02 & 
e1 & \tickNo & \tickYes & Edge of ext. H\,{\sc ii} reg. & {\it $1.5 
\times 10^{-15}$} & 2.5 & 2 & 4\\ 
2002hh$^{5}$ & II-P       & NGC 6946 & 7.0\phantom{0} & 20 34 44.25 & +60 
07 19.4 & 0.38 & k & \tickNo & \tickYes 
& Ext. GH\,{\sc ii} reg. 2$''$.5/85 pc NW & $4 \times 
10^{-14}$ & 4 & 26 & 5\\ 
      &           &           &                  & & & & &(\tickNo) &
 (SNR)&  & & & \\
2003B &  II-P       & NGC 1097 & 14.2\phantom{0} & 02 46 13.78 & --30 13 
45.1& 1.04 & e2 & \tickNo & \tickYes 
& Edge of ext. GH\,{\sc ii} reg. & $6 
\times 10^{-15}$ & 3 & 58 & 5\\ 
2003gd & II-P       & M 74  & 9.0 & 01 36 42.65 & +15 44 19.9 & 0.51 & j1 & 
\tickNo & \tickNo &
H\,{\sc ii} reg. 7$''$/300 pc SW & $2 \times 10^{-15}$ & 1.5 
& 2 & 2\\ 
      &           &           &                  &            &    & 
&&(\tickNo)&
(\tickNo?) & Ext. GH\,{\sc ii} reg. 12$''$/500 pc SW&$2 \times 10^{-14}$ 
& 5 & 20 \\
2004dj & II-P       & NGC 2403 & 3.16 & 07 37 17.02 & +65 35 57.8 & 0.34 &
h$^{4}$ & \tickYes & \tickNo & Ext. H\,{\sc ii} reg. 21$''$/450 pc 
NW & $4 
\times 
10^{-14}$ & 2.5 & 6 & 1\\ 
      &           &           &                  &            &   &&&&
 & Ext. GH\,{\sc ii} reg. 21$''$/450 pc SE & $5 \times 
10^{-13}$ & 7.5 & 73 \\
2004et & II-P       & NGC 6946 & 7.0\phantom{0} & 20 35 25.33 & +60 07 
17.7 & 0.82 & h & \tickNo  &
\tickNo & Ext. H\,{\sc ii} reg. 9$''$/300 pc N & $3 \times 
10^{-15}$ & 
2 & 2 & 2\\ 
      &           &           &                  & & & & &(\tickNo) &
 (\tickNo?) & & & & \\
2005cs & II-P       & M~51a & 8.39 & 13 29 52.85 & +47 10 36.3 & 0.30 & h & 
\tickYes? & 
\tickYes  & Ext. H\,{\sc ii} reg. 1$''$/40 pc E & $1.4 \times 
10^{-15}$ & 
1.5 & 2  & 4\\ 
      &           &           &                  &            & & & & 
(\tickNo) &
(\tickNo) & Ext. GH\,{\sc ii} reg. 13$''$/0.55 kpc E & $1.0 
\times 10^{-14}$ & 2 & 13 \\
2008bk & II-P       & NGC 7793  & 3.61  & 23 57 50.42 & --32 33 21.5 & 0.75 & e2 
& 
\tickNo & \tickNo & 
H\,{\sc ii}  reg. 7$''$/200 pc SW & {\it $1.8 \times 10^{-14}$} & 3 & 5
& 2 \\ 
2009N & II-P        & NGC 4487 &  11.0\phantom{0}  & 12 31 09.46 & --08 02 
56.3 & 0.77 & d1
& \tickNo & \tickNo & Comp. H\,{\sc ii} reg, 3$''$/200 pc NW 
& $8 \times 10^{-16}$ & 1.5 & 2  &  2\\
      &           &           &                  &            & & & &&
 & Ext. H\,{\sc ii} reg. 3$''$/200 pc NE & $3 \times 10^{-15}$ & 2.5 & 
6 \\
2009ib & II-P       & NGC 1559  & 12.6\phantom{0}  & 04 17 39.92 & --62 46 
38.7 & 0.67 & j2 
& \tickNo & \tickNo & Ext. H\,{\sc ii} reg. 1$''$.5/170 kpc SE & $3 
\times 
10^{-15}$ & 1 & 9 & 2\\ 
      &           &           &                  &            & & & & 
(\tickYes) & (\tickNo?) & Ext. GH\,{\sc ii} reg. 6$''$/0.7 kpc SW & $1.3 
\times 10^{-14}$ & 1.5 & 40 \\
2011dh & IIb       & M~51a & 8.39 & 13 30 05.12 & +47 10 11.3 & 0.50 & h & \tickNo 
& \tickNo & 
Bright H\,{\sc ii} 8$''$/0.35 kpc SE & $5 \times 10^{-15}$ & 2 & 7 &  2\\ 
      &           &           &                  &            & & & &  
(\tickNo) &
(\tickNo) & Ext. GH\,{\sc ii} reg. 11$''$/0.5 kpc NE & $1.0 
\times 
10^{-14}$ & 3 & 14 \\
2012A  & II-P       & NGC 3239 & 10.0\phantom{0} & 10 25 07.39 &+17 09 
14.6 
& 0.42 & g & \tickNo & \tickYes & 
Edge of H\,{\sc ii} reg. & $6 \times 10^{-15}$ & 2  & 
9 & 4\\ 
      &           &           &                  &            & & &  & & 
 & Ext. GH\,{\sc ii} reg. 10$''$/0.6 kpc NE & $4 \times 
10^{-13}$ & 4 & 525 \\
2012aw & II-P  
      & M\,95 & 10.0\phantom{0} & 10 43 53.76 & +11 40 17.9 & 0.62 & h & \tickNo 
& \tickNo 
& H\,{\sc ii} reg. 5$''$/260 pc NNE & $4 \times 10^{-16}$ & 2  & 0.6 &  2 \\ 
      &           &           &                  &            & & & 
&(\tickNo)  &
 & H\,{\sc ii} reg. 10$''$/525 pc SW & $1.2 \times 
10^{-15}$ & 2 & 2 \\
 \hline
  \end{tabular}   
\end{small}
  \end{center} 
\begin{small}
\begin{flushleft}
1: SN 1987A is associated with a faint cluster (Panagia et 
al. 2000) that would be not detected at the  typical distance of the other 
ccSNe.\\
2: SN 1992ba is located within or close to a bright H\,{\sc ii} region 
according to Schmidt et al. (1994)\\
3: SN 1995X is located close to the maximum H$\alpha$ brightness of UGC 
12160 according to Anderson et al. (2012) \\
4: KPNO 2.1m H$\alpha$ imaging of Kennicutt et al. (2003) is 
supplemented
by continuum subtracted NOT/ALFOSC H$\alpha$ imaging from Larsen \& 
Richtler (1999)\\
5: Coordinates: SN 1964H (Porter 1993); SN 1970G (Allen et al. 1976); SN 
1986L (McNaught \& Waldron 1986); SN 1999em (Jha et al. 1999); SN 2001X 
(Li et al. 2001); SN 2002hh (Stockdale et al. 2002)
\end{flushleft}
\end{small}
\end{table*}

\subsection{Supernova sample}

Here, we examine the association of ccSNe with H\,{\sc ii} regions in 
their host galaxies. We follow an approach broadly similar to van Dyk 
(1992) and van Dyk et al. (1996). This  technique in complementary to  
qualitative approaches (Crockett 2009; Smartt, priv. comm.)  and the 
cumulative 
distribution technique of James \& Anderson (2006), Anderson \& James 
(2008) and Anderson et al. (2012).

We limit our sample to local, historical, non-type Ia SNe
from the Asiago Catalogue\footnote{http://graspa.oapd.inaf.it}
with a cutoff date of 31 Mar 2012.  Ground-based images  used in this 
study have  a spatial resolution of 0.6--4  arcsec, aside from the LMC, so 
we set an upper  limit of 15 Mpc for ccSNe host galaxies, at which the 
typical image quality (FWHM $\sim$ 1$''$.5) corresponds to the radius of a giant 
H\,{\sc ii}
region ($\sim$100 pc). Distances are uniformly  obtained from the 
Extragalactic Distance Database (EDD, Tully et al. 2009).  For reference, 
Tables~\ref{more_hosts1}-\ref{more_hosts2} in Appendix B lists a ccSNe  host galaxies for 
which EDD distances  lie in the 15--20 Mpc range.

In total 88 ccSNe within 15 Mpc are listed in the Asiago Catalogue,
from which 11 were removed on the basis that they are believed to be SN 
imposters or LBV
eruptions\footnote{SN 1954J, 1961V, 1978K, 1997bs, 1999bw, 
2000ch, 2002kg, 2002bu, 2008S, 2010da and 2010dn}. We also omit 21 ccSNe 
for which merely offsets 
relative to the centre of host galaxy are known, although we retain
historical SNe whose positions are known to a precision of $\sim$1
arcsec. 



Finally, we exclude 15 additional SNe whose host galaxies are
observed at high  inclination ($\geq$65$^{\circ}$)  owing to
the potential for confusion with unrelated 
line-of-sight H\,{\sc ii} regions. Inclinations were obtained 
from HyperLeda\footnote{http://leda.univ-lyon1.fr}, such that
41 ccSNe meet our criteria. Basic properties of these ccSNe are
listed in Tables~\ref{type1_sn} and \ref{type2_sn} for type Ib/c and type 
II ccSNe, respectively, with positions adopted from Asiago except
where noted. 

Uncalibrated H$\alpha$ images were available for NGC 4369 and 
NGC 4691, while no H$\alpha$ observations of either NGC 2082 or UGC 12160 
were available. Therefore, the final sample comprises 39 SNe, subdivided 
into 29 type II and 10 type Ib/c ccSNe.  Basic properties of the SN host 
galaxies are presented in  Table~\ref{hosts},  which include high 
inclination galaxies (shown in bold) plus hosts lacking accurate ccSNe 
coordinates (shown in italics).

\begin{table*}
  \begin{center}
  \caption{Basic properties of host galaxies of ccSNe used in the present 
study (within 15 Mpc), drawn from RC3 or HyperLeda. Hosts viewed at 
unfavourable high inclinations ($\geq$65$^{\circ}$) were excluded from the 
study, but are listed (in bold) separately, as are hosts of ccSNe excluded 
owing to imprecise SN positions (in italics).}
  \label{hosts}
  \begin{tabular}{r@{\hspace{2mm}}
r@{\hspace{2mm}}r@{\hspace{2mm}}
r@{\hspace{3mm}}l@{\hspace{1mm}}
r@{\hspace{3mm}}r@{\hspace{3mm}}
r@{\hspace{3mm}}r@{\hspace{1mm}}
c@{\hspace{1mm}}r@{\hspace{3mm}}
l@{\hspace{3mm}}r@{\hspace{2mm}}
r@{\hspace{3mm}}r@{\hspace{3mm}}
l}
\hline 
PGC & M & NGC & UGC & Type & $cz$        & $i$ & PA& 
\multicolumn{3}{c}{$d$} & Ref & $m_{\rm B_{T}}$ & $A_{\rm 
B}$ & $M_{\rm B_{T}}$ & ccSNe\\
    &  &     &     &      & km\,s$^{-1}$&     & & \multicolumn{3}{c}{Mpc} 
&   mag         & mag         
& mag         & \\
  \hline
03671 &    & 337A  &    & SAB(s)dm & 1074    & 56.1& 8& 11.4\phantom{00} 
&$\pm$& 2.1 & l  & 12.70\phantom{:} & 0.35
& --17.94\phantom{:} & 1998dn \\
05974 & 74 & 628 & 01149 & SA(s)c & 657     & 6.5$^{o}$&25$^{o}$ & 
9.0\phantom{00} && & 
l  & 9.95\phantom{:} & 0.25 & --20.07\phantom{:} & 2002ap, 2003gd \\
10314 &    & 1058 & 02193 & SA(rs)c & 518     & 58.5& 90.4& 9.86\phantom{0} 
&$\pm$& 0.61 & l  & 11.82\phantom{:} & 
0.22 & --18.37\phantom{:} & {\it 1969L}, 2007gr \\ 
10488 &    & 1097 &   & SB(s)b  & 1271    & 55.0& 138.2 & 
14.2\phantom{00} & 
$\pm$ & 2.6 
& l  & 10.23\phantom{:}  & 0.10  & --20.63\phantom{:}  & {\it 1992bd}, 1999eu, 2003B\\
10496 &    & 1087 & 02245 & SAB(rs)c & 1517 & 54.2& 12 & 14.4\phantom{00} 
& 
$\pm$ & 1.8 & 
l & 11.46\phantom{:}   & 0.12  & --19.46\phantom{:}  & 1995V \\ 
14814 &    & 1559 &      & SB(s)cd & 1304    & 60.2&62.8 & 
12.6\phantom{00} 
&$\pm$& 2.5 & l & 11.00\phantom{:} & 0.11 & 
--19.61\phantom{:} & {\it 1984J}, 1986L, 2009ib \\
15821 &    & 1637 &      & SAB(rs)c & 717    & 31.1& 16.3& 
9.77\phantom{0} 
&$\pm$& 1.82 & l  & 11.47\phantom{:} & 0.15 
& --18.63\phantom{:} & 1999em \\
17223 & \multicolumn{3}{c}{----- LMC -----} &SB(s)m &278 & 35.3 &170& 
0.050 
&$\pm$&0.002 & m 
& 0.91\phantom{:} & 0.27 & --17.86\phantom{:} & 1987A \\
17609 &     & 2082 &       &SB(r)b     & 1184 & 26.2& -- & 
13.1\phantom{00} & 
$\pm$ & 1.8 
& l & 12.62\phantom{:}   & 0.21  & --18.18\phantom{:}  & 1992ba \\ 
21396 &     & 2403 & 03918 & SAB(s)cd    & 131 & 61.3& 126& 
3.16\phantom{0} 
&$\pm$& 0.16 & c,d,k,l & 8.93\phantom{:}
&  0.14    & --18.74\phantom{:}       & 2004dj \\ 
27978 &    & 2997 &       & SA(s)c    & 1088 & 54.3 &96.6& 
11.3\phantom{00} 
&$\pm$& 0.8 & l & 10.06\phantom{:} & 0.39
& --20.59\phantom{:} & 2003jg, 2008eh \\
28630 & 81  & 3031 & 05318 &  SA(s)ab         & --34 & 62.7& 157 & 
3.65\phantom{0} &$\pm$& 0.18 & a,d,e,k,l 
& 7.89\phantom{:}   & 0.29   & --20.18\phantom{:}   & 1993J \\
30087 &    & 3184 & 05557 & SAB(rs)cd & 592 &  14.4& 135 & 
13.0\phantom{00}&& 
& l & 10.36\phantom{:} & 0.06 & --20.27\phantom{:} & {\it 
1921B}, {\it 1937F}, 1999gi \\ 
30560 &    & 3239 & 05637 & IB(s)m pec & 753 & 46.8& - & 
10.0\phantom{00}&& 
& l & 11.73\phantom{:} & 0.12 & --18.39\phantom{:} & 2012A \\
32007 & 95 & 3351 & 05850 & SB(r)b & 778 & 54.6& 9.9& 10.0\phantom{00} 
&$\pm$& 1.0 & c,d,h,l & 10.53\phantom{:} & 
0.10 & --19.76\phantom{:} & 2012aw \\
40396 &     & 4369 & 07489 & (R)SA(rs)a&1045 & 18.9&- & 11.2\phantom{00} 
& 
$\pm$ & 1.1 & l & 12.33\phantom{:}  & 0.09  & --18.01\phantom{:}  & 
2005kl\\
41399 &     & 4487 &       & SAB(rs)cd & 1034 & 58.2& 74.2 & 
11.0\phantom{00} & 
$\pm$ & 0.8 & l&  11.69:  &  0.08    & --18.59:      
& 2009N\\ 
42575 &    & 4618 & 07853 & SB(rs)m & 544& 57.6&40.2 & 9.20\phantom{0} 
&$\pm$& 
0.57 & l & 11.22\phantom{:} & 0.08 & 
--18.67\phantom{:} & 1985F \\
43238 &     & 4691  &      & (R)SB0/a(s) pec & 1110 & 38.8& 28.0 & 
12\phantom{.000} &  &  & l 
& 11.66\phantom{:}  &0.10   & --18.84\phantom{:}  & 1997X \\
47404 & 51a & 5194 & 08493 & SA(s)bc pec & 463 & 32.6&163.0 & 
8.39\phantom{0} 
&$\pm$& 0.84 & b,l & 8.96\phantom{:} & 
0.13 & --20.79\phantom{:} & 1994I, 2005cs, 2011dh \\
48082 & 83 & 5236 &        & SAB(s)c & 513 & 14.1&45$^{p}$ & 
4.92\phantom{0} 
&$\pm$& 0.25 & g,l & 8.20\phantom{:} & 0.24
& --20.29\phantom{:} & 1923A, 1968L, 1983N \\
50063 & 101 & 5457 & 08981 & SAB(rc)cd & 241 & 16& & 6.96\phantom{00} 
&$\pm$& 0.35 & 
b,d,h,l & 8.31\phantom{:} 
& 0.03 & --20.99\phantom{:} & {\it 1909A}, {\it 1951H}, 1970G \\
50779 & \multicolumn{3}{c}{----- Circinus -----}&SA(s)b? & 434 & 64.3& 
36.1& 4.21\phantom{0} &$\pm$& 0.78& l &    12.10\phantom{:}
& 2.00  & --18.02\phantom{:}   & 1996cr \\
54849 &     & 5921 & 09824 & SB(r)bc  & 1480 & 49.5 & 
140.0 &14.0\phantom{00} & 
$\pm$ & 3.2 & l &  11.49\phantom{:} & 0.15  & --19.39\phantom{:}  &  
2001X\\
62836 &     & 6744 &       & SAB(r)bc & 841 & 53.5 & 
15.4 &11.6\phantom{00} 
&$\pm$& 0.9 & l & 9.61\phantom{:} & 0.16 & 
--21.34\phantom{:} & 2005at \\
65001 &     & 6946 & 11597 & SAB(rs)cd & 40 & 18.3 & -- 
&7.0\phantom{00}&& & 
l & 9.61\phantom{:} & 1.24 & --20.86\phantom{:} & {\it 1917A}
{\it 1948B}, 1968D\\
       &     &      &       &           &     &     &    && &   &     &      
&        &&
1980K, 2002hh, 2004et \\ 
68941 &     & 7292 & 12048 & IBm       & 986 & 54.5 & 101.0&12.9\phantom{00} 
&$\pm$& 1.0 & l & 13.03\phantom{:} & 0.23 
& --17.75\phantom{:}& 1964H \\
69470 &     &      & 12160 & Scd?      & 1555 & 38.1 &14.8 &14.4\phantom{00} & 
$\pm$ & 3.0 
& l &  14.85:  &  2.04  &  --17.98:  & 1995X \\
70096 &     & 7424 &       & SAB(rs)cd & 939 & 59 & &7.94\phantom{0} 
&$\pm$& 0.77 & l & 10.96\phantom{:} & 0.04 & 
--18.56\phantom{:} & 2001ig \\
73049 &     & 7793 &        &  SA(s)d         & 227  & 53.7$^{n}$ 
&99.3$^{n}$ & 
3.61\phantom{0} &$\pm$& 0.18 & l & 9.63\phantom{:}   &  0.07    &  
--18.42\phantom{:}  & 2008bk \\
\hline
02052 &    & 150   &    & SB(rs)b? & 1584 &  {\bf 66.9} & &
14.9\phantom{00} & $\pm$ & 2.2 & l & 12.00\phantom{:} & 0.05  & --18.92\phantom{:}  & 
1990K \\
09031 &     & 891 & 01831 & SA(s)b? edge& 528 & {\bf 90} &   & 
9.91\phantom{0} &$\pm$& 0.5 & c,e,l &  10.81\phantom{:}  & 0.24   & 
--19.47\phantom{:}      
& 1986J \\
10329 &    & 1073 &02210 & SB(rs)c & 1208 & 52.3 & &12.3\phantom{00} & 
$\pm$ & 1.7 & l & 11.47\phantom{:} & 0.14  & --19.12\phantom{:}  & {\it 
1962L} \\
12286 &    & 1313 &      & SB(s)d  & 470     & 34.8& & 4.25\phantom{0} 
&$\pm$& 0.21 & f,l  &  9.20\phantom{:} & 0.40 & --19.25\phantom{:} & {\it 
1962M}\\
22338 & \multicolumn{3}{c}{--- ESO 209-G009 --- } & SB(s)cd? edge& 
1119 & {\bf 90} &
& 13.4\phantom{00} & $\pm$ & 1.0 & l & 12.68: & 0.94  & --18.89: & 2005ae 
\\
26512 &     & 2841 & 04966 &  SA(r)b?         & 638 & {\bf 65.2} & &
14.1\phantom{00}
 &$\pm$& 1.4 & l 
& 10.09\phantom{:}   &  0.06    & --20.71\phantom{:}       &  {\it 1972R} \\
28655 & 82  & 3034 & 05322 & I0 edge           & 203 & {\bf 76.9}& & 
3.52\phantom{0} &$\pm$& 0.18 & k,l & 9.30\phantom{:}   & 0.58     
&   --18.91\phantom{:}    & 2004am, 2008iz \\
30197 &     &3198 & 05572  &  SB(rs)c         & 663 & {\bf 77.8} & &
13.8\phantom{00} &$\pm$& 1.4 & l &  10.87\phantom{:}  &  0.05    
&--19.87\phantom{:}        
& {\it 1966J} \\ 
33408 &     & 3510 & 06126 & SB(s)m  edge  & 713 & {\bf 78.1} & &
14.7\phantom{00}& $\pm$& 1.7 & l  & 14.30\phantom{:}   & 0.11     
&--16.65\phantom{:}  
& 1996cb \\
34030 & 108 & 3556 & 06225 &   SB(s)cd edge & 699 & {\bf 67.5} & &
9.55\phantom{0} &$\pm$& 1.26 & l &  10.69\phantom{:}  &  0.06    
&  --19.27\phantom{:} & {\it 1969B} \\
34695 & 66  & 3627 & 06346 &  SAB(s)b         & 727 & {\bf 67.5} & &
8.28\phantom{0} &$\pm$& 0.41 & c,d,l & 9.65\phantom{:}  & 0.12    
&  --20.48\phantom{:}      & {\it 1973R}, 2009hd \\ 
39225 &    & 4214 & 07278 & IAB(s)m & 291 & 43.7 && 2.87\phantom{0} 
&$\pm$& 0.14 & k,l & 10.24\phantom{:} & 0.08 & --17.25\phantom{:} & 
{\it 1954A}\\
%
39600 & 106 & 4258 & 07353 &  SAB(s)bc         & 448 & {\bf 68.3}& & 
7.61\phantom{0} &$\pm$& 0.38 & c,e,i,j,l &   9.10\phantom{:} 
& 0.06   &  --20.28\phantom{:}   & 1981K \\
41333 &     & 4490 & 07651 &  SB(s)d pec & 565 & {\bf 79} & &
9.20\phantom{0} &$\pm$& 0.57 & l & 10.22\phantom{:}  & 0.08   &   --19.68\phantom{:} &  
{\it 1982F}, 2008ax \\
42002 &     & 4559 & 07766 & SAB(rc)cd  & 807 & 64.8 & &8.67\phantom{0} 
&$\pm$& 0.57 & l &10.46\phantom{:}    & 0.06     &  --19.29\phantom{:}      
& {\it 
1941A}\\
43451  &     & 4725 & 07989 & SAB(r)sb pec&1206 & 45.4& & 12.4\phantom{00} 
& $\pm$ & 
1.2 & l & 10.11\phantom{:}  &  0.04 &  --20.40\phantom{:}  & {\it 1940B} \\
45279 &     & 4945 &       &  SB(s)cd? edge & 563 & {\bf 90}&   & 
3.36\phantom{0} &$\pm$& 0.17  & l  &  9.30\phantom{:}  & 0.64    
&   --19.25\phantom{:}   & 2005af, 2011ja \\
51106 &     & 5530 &       &  SA(rs)bc         & 1194 & {\bf 66.5}& & 
11.8\phantom{00} &$\pm$& 2.2 & l  & 11.78:   & 0.42    
&  --19.00\phantom{:} & 2007it \\
67671 &     &      & 11861 & SABdm & 1481 & {\bf 75}& & 14.4\phantom{00} & 
$\pm$ & 3.0 
& l &  14.20: & 2.19  & --18.78:  & 1995ag, 1997db \\
68618 & \multicolumn{3}{c}{----- IC 5201 -----}&SB(rs)cd & 915 & {\bf 
66.7}& & 9.20\phantom{0} &$\pm$& 1.71 & l &  11.30:  & 
0.04 & --18.56:   & {\it 1978G} \\
69327 &     & 7331 & 12113 & SA(s)b    & 816  & {\bf 70} & &
14.7\phantom{00} &$\pm$& 1.5   & l  &  10.35\phantom{:}   & 0.33    & --20.82\phantom{:}
&1959D\\
71866 &     & 7713 &       & SB(r)d?         & 692 & {\bf 65.9}& & 
9.95\phantom{0} &$\pm$& 2.07 & l & 11.51\phantom{:}    & 0.06     
&--18.54       & 
{\it 1982L} \\
\hline
  \end{tabular}   
  \end{center}
\begin{small}
\begin{flushleft}
a: Ciardullo et al. (1993); 
b: Feidmeier et al. (1997); 
c: Ciardullo et al. (2002); 
d: Freedman et al. (2001); 
e: Tonry et al. (2001); 
f: M\'{e}ndez et al. (2002); 
g: Thim et al. (2003); 
h: Sakai et al. (2004); 
i: Macri et al. (2006); 
j: Mager et al. (2008); 
k: Dalcanton et al. (2009); 
l: Tully et al. (2009);
m: Schaefer (2008);
n: Carignan \& Puche (1990);
o: Kamphuis \& Briggs (1992)
p: Danver (1942)
\end{flushleft}
\end{small}
\end{table*}

\begin{table*}
  \begin{center}
  \caption{Source of ground-based H$\alpha$ and continuum imaging used in 
this study.}
  \label{imaging}
  \begin{tabular}{lllrrcccl}
\hline 
Tel & Telescope &Instrument & CCD scale & FWHM & 
\multicolumn{2}{c}{----- H$\alpha$ -----} &  Continuum & Reference \\
 ID &                    & & arcsec/pix & arcsec & $\lambda_c$ (\AA) & 
FWHM 
(\AA) & & 
\\
  \hline
a1&Nikon Survey Camera &2K CCD & 12.0\phantom{00} & 30 & 6570 & 15 & HaC 
(6676, 55) & Murphy \& Bessell (2000)\\
a2&Parking Lot Camera & 800$\times$800  & 36.9\phantom{00} & 80 & 6571 & 
14 & R & Kennicutt et al. (1995)\\ 
b & KPNO 0.6m Schmidt &Tek 2K CCD& 2.03\phantom{0} & 4.4 & 6573 & 67 & R & Hoopes et 
al. (2001)\\
c & CTIO 0.9m &Tek 2K CCD& 0.792 & $\sim$3.0 &6563 & 75 & R & Kennicutt et 
al. (2008) \\
d1 & JKT 1.0m & CCD & 0.241  & 1.5--2 &6594 & 44 & R & Knapen et al. 
(2004) \\
d2 &          &      &  0.331 & 1.5--3 & 6594 & 44 & R & James et al. (2004) \\
e1& CTIO 1.5m & CCD & 0.434 & 1.0 &6568 & 30 & R & Meurer et al. (2006) \\
e2&           &      &       & 1.2 & 6568 & 20 & R & Kennicutt et al. 
(2003) \\
f & Danish 1.54m &DFOSC & 0.400 & 1.8 &6565  & 114 & R & Larsen \& Richtler (1999)\\
g & VATT 1.8m &CCD     & 0.400 & $\sim$1.8 &6580 & 69 & R & Kennicutt et 
al. (2008)\\
h & KPNO 2.1m &CFIM  &  0.305 & 1.0 & 6573 & 67 & R & Kennicutt et al. 
(2003)\\
i & Bok 2.3m & CCD21  & 0.432 & $\sim$1.5 & 6575 & 69 & R & Kennicutt et 
al. (2008) \\
j1$^{1}$& VLT 8.0m & FORS1+2       & 0.126 & 0.8 & 6563 & 61 & HaC (6665, 
65)  & Hadfield et 
al. (2005)\\
j2$^{2}$& & & 0.200 & 0.8 & 6563 & 61 & V  &  \\
k$^{3}$ & Gemini-N 8.1m & GMOS     & 0.145 & 0.6 & 6560 & 70 & HaC (6620, 
60) & 
\\
l1 & INT 2.5m & WFC & 0.333 & 1.8 & 6568 & 95 & R & Anderson \& James 
(2008)\\
l2 &          &      &       & 1--2  & 6568 & 95 & HaC (6657, 79), r & \\
m & LT 2.0m & RATCam & 0.278 & 1.2 & 6557 & 100 & r$'$ & Anderson \& James 
(2008)\\
n & CFHT 3.6m & CFH12K & 0.206 & 0.9 & 6584 & 96 & R & Crockett et al. 
(2007)\\
  \hline
  \end{tabular}   
  \end{center}
\begin{small}
\begin{flushleft}
1: 067.D-0006(A), 069.B-0125(A), 380.D-0282(A), 081.B-0289(C); 2: 
075.D-0213(A); 3: GN-2009B-Q-4
\end{flushleft}
\end{small}
\end{table*}

\subsection{H$\alpha$ datasets of ccSNe host galaxies}


We have examined the immediate environment of these 41 ccSNe in their
30 host galaxies using (primarily) archival, continuum subtracted 
H$\alpha$ (+[N\,{\sc ii}]) imaging. Calibrated images are available
for 37 ccSNe, uncalibrated images are available for SN 1997X (NGC 4691) 
and SN 2005kl (NGC~4369) from the 2.0m Liverpool Telescope and 2.5m Isaac 
Newton Telescope, respectively. We exclude SN 1992ba (NGC~2082) and SN 
1995X (UGC 12160) from our global 
statistics since no H$\alpha$ images of their host galaxies are 
publically available, although we discuss literature descriptions for 
these cases. Therefore in  39 cases we
have examined the association between ccSNe with H\,{\sc ii} regions, and 
in 37 cases measured nebular fluxes and converted these into 
H$\alpha$ luminosities. In view of the possibility that the SN progenitor 
may have been ejected from its birth cluster, we examine the 
environment to typical projected distances of 0.5 kpc.

Table~\ref{imaging} indicates the source of the continuum subtracted 
H$\alpha$ images used in our study. The majority of archival images were
provided in flux calibrated format, for which the 11Mpc  H$\alpha$ and UV 
Survey (11HUGS, Kennicutt et al. 2008) and SINGS 
(Kennicutt et al. 2009)\footnote{Available from Local Volume
Legacy Survey (LVLS) at http://www.ast.cam.ac.uk/research/lvls} 
alone enabled the nebular environment of 40\% of 
the  ccSNe to be assessed. The majority of the remaining host galaxies 
were  included in other H$\alpha$  surveys of nearby galaxies, namely 
H$\alpha$GS  (James et al. 2004), SINGG  (Meurer et al. 2006) and those by 
Hoopes et al. (2001) and Knapen et al. (2004). For M~101, we used the 
low-resolution, flux calibrated 0.6m  KPNO Schmidt imaging by Hoopes et 
al. (2001) together with higher spatial resolution, uncalibrated 2.5m 
INT/Wide Field Camera imaging obtained from the ING 
archive\footnote{http://casu.ast.cam.ac.uk/casuadc/archives/ingarch}, 
which was also used in several other cases (e.g. NGC 7292).
We have also inspected Hubble Space Telescope (HST) WFPC2, ACS and WFC3 
imaging
from the ESA Hubble Science 
Archive\footnote{http://archives.esac.esa.int/hst/}, which is available
for a subset of the ccSNe. The Hubble Heritage Team ACS/WFC mosaic of M~51a 
was obtained from a dedicated 
website\footnote{http://archive.stsci.edu/prepds/m51/}. Narrow-band 
H$\alpha$ imaging is  only available for 8 ccSNe from our sample, while 
broad band images have been obtained in 22 cases, the latter relevant
for the potential association with compact star clusters.

NGC~2997, NGC~1559 and the LMC host the remaining four ccSNe 
omitted from these H$\alpha$ surveys. For the LMC we employ continuum 
subtracted H$\alpha$  imaging obtained with a Nikon survey camera plus 2K 
CCD (M.S. Bessell, priv. comm.), lower resolution Parking Lot Camera 
(PLC) H$\alpha$ and R-band images from Bothun \& Thompson (1988) and  
Kennicutt  et al. 1995), plus higher resolution MCELS imaging of 30 
Doradus obtained using the CTIO Curtis Schmidt telescope (Smith et al.  
2000). For NGC~2997 we have resorted to the Danish 1.5m observations of 
Larsen \& Richtler (1999). For NGC~1559, we have used high spatial 
resolution archival VLT/FORS1 imaging (from 075.D-0213(A), PI D.~Baade). 
In addition, high spatial resolution H$\alpha$ imaging of several 
ccSNe host galaxies has been included in our extragalactic Wolf-Rayet 
surveys. These comprise VLT/FORS2 imaging of M~83 (Hadfield et al. 2003) 
plus unpublished VLT/FORS1 imaging of M~74 (from 380.D-0282(A), PI
P.~Crowther and NGC~6744 (081.B-0289(C), PI P.~Crowther) 
and unpublished Gemini-N GMOS H$\alpha$ imaging of NGC~6946 (GN-2009B-Q-4, 
PI J.~Bibby). The relatively small field of view of these instruments 
excluded the investigation of some ccSNe environments in these galaxies
from out VLT or Gemini imaging, although (non-calibrated) archival 
VLT/FORS1 imaging is available for the SN 2008bk in NGC~7793 
(067.D-0006(A), PI W. Gieren).

Flux calibration was necessary in these cases. The LMC Nikon survey camera 
image was calibrated against Parking Lot Camera datasets for 30 Doradus 
(Kennicutt et al. 1995). 
The Danish 1.54m datasets of Larsen \& Richtler (1999) were calibrated 
with respect to 5 galaxies in common with 11HUGS (NGC~300, NGC~1313, M~83, 
NGC~6946, NGC~7793). VLT/FORS1+2 and Gemini-G GMOS images were calibrated 
against imaging of spectrophotometric standard stars (LTT 4816, LTT 1020, 
BD+28 4211). 

Finally, where necessary (e.g. H$\alpha$GS), astrometric 
calibration was performed  using the Starlink {\tt gaia}  
package\footnote{http://star-www.dur.ac.uk/~pdraper/gaia/gaia.html} using 
the USNO-A2 catalogue, with typical RMS of $<$1 pix, corresponding to 
$<$0.5 arcsec in the majority of instances.

\begin{table*}
  \begin{center}
  \caption{Integrated H$\alpha$ + [N\,{\sc ii}] fluxes (H$\alpha$ in
italics) of host galaxies of ccSNe used in the present study, from which 
star formation rates (SFR) and star  formation intensities ($\Sigma$) are 
obtained. Galaxy radii are  from RC3 or HyperLeda ($R_{\rm 25}$), or Bothun \& 
Thompson (1998, $R_{\rm  D}$). Uncertainties, where known, are indicated, 
while the [N\,{\sc ii}]/H$\alpha$ calibration from Lee et al.  (2009b) is reliable
to a factor of two, and the $A_{\rm  H\alpha}$ calibration is robust to $\pm$50\%, such that
global H$\alpha$ luminosities (SFR and $\Sigma$) should be reliable to between $\pm$20\% (direct
measurements of [N\,{\sc ii}]/H$\alpha$ and $A_{H\alpha}$) and $\pm$40\% (calibrations).}
  \label{hosts_sfr}
  \begin{tabular}{
l@{\hspace{2mm}}r@{\hspace{2mm}}
r@{\hspace{2mm}}c@{\hspace{2mm}}l@{\hspace{3mm}}l@{\hspace{3mm}}
l@{\hspace{3mm}}l@{\hspace{3mm}}l@{\hspace{3mm}}
r@{\hspace{3mm}}r@{\hspace{3mm}}r@{\hspace{3mm}}l@{\hspace{3mm}}
l@{\hspace{3mm}}l}\hline 
PGC & Alias & $d$ & $\log F$(H$\alpha$+[N\,{\sc ii}]) & Ref & [N\,{\sc 
ii}]/H$\alpha$
& $A_{H\alpha}$ & Ref & L(H$\alpha$)  & SFR                    & $R_{\rm 25}$ ($R_{\rm D}$) & $R_{\rm 25}$ 
($R_{\rm D}$) & $\Sigma_{R_{25}}$ 
($\Sigma_{R_{D}}$)      \\
     &  &Mpc& erg\,s$^{-1}$\,cm$^{-2}$     &     &   &     mag                 &     & erg\,s$^{-1}$ & 
$M_{\odot}$\,yr$^{-1}$ & arcmin & kpc & $M_{\odot}$\,yr$^{-1}$\,kpc$^{-2}$ \\
  \hline
03671&NGC 337A &11.4\phantom{0} & $-11.73 \pm 0.10$ & 6, 9 (d1)  & 
0.16 & 0.71 & 9($M_{B}$) & $4.8 \times 10^{40}$ & 0.38 & 2.9 & 9.7 & $1.3 
\times 10^{-3}$ \\
05974&M 74     & 9.0\phantom{0} & $-10.84 \pm 0.04$ &1 & $0.35 \pm 0.05$ & 0.30 $\pm$ 0.18 & 3 & 
$1.4 \times 10^{41}$ & 1.1\phantom{0}  & 5.2 & 13.7 & $1.8 \times 10^{-3}$ 
\\
10314&NGC 1058 & 9.86& $-10.63 \pm 0.05$ & 1 & 0.48 $\pm$ 0.05 & 0.69 $\pm$ 0.22 & 5 & 
$3.5 \times 10^{40}$ & 0.27 & 1.5 & 4.3 & $4.7 \times 10^{-3}$ \\ 
10488&NGC 1097 & 14.2\phantom{0}& $-10.95$ \phantom{$\pm$ 0.00} & 3 & 0.46 & 1.50 $\pm$ 0.08
&  9($M_{B}$),3 & $7.3 \times 10^{41}$ & 5.8\phantom{0} & 4.7 & 19.2 & $5.0 \times 
10^{-3}$ \\     
10496&NGC 1087& 14.4\phantom{0} & $-11.30 \pm 0.04$ & 2 & 0.40 $\pm$ 0.03 & 0.75 $\pm$ 0.20 & 5
& $1.8 \times 10^{41}$ & 1.4\phantom{0} &1.9 & 7.8 & $7.4 \times 10^{-3}$ 
\\
14814&NGC 1559 &12.6\phantom{0} & $-10.81 \pm 0.10$ & 9 (j2)& 0.31 & 0.86 & 9($M_{\rm 
B}$) & $4.9 \times 10^{41}$ & 3.9\phantom{0} & 1.7 & 6.3 & $3.1 \times 
10^{-2}$ \\
15821&NGC 1637 & 9.77& $-11.59 \pm 0.07$ &4, 9 (e1) & 0.85 & 0.69 
& 8, 9($M_{\rm B}$)  & 
$3.0 \times 10^{40}$ & 0.24 & 2.0 & 5.6 & $2.4 \times 10^{-3}$ \\
17223&LMC      & 0.05& \phantom{0}$-6.96 \pm 0.05$  &1 & 
0.15 & 
0.64 & 9($M_{\rm B}$)  & 
$5.0 \times 10^{40}$ & 0.40 & 323\phantom{.0} & 4.7 & $5.8 \times 
10^{-3}$ \\
     &         &     &                       &  &      &   &      &    &
                     &  (103)\phantom{0}    &(1.5)& ($5.6 \times 
10^{-2}$) \\
%
21396&NGC 2403 & 3.16 & $-10.25 \pm 0.04$ & 1 & 0.22 $\pm$ 0.04 & 0.45 $\pm$ 0.20 & 3  &
$8.5 \times 10^{40}$ & 0.67 & 10.9 & 10.1 & $2.1 \times 10^{-3}$ \\
27978&NGC 2997 & 11.3\phantom{0}& $-10.80 \pm 0.10$ & 7, 9 (f) & 0.46 
& 1.25 & 9($M_{\rm 
B}$)  
& $5.1 \times 10^{41}$ & 4.1\phantom{0} & 4.5 & 14.6    & $6.1 \times 
10^{-3}$ \\
28630&M 81      & 3.65 & $-10.31 \pm 0.02$ & 1 & 0.55 $\pm$ 0.08 & 0.15 $^{+0.30}_{-0.15}$ & 3
& $5.8 \times 10^{40}$ & 0.46 & 13.5  & 14.2 & $7.1 \times 10^{-4}$ \\
30087&NGC 3184 & 13.0\phantom{0}& $-11.12 \pm 0.05$  &3 & 0.52 $\pm$ 0.05 & 0.63 $\pm$ 0.18 & 3 & 
$1.8 \times 10^{41}$ & 1.4\phantom{0} & 3.7  & 14.0 & $3.1 \times 10^{-3}$ 
\\ 
30560&NGC 3239 & 10.0\phantom{0}& $-11.32 \pm 0.03$ &1 & 0.09 $\pm$ 0.01 & 0.30 $\pm$ 0.21 & 5 & 
$6.9 \times 10^{40}$ & 0.55 & 2.5 & 7.3  & $3.3 \times 10^{-3}$\\
32007&M 95     & 10.0\phantom{0}& $-11.24 \pm 0.08$ &1 & 0.66 $\pm$ 0.03 & 0.73 $\pm$ 0.17 & 3 & 
$8.1 \times 10^{40}$ & 0.64 &  3.7 & 10.7 & $1.8 \times 10^{-3}$ \\
41399& NGC 4487& 11.0\phantom{0} & $-11.93 \pm 0.10$ & 6, 9 (d1)& 0.21 & 0.65 & 9($M_{\rm 
B}$)& $2.9 \times 10^{40}$ & 0.23 & 2.1   & 6.7 & $1.7 \times 10^{-3}$ \\ 
42575&NGC 4618 & 9.2\phantom{0} & $-11.36 \pm 0.04$ &1 & 0.28 $\pm$ 0.03 & 0.13 $^{+0.21}_{-0.13}$ & 5 & 
$3.9 \times 10^{40}$ & 0.31 &   2.1  &  5.6 & $3.2 \times 10^{-3}$ \\ 
%
%
47404&M 51a    & 8.39& $-10.42 \pm 0.08$ &1 & 0.59 $\pm$ 0.01 & 1.05 $\pm$ 0.21 & 5 & 
$5.3 \times 10^{41}$ & 4.2\phantom{0} & 5.6 & 13.6 & $7.1 \times 10^{-3}$ 
\\
48082&M 83     & 4.92& $-10.00 \pm 0.04$ &1 & 0.40 & 1.08 & 9($M_{\rm B}$) & 
$5.6 \times 10^{41}$ & 4.4\phantom{0} &  6.4  & 9.2 & $1.7 \times 10^{-2}$ 
\\ 
50063&M 101    & 6.96& $-10.22 \pm 0.13$ &1 & 0.54 & 1.10 & 9($M_{\rm B}$)  & 
$6.3 \times 10^{41}$ & 5.0\phantom{0}  & 14.4 & 29.1 & $1.9 \times 
10^{-3}$ \\
50779&Circinus & 4.21 & $-11.19 \pm 0.06$ & 1 & 0.16& 1.74 & 9($M_{\rm B}$) &
$5.8 \times 10^{40}$ & 0.46 & 3.5 & 4.2 & $8.2 \times 10^{-3}$ \\
54849 & NGC 5921 & 14.0\phantom{0} & $-11.63 \pm 0.05$ & 2 & 
0.28 & 0.83 & 9($M_{\rm B}$) 
& $9.3 \times 10^{40}$ & 0.73 & 2.5   &  9.9 & $2.4 \times 10^{-3}$ \\
62836&NGC 6744 & 11.6\phantom{0}& $-11.33$\dag \phantom{00.00} & 7, 9 (f) & 0.61 & 1.28 & 
9($M_{\rm B}$) & $1.5 \times 10^{42}$\dag & 12.0\dag & 10.0  
& 33.6 & $3.4 \times 10^{-3}$\dag\\
65001&NGC 6946 &  7.0\phantom{0}& $-10.42 \pm 0.06$ & 3 & 0.45 $\pm$ 0.09 & 0.45 $\pm$ 0.30 & 3 & 
$2.3 \times 10^{41}$ & 1.8\phantom{0} &  5.7 & 11.7 & $4.3 \times 
10^{-3}$\\
68941&NGC 7292 & 12.9\phantom{0}& $-11.78 \pm 0.04$ &2 & 0.15 & 0.60 & 9 ($M_{\rm B}$)  & 
$5.0 \times 10^{40}$ & 0.39 &  1.1  & 4.0& $7.8 \times 10^{-3}$ 
\\
70096&NGC 7424 & 7.94& $-11.28 \pm 0.07$ &4, 9 (e1) & 0.20 
& 0.62 & 9($M_{\rm B}$)  & 
$5.7 \times 10^{40}$ & 0.45 & 4.8   & 11.0 & $1.2 \times 
10^{-3}$\\
73049&NGC 7793 & 3.61& $-10.60 \pm 0.08$ &3 & 0.31 $\pm$ 0.07 & 
0.67 $\pm$ 0.16 & 3 & $5.5 \times 10^{40}$ & 0.44 &  4.7   & 4.9 & $5.8 \times 
10^{-3}$ \\
  \hline
%
%
  \end{tabular}   
  \end{center}
\begin{small}
\begin{flushleft}
1: Kennicutt et al. (2008); 
2: James et al. (2004); 
3: Kennicutt et al. (2009); 
4: Meurer et al. (2006); 
5: Moustakas \& Kennicutt (2006); 
6: Knapen et al. (2004); 
7: Larsen \& Richtler (1999); 
8: Kennicutt \& Kent (1983);
9: this work (note) \\
\dag: Lower limit using aperture of radius 6$'$.4 (0.64 $R_{25}$)
\end{flushleft}
\end{small}
\end{table*}

For the present sample, 8 (19\%) of the ccSNe originate from early-type 
spirals (S0/a/b), 31 (76\%) from late-type spirals (Sc/d/m) and 2 (5\%) 
from irregulars (Im). 25 (61\%) ccSNe are from high luminosity ($M_{\rm 
B}$ $<$ --19 mag) hosts, with 16 (39\%) from low luminosity galaxies, with 
a similar fraction of type II and type Ib/c ccSNe from dwarf hosts.

Table~\ref{hosts_sfr} presents star formation rates and star formation 
intensities for all host galaxies, excluding those lacking calibrated 
H$\alpha$ imaging (NGC~2082, 4369, 4691, UGC 12160). H$\alpha$ fluxes are 
adjusted for the contribution of [N\,{\sc ii}] $\lambda\lambda$6548--6583 
preferentially from integrated spectrophotometry (e.g. Moustakas \& 
Kennicutt 2006) or the Lee et al.  (2009b) $M_{\rm B}$--calibration. 
Similarly, extinction corrections, $A_{H\alpha}$, are preferentially 
obtained from measured integrated nebular H$\alpha$/H$\beta$ ratios (e.g. 
Moustakas \& Kennicutt 2006, via eqn.~4 from Lee et al. 2009b), or are the 
sum of (measured) foreground and (estimated) internal extinctions. 
Foreground extinctions are from Schlafly \& Finkbeiner (2011, 
recalibration of Schlegel et al. 1998), for which we assume $A_{H\alpha}$ 
= 0.62 $A_{\rm B}$, while internal extinctions are estimated from a 
scaling 
relation between extinction and $M_{\rm B}$ (Lee et al. 2009b). In the 
case of Circinus, an extinction of $A_{B}$ = 2.0 mag was adopted (Freeman 
et al. 1977) owing to its low galactic latitude.

Integrated fluxes largely confirm previous results (e.g. Lee et al. 
2009b), aside from differences in distances and the source of fluxes. One
notable exception is NGC 6744, for which the newly calibrated Larsen \& 
Richtler (1999)  datasets reveal a lower limit of 4.7$\times 10^{-12}$ 
erg\,s$^{-1}$\,cm$^{-2}$ (6.4$'$ radius) to the H$\alpha$+[N\,{\sc ii}] 
flux, significantly higher than 2.0$\times 10^{-12}$ 
erg\,s$^{-1}$\,cm$^{-2}$ from Ryder \& Dopita (1993), as reported in 
Kennicutt  et al. (2008). A net flux of 1.2$\times 10^{-12}$  
erg\,s$^{-1}$\,cm$^{-2}$  within the central region (3.5$'$ radius) from 
Larsen \& Richtler (1999)  matches that from the calibrated VLT/FORS1 
dataset to within 2\%. Uncertainties for hosts in which fluxes, 
[N\,{\sc ii}]/H$\alpha$ and reddenings have been measured, such as M~74, are
typically  $\pm$20\%, whereas cases for which calibrations have been adopted 
for [N\,{\sc ii}]/H$\alpha$ (factor of two) and $A_{H\alpha}$ ($\pm$50\%), 
such as NGC 337A, are typically $\pm$40\$.

Star formation intensities are uniformly based upon $R_{\rm 25}$, although 
the intensity is also calculated for the LMC from its V-band scale length, 
$R_{\rm D}$ (Hunter \& Elmegreen 2004) following Bothun \& Thompson 
(1988).

\subsection{Calculation of H$\alpha$ luminosities}

We have calculated H$\alpha$ luminosities of H\,{\sc ii} regions in close 
proximity to the ccSNe location in the following way. Fluxes were measured 
using apertures no smaller than the image FWHM. Aside from the LMC Nikon 
Survey Camera and CTIO 1.5m/CFCCD imaging, H$\alpha$ filters include the 
contribution of [N\,{\sc ii}] $\lambda$6548--84. To adjust for this 
contribution, we either selected the [N\,{\sc ii}]/H$\alpha$ ratio 
measured from spectrophotometry of their host galaxies, or estimated the 
ratio from an empirical scaling relation between [N\,{\sc ii}]/H$\alpha$ 
and the absolute B-band magnitude $M_{\rm B}$ (Kennicutt et al. 2008), as 
above for global star formation rates. Of course, this adds an additional 
uncertainty, namely the radial metallicity gradient. By way of example,
Bresolin et al. (2004) measured 0.31 $\leq$ [N\,{\sc 
ii}]/H$\alpha$ $\leq$ 0.64 for 10 H\,{\sc ii} regions spanning the full 
radial extent of the disk of M~51a. Here, we adopt [N\,{\sc ii}]/H$\alpha$ 
= 0.59 (Moustakas \& Kennicutt 2006), such that H$\alpha$ luminosities should be 
underestimated by, at most, $\sim$20\% in the outer disk where 
metallicities are lower than average. However, no correction is attempted 
due to the azimuthal variation in metallicity, as shown from integral 
field spectroscopy of M~74 by S\'{a}nchez et al. (2011, their fig.~17).

\begin{table*}
  \begin{center}
  \caption{Summary of the present work and related studies, 
indicating sample size, $N_{\rm SNe}$, mean distances, $\overline{d}$, 
the number of ccSNe associated with H\,{\sc ii} regions, $N_{\rm HII}$
or the normalized cumulative rank (NCR) pixel (see James \& 
Anderson 2006). Distances are obtained from individual SN 
hosts, drawn preferentially from the Extragalactic Distance  Database 
($\leq$ 3000 km\,s$^{-1}$, Tully et al. 2009)  or otherwise NASA 
Extragalactic Database.}
  \label{compare}
  \begin{tabular}{l
@{\hspace{5mm}}r@{\hspace{2mm}}c@{\hspace{2mm}}l@{\hspace{2mm}}
@{\hspace{5mm}}r@{\hspace{2mm}}c@{\hspace{2mm}}l@{\hspace{2mm}}
@{\hspace{5mm}}r@{\hspace{2mm}}c@{\hspace{2mm}}l@{\hspace{2mm}}
@{\hspace{5mm}}r@{\hspace{2mm}}c@{\hspace{2mm}}l}
\hline 
SN      &         \multicolumn{3}{c}{van Dyk et al. (1996)}        & \multicolumn{3}{c}{Anderson et al. (2012)} & \multicolumn{3}{c}{Smartt (priv. comm.)} &
\multicolumn{3}{c}{This work} \\
type    & $N_{\rm SNe}$ & $\overline{d}$ (Mpc) & $N_{\rm HII}/N_{\rm 
SNe}$& 
$N_{\rm SNe}$ & $\overline{d}$ (Mpc) & $\overline{NCR}^{a}$ & $N_{\rm 
SNe}$ & 
$\overline{d}$ (Mpc) & 
$N_{\rm 
HII}$ & $N_{\rm SNe}$ & $\overline{d}$ (Mpc) & $N_{\rm HII}/n_{\rm SNe}$ 
\\
  \hline
II      & 32 & 16.2 & 72$\pm$10\% & 163.5 & 32 & 0.25$\pm$0.02     & 17 & 14.6 & 0 & 29 & 8.7 & 38$\pm$11\% \\
Ib      & \multirow{2}{*}{17} & \multirow{2}{*}{17.7} & 
\multirow{2}{*}{68$\pm$12\%} & 39.5 & 40 & 0.32$\pm$0.04 & \multirow{2}{*}{9} & \multirow{2}{*}{19.3} & 
\multirow{2}{*}{1} & \multirow{2}{*}{10} & \multirow{2}{*}{9.9} & \multirow{2}{*}{70$\pm$26\%} \\
Ic      &    &      &             &   52 & 42  & 0.47$\pm$0.04    \\
 \hline
\end{tabular}
  \end{center}
\begin{flushleft}
(a) NCR = 0 if the ccSN site is not associated with any H$\alpha$ 
emission and NCR = 1 if it is associated with the brightest H$\alpha$ 
emission in its host. 
\end{flushleft}
\end{table*}

Corrected H$\alpha$ fluxes were converted into intensities, using the 
method set out above for global star formation rates. Once again, we 
neglect spatial variations of internal extinctions. For example, a global 
average of $A_{H\alpha}$ = 1.05 mag is adopted for M~51a (Moustadas \& 
Kennicutt 2006), whereas Bresolin et al. (2004) obtained 0.09 $\leq 
A_{H\alpha} \leq$ 1.05 mag for 10 H\,{\sc ii} regions distributed 
throughout its disk\footnote{We have converted the $c$(H$\beta$) values 
from Bresolin et al. (2004) to $A_{H\alpha}$ via E(B-V) $\sim$ 0.7 $c$(H$\beta)$ 
and $A_{H\alpha}$ $\sim$ 2.5 E(B-V)}. 
For M~51a, the luminosity of individual H\,{\sc ii} regions 
may be overestimated by up to a factor of 2.5. Once again, no 
radial correction is attempted owing to the clumpy nature of dust attenuation, 
as shown in integral field spectroscopy of M~74 by S\'{a}nchez et al. (2011, 
their fig.~11).

\begin{figure}
\begin{center}
\includegraphics[bb=25 5 195 360, width=0.8\columnwidth]{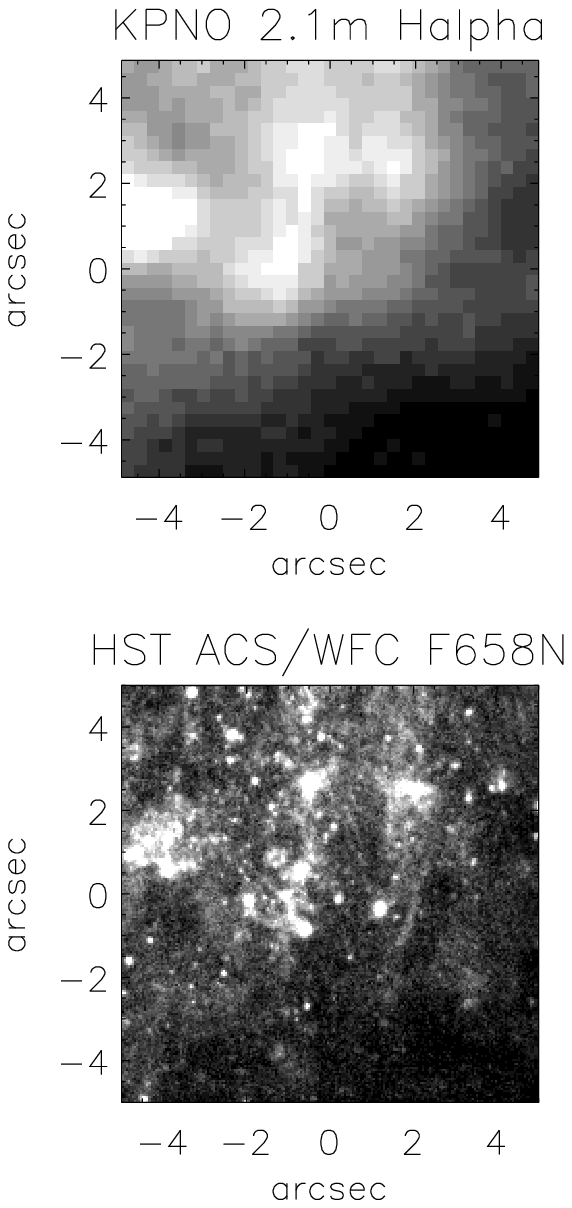}
\caption{(top) 10$\times$10 arcsec$^{2}$ KPNO 2.1m H$\alpha$ image (from 
Kennicutt et al. 2003) showing the environment of SN 2005cs in M~51a,
corresponding to 400$\times$400 pc at a distance of 8.39 Mpc; 
(bottom) MultiDrizzled HST ACS/WFC F658N image from GO 10452 (PI S. 
Beckwith, see discussion in Li et al. 2006). North is up and east is to the 
left.}
\label{sn2005cs_2}
\end{center}
\end{figure}

Overall, our adoption of global [N\,{\sc ii}]/H$\alpha$ and $A_{H\alpha}$ 
values should have a negligible effect for the majority of sources, such 
that the 20--40\% uncertainties quoted above will apply to individual 
H\,{\sc ii} regions. However, H$\alpha$ luminosities of regions far from 
large H\,{\sc ii} complexes -- typically those at large galactocentric 
radii -- may be overestimated by up to a factor of two ([N\,{\sc ii}]/H$\alpha$ 
and $A_{H\alpha}$ corrections act in opposite senses). The average 
galactocentric distance is $R_{\rm SN}/R_{\rm 25}$ = 0.47 
($\sigma$=0.30),
although H$\alpha$ luminosities of nebulae located in the extreme outer 
disks of their hosts may be overestimated (e.g. SN~1980K, SN~2001ig, 
SN~2002ap, SN~2004et). Nevertheless, such adjustments would not affect our 
main conclusions.

\subsection{Results}

Following
van Dyk (1992), a  ccSNe is considered to be associated with a H\,{\sc ii} 
region if it is offset by an amount less than or equal to 
the radius of the H\,{\sc ii} region. For example, SN 1923A lies 4$''$ 
from a H\,{\sc ii} region whose radius is $\sim$4$''$, while SN 1964H lies 
5$''$ from a H\,{\sc ii} region, whose radius is 2$''$. Therefore the
former is considered to be associated with a H\,{\sc ii} region while
the latter is not.   More details notes relating to  individual ccSNe are 
provided  in Appendix A. 

Tables~\ref{type1_sn}-\ref{type2_sn} present basic properties of
individual type Ib/c and type II ccSNe, respectively, including the 
radius and H$\alpha$ flux/luminosity of the nearest H\,{\sc ii} region.
Overall, approximately half (18 of the 
39) of our sample of ccSNe are associated with H\,{\sc ii} regions. 
Of these,  the mean H$\alpha$ luminosity is $3 \times 10^{38}$ 
erg\,s$^{-1}$,
excluding SN 1970G which lies at the periphery of the supergiant H\,{\sc 
ii} region NGC~5455 within M~101. If we now 
separate these  ccSNe into their main types, a much higher fraction of 
type Ib/c ccSNe (7 out of 10), namely 70 $\pm$ 26\%, are associated with 
nebular emission than  type II ccSNe (11 out of 29), for which the 
fraction is 38 $\pm$ 11\%.  This is qualitatively in agreement with 
the ground-based studies of Anderson \& James (2008) and Anderson
et al. (2012).


However, recall the lack of an association between H\,{\sc 
ii}  regions and nearby ($cz <$ 2000 km\,s$^{-1}$) ccSNe 
(Smartt et al. 2009) discovered between 1998--2008.5 
observed at high spatial resolution (Crockett 2009; Smartt, 
priv. comm.). Only 4 type II ccSNe are in common between the present study 
and the subset of ccSNe from Smartt et al. (2009) that have been observed
with HST. Of these, the ground-based study reveals consistent results,
except for SN  2005cs (II-P).  Fig.~\ref{sn2005cs_2} shows that SN 
2005cs appears to be associated with nebular emission from  ground-based 
KPNO 2.1m  imaging, yet HST  ACS/WFC imaging reveals that 
it is offset by $\sim$1$''$ (40 pc at 8.4  Mpc), as discussed by Li et al. 
(2006).

Turning to Ib/c ccSNe, only 3 are in common between the present study 
and Crockett (2009).  Broadly consistent results are obtained, although SN 
2007gr (Ic) merits discussion, since it is the only Ib/c 
associated with a star forming region at the spatial resolution of HST. 
SN 2007gr is formally associated with H$\alpha$  emission on the basis of 
our  ground-based imaging (Fig.~\ref{sn2007gr}),  whereas Crockett et al. 
(2008)  indicate a small offset from H$\alpha$  emission (from INT/WFC). 
More  recent HST WFPC2/F675W and WFC3/F625W  suggest faint nebulosity is  
spatially coincident with the SN position.

In Table~\ref{compare} we provide a summary of the present results,
together with previous related studies. Despite the low number statistics, 
our results have the advantage over previous ground-based studies owing
to smaller positional uncertainties with respect to van Dyk et al. (1996)
and significantly smaller average distances than Anderson et al. (2012).
For a nominal ground-based imaging quality of FWHM$\sim$1.5\AA, the 
typical spatial  resolution achieved is 125\,pc (van Dyk et al. 
1996), 260\,pc (Anderson et al. 2012) and 70\,pc in the present work. 
A characteristic scale of $\sim$10\,pc -- an order of magnitude higher 
-- is achieved from the HST-selected  sample of Smartt et al. (2009) and 
Crockett (2009). This is likely the origin of the very different 
statistics with respect to the ground-based studies. We shall 
return to the issue of spatial resolution in Sect.~\ref{discussion}.




\section{Implications for progenitor masses of ccSNe}\label{implications}

\subsection{Core-collapse SN environments}

We now attempt to combine our results with the earlier discussion to place 
constraints upon ccSNe progenitors, recognising that this approach is 
inferior to than methods involving photometric detection of immediate 
ccSNe  progenitors (e.g. Smartt et al. 2009). 


We have examined the environment of each  ccSNe and assign each case to 
one of the five classes set out in  Sect.~\ref{clusters}. 
For our previous examples, SN 1923A narrowly lies within the radius
of a luminous H\,{\sc ii} region (Fig.~\ref{sn1923a}) so it is assigned to 
class 5, whereas SN 1964H lies far from nebular emission
(Fig.~\ref{sn1964h}),  so it is assigned to {\bf Class} 2 since it is not 
coincident with a bright star cluster.

Unsurprisingly, owing to  the  poor  spatial  resolution and sensitivity 
of the  ground-based  imaging, only one  of the  ccSNe, SN  2004dj (II-P), 
was assigned to  Class 1, i.e.   nebular emission  absent but (young) 
cluster detected, as shown in Fig.~\ref{sn2004dj}. A massive 
progenitor ($>$20  $M_{\odot}$) might be expected for SN 2004dj, although 
detailed studies of the cluster indicate a lower progenitor mass $<$20 
$M_{\odot}$ (Vink\'{o} et al. (2006, 2009). As discussed above for the
case of SN  2005cs (Fig.~\ref{sn2005cs_2}), more ccSNe from our 
sample would  have been  included in this  category on the basis of HST 
imaging (e.g. SN 2009ib). 

Half of type II ccSNe (15 from 29) were assigned to 
class 2, i.e. ccSNe lacking nebular emission, an associated 
(bright) cluster or  a nearby giant H\,{\sc ii} region. The progenitor 
masses of  such are expected to be $< 20 M_{\odot}$, in accord with Smartt 
et al.  (2009)  since the majority of these H-rich ccSNe either have 
unknown  subclass or  are type II-P -- exceptions are type IIb ccSNe SN 
1993J (Fig.~\ref{sn1993j}) and SN 2011dh (Fig.~\ref{sn2011dh}). 

Two
H-deficient ccSNe also fall in this category (SN 2002ap, SN 
2003jg), of which SN 2002ap represents an archetypal case
(Fig.~\ref{sn2002ap}). The inclusion of type Ic ccSNe in this category 
favour an interacting binary scenario over a single star progenitor for 
these cases, as  discussed by  Crockett et al. (2007) for SN 2002ap. 
Indeed, Mazzali et al. 
(2002, 2007) proposed a binary scenario for SN 2002ap involving a 
progenitor with 15--25 $M_{\odot}$. The case of SN 2003jg is marginal 
since it lies 2$''$ away (140 pc deprojected) from a H\,{\sc ii} region, 
whose radius is 1$''$.5 on the basis of ground-based Danish 1.5m imaging
(Fig.~\ref{sn2003jg}) such that  it narrowly fails to meet our 
threshold for association.



From our sample we identify 3 potential runaways from nearby giant 
H\,{\sc ii} regions (Class 3), namely SN 1986L (II-L),  
SN 1997X (Ib) and SN 2001X (II-P), of which the former
serves as a useful example (Fig.~\ref{sn1986l}). Here we consider
possible runaways if they lie at deprojected distances of up to 0.4 kpc
from luminous H\,{\sc ii} regions ($L(H\alpha) \geq 10^{38}$ 
erg\,s$^{-1}$). If  this were so,  we are  are unable  to  assign  
progenitor masses, although  high runaway masses  are believed to be 
favoured (Fujii \& Portegies Zwart 2011). Of course, the presence of a  nearby giant 
H\,{\sc ii} region does not require a high mass cluster that is 
sufficiently dense for runaways via dynamical interactions during the 
cluster formation. In addition, the close proximity of a giant H\,{\sc ii} 
region to the ccSNe does not necessarily imply the progenitor originated 
from this region (Class 2 is also likely). 



Turning to ccSNe which are associated with H\,{\sc ii} regions, 
we find six cases matching Class 4, involving a relatively faint 
H\,{\sc ii} region in close proximity to the ccSN. Of these, every
H\,{\sc ii} region is spatially extended, namely: SN 1983N (Ib), SN 
1987A (IIpec), SN 2001ig (IIb), SN 2005at (Ic), SN 2005cs (II-P) and SN 
2012A (II-P), indicating a  lower limit to the progenitor mass of 12 (20) 
$M_{\odot}$ for a duty cycle  of 20 (10) Myr. We cannot exclude the 
possibility of higher mass  progenitors since the H\,{\sc ii} regions 
identified as extended from ground-based imaging may involve multiple 
compact H\,{\sc ii}  regions in some instances. Still, close binary 
(accretion or merger) predictions favour a mass of $\sim$15 $M_{\odot}$ 
for the progenitor of SN 1987A (Podsiadlowski 1992). Panagia et al. (2000) 
discuss a loose 12$\pm$2 Myr cluster likely to be  associated with the 
progenitor of SN~1987A, 
although this would not be detected in ground-based imaging 
at the typical distance of other ccSNe in our sample. 

The spatial resolution of ground-based observations certainly limits the 
potential association with H\,{\sc ii} regions (recall 
Table~\ref{compare}). Fig.~\ref{sn2005cs_2} contrasts (non-Adaptive 
Optics) ground-based and space-based H$\alpha$ imaging of the immediate 
environment of SN 2005cs. HST imaging reveals several compact H\,{\sc ii} 
regions that are in close proximity to the ccSNe, albeit none 
spatially coincident with it, although In addition, bright H\,{\sc ii} 
regions are 
often in close proximity to the SN site, such that a runaway status cannot 
be excluded either (e.g. SN 1983N, Fig.~\ref{sn1983n}).


Finally, 12 of the 39 ccSNe are associated with giant H\,{\sc ii} regions
(Class 5), comprising 50$\pm$22\% (5/10) of the type Ib/c SNe, 
though only 
24$\pm$9\% (7/29) of the type II ccSNe sample. Solely SN 1985F (Ib/c) is 
spatially  coincident with a bright cluster (Fig.~\ref{sn1985f}), while 
some others are found in complexes (e.g. SN 2007gr, Fig.~\ref{sn2007gr}). Age 
estimates for individual stellar populations  within each region are not 
available, so we consider a characteristic duty cycle  of (super)giant  
H\,{\sc ii} regions of 20 (10) Myr, from which lower progenitor mass 
limits of  12 (20) $M_{\odot}$ are implied for the initial stellar 
generation, with higher
limits for subsequent episodes of star formation. Unfortunately, no limit 
to the  progenitor  mass of SN 1968L (II-P) can be assigned since it lies 
within  the nuclear starburst of M 83 (Fig.~\ref{sn1968l}). 

Detailed studies based either on either pre-supernova 
imaging (e.g. Aldering et al. 1994) or post-supernova light curves 
(e.g. Chugai 
\& Utrobin 2000)  have been carried out in some cases. For example, the 
former technique was used by Leonard et al.  
(2002) to obtain an upper limit of 15$^{+5}_{-3} M_{\odot}$ for the 
progenitor mass of SN 1999gi (II-P, see also Hendry 2006)  while the latter 
approach enabled Iwamoto et al. (1994) to  estimate a progenitor mass of $\sim$15 
$M_{\odot}$ for SN 1994I (Ic). Therefore a giant H\,{\sc ii} region duty 
cycle of 20 Myr is the most realistic case (recall also 
Sect.~\ref{lifetimes}),  which naturally provides only weak limits upon 
progenitor masses for these II-P, IIn and Ib/c types of ccSNe (e.g. Smartt 2009). 
\begin{table*}
  \begin{center}
  \caption{Summary of expected association between H\,{\sc ii} 
regions and ccSNe (+ long GRBs) in host galaxies of differing 
star formation intensities ($\Sigma_{R}$), following Kennicutt et al. 
(1989) and Gieles (2009). Star formation intensities are obtained
from $R_{\rm  25}$ (spirals) or $R_{\rm D}$ (irregulars). A 20 Myr duty 
cycle is adopted for giant  H\,{\sc ii} regions, versus $\sim$4 Myr for 
isolated H\,{\sc ii} regions.}
  \label{summary}
  \begin{tabular}{llccccrc}\hline 
  Host &Star Formation &$\Sigma_{R_{25}}$ ($\Sigma_{R_{\rm D}}$) & Cluster 
range & 
Characteristic      & SN/GRB-H\,{\sc ii} & 
Example\\
       &Intensity&($M_{\odot}$ yr$^{-1}$ kpc$^{-2}$)  & ($M_{\odot}$)  & 
H\,{\sc ii} 
region & association?        &\\
\hline
Sab & Low & 4$\times 10^{-5}$  & 10$^{2-4}$ & 
Isolated & \tickNo (all types)      & M\,31 \\
Scd & High & $2 \times 10^{-3}$ & 10$^{2-6}$& Giant; Isolated   
& 
\tickNo ($\leq 12 M_{\odot}$), ? (12--85 $M_{\odot}$), \tickYes ($\geq 85 
M_{\odot})$ & M\,101\\
Irr & Low & ($5 \times 10^{-3}$)  & 10$^{2-4}$ & 
Isolated 
& \tickNo (all types)      & SMC \\
Irr & High & (1.5) & 10$^{2-6}$& Giant; Isolated   
& 
\tickNo ($\leq 12 M_{\odot}$), ? (12--85 $M_{\odot}$), \tickYes ($\geq 85 
M_{\odot})$ & NGC 1569\\
  \hline
  \end{tabular}   
  \end{center}
\begin{small}
\begin{flushleft}
Star formation intensities: M\,31 (Lee et al. 2009b);  M\,101: This work; SMC: 
Massey et al. (2007); NGC 1569: Kennicutt et al. (2008).
\end{flushleft}
\end{small}
\end{table*}

\subsection{Previous ccSNe environmental studies in context}

We have attempted to constrain progenitor masses from the association 
between ccSNe and H\,{\sc ii} regions, or lack thereof. Overall, 
establishing progenitor masses in either case is challenging, given the
{\it short} duration of the H\,{\sc ii} region phases for isolated, 
compact clusters and the {\it long} duration of the giant H\,{\sc ii} 
region phase for  extended, star forming complexes, which form the 
majority of the sample. SN 2004dj (II-P) ought to provide the
strongest indirect (young) age constraint, since this is spatially 
coincident with  a star cluster though not a H\,{\sc ii} region. However,
direct analysis of  its light curve suggests an old age/low mass
(Vink\'{o} et al. 2006, 2009). Therefore, ccSNe progenitor inferences from 
H\,{\sc ii} regions should be treated with caution, especially for high 
inclination hosts and/or low spatial resolution observations.

How, then, should one interpret previous interpretations of the 
association between H\,{\sc ii} regions and ccSNe, or lack thereof? 
According to Anderson \& James (2008), a progenitor mass  sequence II  
$\rightarrow$ Ib $\rightarrow$ Ic was proposed, with H-rich ccSNe 
further subdivided into IIn  $\rightarrow$ II-P $\rightarrow$ II-L 
$\rightarrow$ IIb by Anderson et al. (2012).

We concur with Anderson et al. that the higher frequency of type Ibc with 
H\,{\sc ii} regions than type II ccSNe arises from the relative lifetimes 
of their progenitors with respect to H\,{\sc ii} regions. However, we 
disagree with their implications since the H\,{\sc ii} regions 
detected at distances typical of their sample ($\overline{d}$ = 35 Mpc,
Table~\ref{compare}) are extended, giant, multi-generation H\,{\sc ii} 
regions. Let us assume that type II ccSNe result from progenitors with 
8--20 $M_{\odot}$ (Smartt 2009) from the first stellar population. A 
duty cycle of $\sim$20 Myr  would imply  that 40\% of type II progenitors 
with 12--20 $M_{\odot}$ are associated  with H\,{\sc ii} regions, while 
60\% of progenitors with 8--12 $M_{\odot}$ 
are not, based upon a standard Salpeter IMF slope for massive stars. 
Although approximate, such expectations agree well with the 34$\pm$11\% of 
type II ccSNe that are associated with H\,{\sc ii} regions in our study.

To illustrate the restricted progenitor mass limits that 
can be  achieved from this approach, the only case of a type IIn 
supernova in our 
present study is SN 1996cr. It has been proposed that these arise from
massive Luminous Blue Variables (Smith 2008), yet Anderson et al. (2012)
claim type IIn SN possess the lowest mass progenitors of all massive 
stars. SN 1996cr is associated with a giant H\,{\sc ii} region, although 
is not coincident with a bright cluster, so one can merely set a 
lower mass limit of 12 $M_{\odot}$ to the progenitor, with no robust 
upper limit, so one cannot argue against a high mass progenitor
on the basis of its immediate environment.

Of course, a higher fraction (70$\pm$26\%) of type Ib/c  ccSNe are 
associated with H\,{\sc ii} regions. This suffers from small number 
statistics, but likely reflects the shorter lifetime of stars with
$\geq$12 $M_{\odot}$, 
the majority of which will be associated with a $\sim$20 Myr duty
cycle of giant H\,{\sc ii} regions. In general, this fails to discriminate  
between most single star (Crowther 2007; Georgy et  al. 2012), 
and close binary progenitor scenarios (Podsiadlowski et al. 1992; 
Yoon et al. 2010), aside from a higher mass threshold ($\geq 12 M_{\odot}$?)
for Ib/c than type II ccSNe ($\geq 8 M_{\odot}$).  Still, for the 
two cases lacking any associated H\,{\sc ii} region or nearby giant
H\,{\sc ii} region, SN  
2002ap (Ic), SN 2003jg (Ib/c), a close binary scenario is favoured.

\section{Discussion}\label{discussion}

\subsection{Can local H\,{\sc ii} regions constrain ccSNe progenitor 
masses?}

We have assessed the immediate nebular environment of ccSNe in nearby
star forming galaxies, and confirm the results from Anderson \& James
(2008) and Anderson et al. (2012) that type Ib/c ccSNe are more likely
to be associated with a H\,{\sc ii} region than type II ccSNe. However,
the typical H\,{\sc ii} regions identified in H$\alpha$ imaging from 
ground based telescopes are extended, giant H\,{\sc ii} regions with
long duty cycles. Indeed, the issue of differing duty cycles for compact, 
isolated  H\,{\sc ii}  regions versus extended, H\,{\sc ii} complexes is 
particularly relevant  for late-type spirals and irregulars. Such hosts
dominate the statistics of Anderson  et al. (2012), for which
large star forming complexes -- ionized by multiple  generations of star clusters
-- are common (Kennicutt et al. 1989).

Nevertheless, firm limits upon the lifetime  ($\leq$4 Myr) and mass 
($\geq$85  $M_{\odot}$) of ccSNe progenitors would be possible if 
examples  of isolated, classical H\,{\sc ii} regions could be identified.
Resolving the specific  location of the ccSNe within such a region is 
only realistic at much higher spatial resolution than typically achieved 
here, whether from space with HST or using Adaptive Optics with large 
ground-based telescopes (recall Fig.~\ref{sn2005cs_2}). From 
Table~\ref{compare}, ccSNe are rarely associated with compact H\,{\sc ii} 
regions, although this is unsurprising in view of the small numbers of
very high mass stars within nearby star-forming galaxies.

Overall, we confirm previous findings  by Anderson \& James (2008) and
Anderson et al. (2012) that the association between 
different SNe flavours  and H\,{\sc ii} regions does vary between 
H-rich and H-poor ccSNe. Unfortunately, minimal implications for progenitor 
masses can be drawn which prevents discrimination between the single
versus close binary progenitor scenarios proposed for Ib/c (and IIb) ccSNe 
(Anderson \&  James 2008). Our findings fail to support claims that
the progenitors of IIn ccSNe possess relatively low masses (Anderson et 
al. 2012). In only a few cases does the lack of nebular emission provide  
limits upon  progenitor masses, i.e. favouring the close binary scenario 
for 2 type Ib/c ccSNe.

\subsection{Core-collapse SNe beyond the local universe}

Mindful of the spatial resolution issue, let us re-assess the Kelly
et al. (2008) study of SNe locations, with respect to the continuum 
($g'$-band) light of their low redshift ($z<0.06$) host galaxies.
Kelly  et al.  revealed that Ic SNe are much more likely to be 
found in  the  brightest regions of their hosts than type II SNe, with 
intermediate properties for type Ib SNe. Kelly et  al. (2008) argued that 
if the brightest locations correspond to  the  largest star-forming 
regions, type Ic SNe are  restricted  to the most massive  stars, while 
type Ib and especially type  II-P SNe are  drawn from stars with more 
moderate masses.

The SN host sample of Kelly et al. (2008) is not expected to be 
significantly different from our present sample. From 
Table~\ref{hosts_sfr}, star formation intensities from the 
present (spiral) hosts are uniformly high, with an  average value 
of $\Sigma_{R_{25}}$ = $5 \times  10^{-3} M_{\odot}$\,yr$^{-1}$\,kpc$^{-2}$. This is
more  representative of 'typical' late-type (Scd) spirals, even for Sab 
hosts ($\Sigma_{R_{25}} \sim 8 \times 10^{-3} M_{\odot}$\,yr$^{-1}$\,kpc$^{-2}$ for 
Circinus). For the  irregular galaxies, comparisons with the reference 
galaxies are more  difficult due to the lack of V-band scale lengths for 
NGC~3239 and  NGC~7292. If intensities are typically an order of magnitude 
higher (on  the basis of the LMC), these are intermediate between the 
extremes of the SMC and NGC~1569.

Therefore, how should the difference between type Ib/c and type II ccSNe 
identified by Kelly et al. (2008) be explained? In the $g'$-band, a 
star cluster will fade by 1 mag at young ages (5 to 10 Myr), with a further
1 mag dimming at intermediate ages (10 to $\sim$60 Myr), as shown in fig.~5
from Bik et al. (2003). Still, individual star clusters are not spatially resolved
in ground-based imaging so it is more likely that the difference relates to 
the different frequencies of type Ib/c and type II ccSNe in large star forming complexes, 
as discussed above for our local sample. 

Very high mass ($\sim$ 50 $M_{\odot}$) stars are anticipated to be
limited to massive (bright)  clusters, whereas lower mass ccSNe 
progenitors ($\sim$10 $M_{\odot}$) will be found in clusters spanning a 
 broad range of masses. The former are typically found in large 
(bright) star forming complexes (e.g. Carina Nebula, 30 Doradus). 
Therefore the higher frequency for H-deficient ccSNe in bright regions of 
their hosts with respect to H-rich ccSNe does suggest that a 
non-negligible fraction of type Ib and especially Ic ccSNe originate 
from higher mass  stars than type II ccSNe. In reality, a mixture of close 
binaries and  higher mass single stars are likely to be responsible 
for type Ib/c ccSNe (Bissaldi et al. 2007; Smith et al. 2011).

\subsection{Long Gamma Ray Bursts}

From an analysis of high redshift galaxies, Fruchter et al. (2006) 
revealed  that long GRBs ($<z>$ = 1.25) were also strongly biased towards 
the brightest part of their hosts, in contrast to core-collapse SNe ($<z>$ 
= 0.63, most presumably type II-P) which merely traced the light from their hosts. 

One significant difference between the low-redshift Kelly et al. (2008)
SN study  and the high-redshift GRB study of  Fruchter et al. (2006) is 
that hosts of the former are relatively high mass, metal-rich spirals, 
while those of the latter are low mass,  metal-poor dwarfs.  In normal 
star-forming galaxies the cluster mass distribution follows a 
power law with index $-$2, albeit truncated at high mass depending upon 
the star formation intensity ($\Sigma_{R}$, Gieles 2009). Consequently, similar 
absolute  numbers of stars are formed in low mass ($M_{\rm cl} \sim 10^{2} 
M_{\odot}$), intermediate mass ($\sim 10^{3} M_{\odot}$)  and high mass 
($\sim 10^{4} M_{\odot}$)  clusters, albeit with the former deficient
in stars at the extreme upper end of the IMF.

This star cluster mass function is repeated in nearby dwarf galaxies (Cook 
et al. 2012), but galaxy-wide triggers may induce intense, concentrated 
bursts of star formation (e.g. NGC 1569, Hunter et al. 2000), leading to 
disproportionately numerous massive star clusters (Billett et al. 2002; 
Portegies Zwart et al. 2010)\footnote{Of course, not all dwarf galaxies 
are starbursting. Within the local volume ($<$11 Mpc) only a quarter of 
the star formation from dwarf galaxies is formed during starbursts (Lee et 
al. 2009a)}.

We have attempted to set out the potential association between H\,{\sc ii} 
regions, ccSNe (and long GRBs) for star forming spirals and irregulars in 
Table~\ref{summary}. Here, the full spectrum of galaxy types has been 
distilled down to two dominant types for both spirals and irregulars, 
depending upon the rate of star formation, or more strictly the star 
formation intensity. For spiral galaxies we quote intensities with respect 
to $R_{\rm 25}$, while the $V$-band scale length $R_{D}$ (Hunter \& 
Elmegreen 2004) is used for irregulars.
Of course, intensities depend upon the diagnostic used to calculate star 
formation rates (Calzetti et al. 2007; Lee et al. 2009b;
Botticella et al. 2012), and very different intensities would follow from 
the use of alternative radii (e.g. Kennicutt et al. 2005).


From Table~\ref{summary} one would not expect ccSNe to usually be 
associated with bright regions in low intensity environments, such as 
typical early-type spirals (e.g. M~31) or dwarf irregulars (e.g. SMC), 
owing to the scarcity of giant H\,{\sc ii} regions in such hosts. Indeed, 
isolated H\,{\sc ii} regions in such galaxies would rarely produce high 
mass stars (recall Fig.~\ref{fig1}a). Exceptions do exist of course, 
including early-type spiral galaxies that possess high star formation 
intensities (e.g. M~81), plus abnormal regions within non-starburst 
irregulars such as NGC~346 in the SMC.

In contrast, the high star formation intensity of late-type spirals (e.g. 
M 101) and starburst irregulars (e.g. NGC 1569) will produce many large, 
star-forming complexes. Consequently, ccSNe will frequently be associated 
with bright star forming regions within their host galaxies. Of the 
present ccSNe sample, whose hosts are typical of high intensity late-type 
spirals, 18/37 ccSNe lie  within $\sim$300 pc of a bright (L(H$\alpha$) $> 
10^{38}$ erg\,s$^{-1}$) star forming region.
Our sample includes only two irregular galaxies, so we are unable to 
assess the situation for starburst versus non-starburst dwarf galaxies.

Nevertheless, we can re-assess the likelihood that long GRBs arise from 
moderate ($\sim$15 $M_{\odot}$) or high ($\sim$50 $M_{\odot}$) mass stars 
if the  local volume is fairly representative of metal-poor star formation 
(Lee et  al. 2009a). In the former case, long GRBs would be dominated by 
quiescent  star formation from non-starburst dwarfs, whose H\,{\sc ii} 
regions would  be isolated (and faint) since they would lack the high mass 
clusters  necessary for very massive stars, and their corresponding bright 
H\,{\sc  ii} regions). In the latter case, long GRBs would be associated 
with stars  formed in very massive clusters, since where localised 
activity takes place in dwarf galaxies, it can be very intense (Billett 
et al. 2002). 

The  tight correlation between long GRBs and the brightest regions of 
their  hosts (Fruchter et al. 2006), does not {\it prove} a link between 
high  mass  stars and long GRBs but it is certainly highly {\it suggestive}. 
Recall the  preference for broad-lined type Ic ccSNe towards dwarf galaxies 
(Arcavi  et al.  2010), the broad lined Ic-GRB connection (Woosley \& Bloom 
2006) plus the preference of long GRBs for metal-poor hosts (Levesque et al. 
2010).

Of course, the formation of dense star clusters will lead to a significant 
number of high mass runaway stars, either dynamically ejected during the 
formation process or at later stages after receiving a kick following a 
supernova explosion within a close binary system (e.g. Fujii \& Portegies 
Zwart et al. 2011). Still, the majority of high mass runaways will remain 
relatively close to their birth cluster in view of their short lifetimes 
and typical ejection velocities of $\sim$100 km\,s$^{-1}$. 

For example, 
the progenitor of the nearby GRB 980425/SN 1998bw was located 
close to another bright H\,{\sc ii} region in its (late-type) host 
galaxy, albeit offset by 0.8 kpc from a 30 Dor-like giant H\,{\sc ii} 
region (Hammer et al. 2006). Either the SN/GRB progenitor was dynamically 
ejected from the H\,{\sc ii} region at relatively high velocities, or it 
was formed in situ in a more modest star forming region, which 
is comparable to the  Rosette Nebula (Table~\ref{table2}). Evans et al. 
(2010) have identified VFTS \#016 as a high mass runaway from R136 
in the  30 Doradus, having traversed 0.12 kpc in the 1--2 Myr since the 
formation of the cluster.

Therefore, from environmental considerations one can understand the 
preference for certain flavours of ccSNe towards the brightest regions of 
their host galaxies. Still, there is little predictive power regarding 
progenitor masses, other than potentially a high likelihood for very 
massive  stars to produce long GRBs/broad-lined Ic ccSNe. 

\section{Conclusions}\label{conclusions}

We have reexamined the immediate H\,{\sc ii} environment of ccSNe from 
nearby ($\leq$15 Mpc) low inclination ($\leq$ 65$^{\circ}$), host 
galaxies. A total of 41 ccSNe have good position accuracy, of 
which ground-based H$\alpha$ imaging is available in 39 cases. Our 
findings can be summarised as follows: 
\begin{enumerate}

\item Overall, half of the ccSNe are associated with nebular emission, 
in close  agreement with van Dyk (1992). Separating these into type II and 
type Ib/c  ccSNe, 11 of the 29 hydrogen-rich ccSNe are associated with 
nebular  emission (38$\pm$11\%), versus 7 of the 10 hydrogen-poor ccSNe 
(70$\pm$26\%), supporting previous studies of Anderson \& James (2008) and 
Anderson et al. (2012).

\item Of the 18 ccSNe associated with star forming regions, 12 are 
associated with giant H\,{\sc ii} regions, with the remaining 6 associated 
with  low luminosity, extended H\,{\sc ii} regions. Overall, the mean 
H$\alpha$ luminosity of star forming regions associated with ccSNe is 
typical of a modest giant H\,{\sc ii} region, $3 \times 10^{38}$ 
erg\,s$^{-1}$, if we were to exclude SN 1970G which lies at the 
periphery of the supergiant H\,{\sc  ii} region NGC~5455 within M~101. Both 
categories have  multiple  sites of star formation, and so long duty 
cycles ($\sim$20 
Myr), implying only weak limits upon progenitor masses ($\geq$12 
$M_{\odot}$). 

\item Of the 21 ccSNe not associated with star forming regions, only one
case is coincident with a bright cluster (SN 2004dj), 
from which a 
massive progenitor ($>$20 $M_{\odot}$) would be expected. More detailed 
studies of the cluster indicate a lower progenitor mass $<$20 $M_{\odot}$ 
(Vink\'{o} et al. 2006, 2009) in common with most of the other ccSNe
that are not associated with a star forming region. In a few instances,
nearby giant H\,{\sc ii} regions indicate the possibility that the 
progenitor was a (high-mass) runaway from a putative dense star cluster.

\item 
Our primary result is that the different frequency of association with 
H\,{\sc ii} regions for hydrogen-rich (mostly II-P) and hydrogen-poor 
(Ib/c) ccSNe is attributed simply to different minimum progenitor  
mass thresholds, $\sim 8 
M_{\odot}$ and $\sim 12 M_{\odot}$, respectively, since they correspond to 
upper age limits of $\sim$50 Myr and $\sim$20 Myr. Among the type Ib/c 
ccSNe, only two cases lacked both nebular emission and a nearby giant 
H\,{\sc ii} region (SN 2002ap, SN 
2003jg), favouring the interacting binary channel ($< 20 M_{\odot}$).

\item For the present sample, 8 of the ccSNe originate from early-type spirals 
(S0/a/b), 31 from late-type spirals (Sc/d/m) and 2 from irregulars (Im), with
the majority (61\%) arising from high luminosity ($M_{\rm B} < -19$ mag) hosts.
Giant H\,{\sc ii} regions are common in these hosts because star formation 
intensities are uniformly high, whereas isolated, compact H\,{\sc ii} 
regions would be expected to dominate in low star formation intensity 
hosts. Core-collapse SNe from isolated, classical H\,{\sc ii} regions would 
provide  firm limits upon the lifetime  ($\leq$3--5 Myr) and mass 
($\geq$50--100  $M_{\odot}$) of the progenitor owing to the brief lifetime of 
such H\,{\sc ii} regions, although this would require higher
spatial resolution (HST or ground-based Adaptive Optics). An 
association between a ccSN and compact H\,{\sc ii} region has
not been observed with HST to date (Table~\ref{compare}).

\item We have also qualitatively reassessed the preference for type Ib/c 
towards the 
brightest regions of their host galaxies (Kelly et al. 2008). This is 
suggestive that a fraction of H-poor ccSNe originate from  
significantly higher mass 
stars than type II ccSNe, since high mass stars are more likely to be 
associated
with high mass clusters within large (bright) star forming complexes. 
The preference for long GRBs towards the brightest 
regions of their metal-poor hosts (Fruchter et al. 2006) is also 
suggestive of very high mass progenitors. This is because low intensity 
star forming dwarfs do not form very high mass stars, yet dominate the
overall metal-poor star formation in the local volume (Lee et al. 2009a).

\end{enumerate}

\section*{Acknowledgements}

I wish to thank the referee for suggestions which 
helped to clarify some aspects of this work. In addition, I am especially 
grateful to John Eldridge for alerting me to this issue, and to Mark 
Gieles and Rob Kennicutt for useful discussions. Janice Lee and Johan 
Knapen kindly provided me with electronic access to archival imaging, 
while Maryam Modjaz kindly provided updated SN spectral types
prior to publication. 

This study was based in part on observations made with (a)
ESO Telescopes at the La Silla 
Paranal Observatory under programme IDs 067.D-0006(A), 
069.B-0125(A), 069.D-0453(A), 075.D-0213(A), 380.D-0282(A) and 
081.B-0289(C)); (b) Gemini 
Observatory under programme GN-2009B-Q-4, which is 
   operated by the Association of Universities for Research in Astronomy, 
   Inc., under a cooperative agreement with the
    NSF on behalf of the Gemini partnership: the National Science
    Foundation (United States), the Science and Technology Facilities
    Council (United Kingdom), the National Research Council (Canada),
    CONICYT (Chile), the Australian Research Council (Australia),
    Minist\'{e}rio da Ci\^{e}ncia, Tecnologia e Inova\c{c}\~{a}o (Brazil)
    and Ministerio de Ciencia, Tecnolog\'{i}a e Innovaci\'{o}n Productiva
    (Argentina); (c) NASA/ESA Hubble 
Space Telescope,  obtained from the data archive at the Space Telescope 
Institute. STScI is  operated by the association of Universities for 
Research in Astronomy, Inc. under the NASA contract NAS 5-26555. This 
research has also used the facilities of the CADC 
operated by  the National Research Council of Canada with the support of 
the Canadian Space Agency; (d) Isaac Newton Group 
Archive which is maintained as part of the CASU  Astronomical Data Centre 
at the Institute of Astronomy, Cambridge; (e) Liverpool Telescope
Data Archive, provided by the Astrophysics Research Institute at Liverpool 
John Moores University. 

This research has made extensive use of the 
NASA/IPAC  Extragalactic Database (NED) which is operated by the Jet 
Propulsion  Laboratory, California Institute of Technology, under contract 
with the  National Aeronautics and Space Administration. We 
also acknowledge the usage of  the HyperLeda database 
(http://leda.univ-lyon1.fr).

\clearpage

\appendix

\section{Description of individual ccSNe environments}\label{description}

A brief description of the immediate environment of each ccSNe in its
host galaxy is presented, together with ground-based net H$\alpha$ and 
continuum images on a (projected) scale of 1$\times$1 kpc (2$\times$2 
kpc or 4$\times$4 kpc in some cases). A brief description of the immediate environment of each ccSNe in its
host galaxy is presented, together with ground-based net H$\alpha$ and 
continuum images on a (projected) scale of 1$\times$1 kpc (2$\times$2 kpc 
or 4$\times$4 kpc in some cases). An illustrative case is included 
below, with discussions and figures for other ccSNe presented in the 
on-line only Appendix A.

\subsection{SN 1923A in M 83}

\begin{figure*}
\includegraphics[bb=20 10 500 235, width=0.75\textwidth]{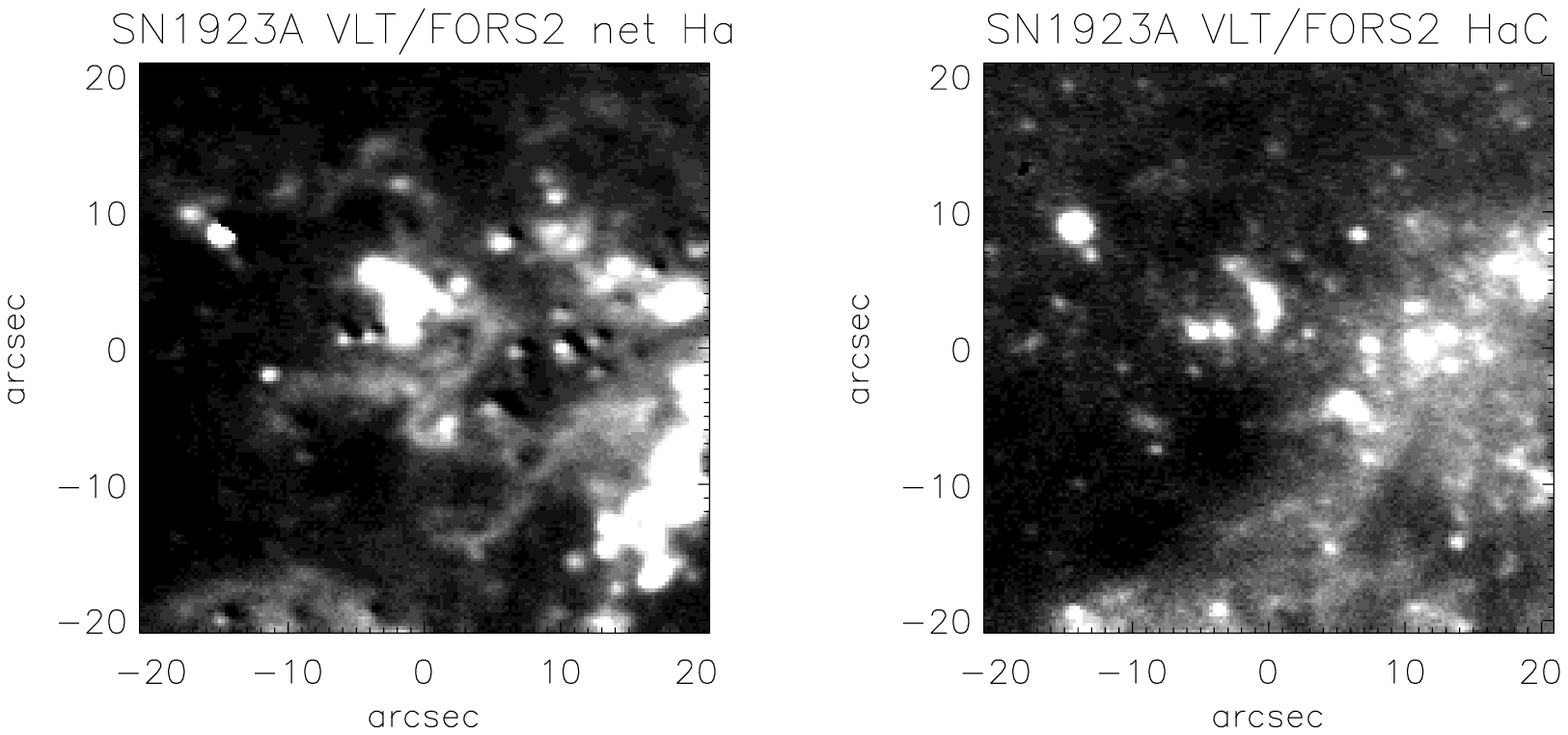}
\caption{(left) VLT/FORS2 net H$\alpha$  image (from Hadfield et 
al. 2005) showing the nebular environment of SN 1923A (at centre of 
image, Class 5). The 42$\times$42 arcsec$^{2}$ field of view 
projects to 1$\times$1 
kpc$^{2}$ at the 4.9 Mpc distance of M 83; (right) 
Continuum image ($\lambda_{c}$ = 6665\AA).  
North is up and  east is to the left  for 
these and all subsequent images.}
\label{sn1923a}
\end{figure*}

SN 1923A (II-P) was discovered in May 1923, 
2.1$'$ (0.33 $R_{\rm 25}$) NE from  the centre of M 
83 (NGC 5236, Pennington et al. 1982). 
The low inclination of M~83
implies negligible projection effects, so this corresponds
to 3.0 kpc for the adoped 4.9 Mpc distance to M 83
(1$''$ closely approximates to 25  pc). 
As illustrated in Fig.~\ref{sn1923a}, the SN position is immediately to 
the south of a bright, extended, star-forming region in our VLT/FORS2 
imaging from June 2002, \# 59 from the H\,{\sc ii} region catalogue of 
Rumstay \& Kaufman (1983). 
A giant H\,{\sc ii} region within the complex lies 4$''$ (100 pc) to 
the N of the SN position, although extended emission extends
significantly closer in our VLT/FORS2 imaging. The luminosity of 
the H\,{\sc ii} region is comparable to N66 (SMC), for which we 
measure $1.7 \times  10^{38}$ ($7.8 \times 10^{38}$)  erg\,s$^{-1}$ using 
a  1$''$ (4$''$) radius  aperture. HST WFC3 imaging (GO 11360, PI R.W. 
O'Connell) using the F657N filter provides a higher spatial view of 
the region, and reveals several point sources within the error 
circle of the SN position, plus a faint arc coincident with the SN
that extends further to the SW. A more extended star forming region,
\#79 from Rumstay \& Kaufman (1983), lies $\sim$23$''$ (0.55 kpc)  to the 
W, at the edge of Fig.~\ref{sn1923a}.

\subsection{SN 1964H in NGC 7292}

\begin{figure*}
\includegraphics[bb=20 10 500 235, width=0.75\textwidth]{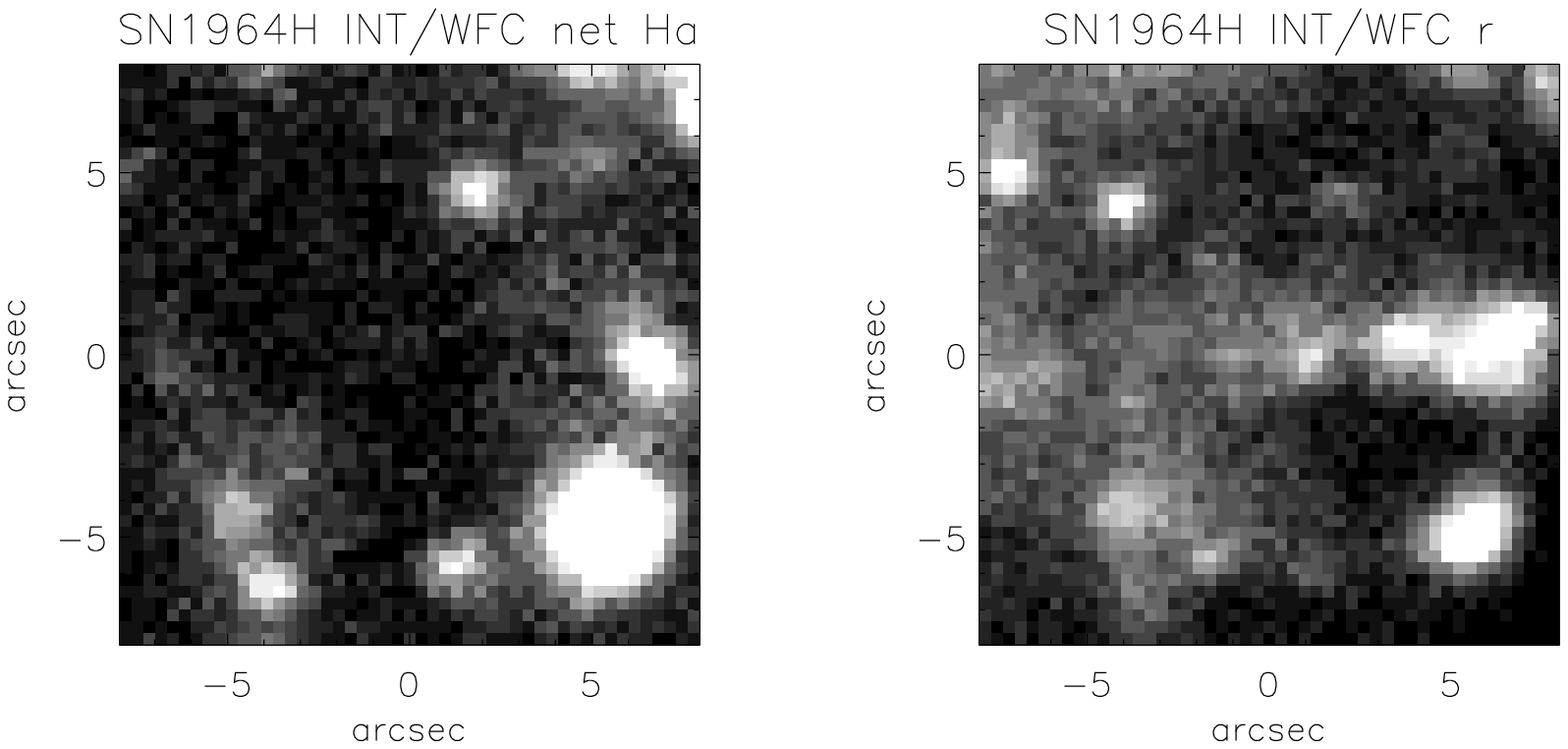}
\caption{(left) INT/WFC net H$\alpha$  image showing 
the nebular environment of SN 1964H (at centre of image, Class 2). 
The  16$\times$16 arcsec$^{2}$ field of view projects to 1$\times$1 
kpc$^{2}$ at the 12.9 Mpc distance of NGC 7292; (right) 
r-band image.}
\label{sn1964h}
\end{figure*}

This type II supernova occurred  29$''$ to the SW of the nucleus
of NGC 7292  (Porter  1993), corresponding to a deprojected distance of 
32$''$ (0.51 $R_{\rm 25}$) or 2.0 kpc for the adopted distance of 12.9 Mpc 
to NGC 7292. Calibrated JKT imaging of NGC 7292 from
Jul 2000 (James et al. 2004) has been supplemented by INT/WFC imaging
from Jul 1999. The latter are presented in Fig.~\ref{sn1964h},
indicating that the SN location is devoid of nebular emission, with the 
closest Orion-like H\,{\sc ii} region to the SN offset 5$''$ to the NW, a 
deprojected distance of 350 pc away. Brighter, giant H\,{\sc ii} regions lie 
7$''$ (0.5 kpc) to the  W 
and 8$''$ (0.55 kpc)  to the SW, with H$\alpha$-derived luminosities of 1.3 
$\times 10^{38}$  erg\,s$^{-1}$ and $4.8 \times 10^{38}$ erg\,s$^{-1}$, 
respectively. From inspection of HST WFPC2 imaging with the F300W filter 
(GO 8632, PI M.Giavalisco), no source is detected either at the position 
of the SN or the closest H\,{\sc ii} region.



\subsection{SN 1968D in NGC 6946}

\begin{figure*}
\includegraphics[bb=20 10 500 235, width=0.75\textwidth]{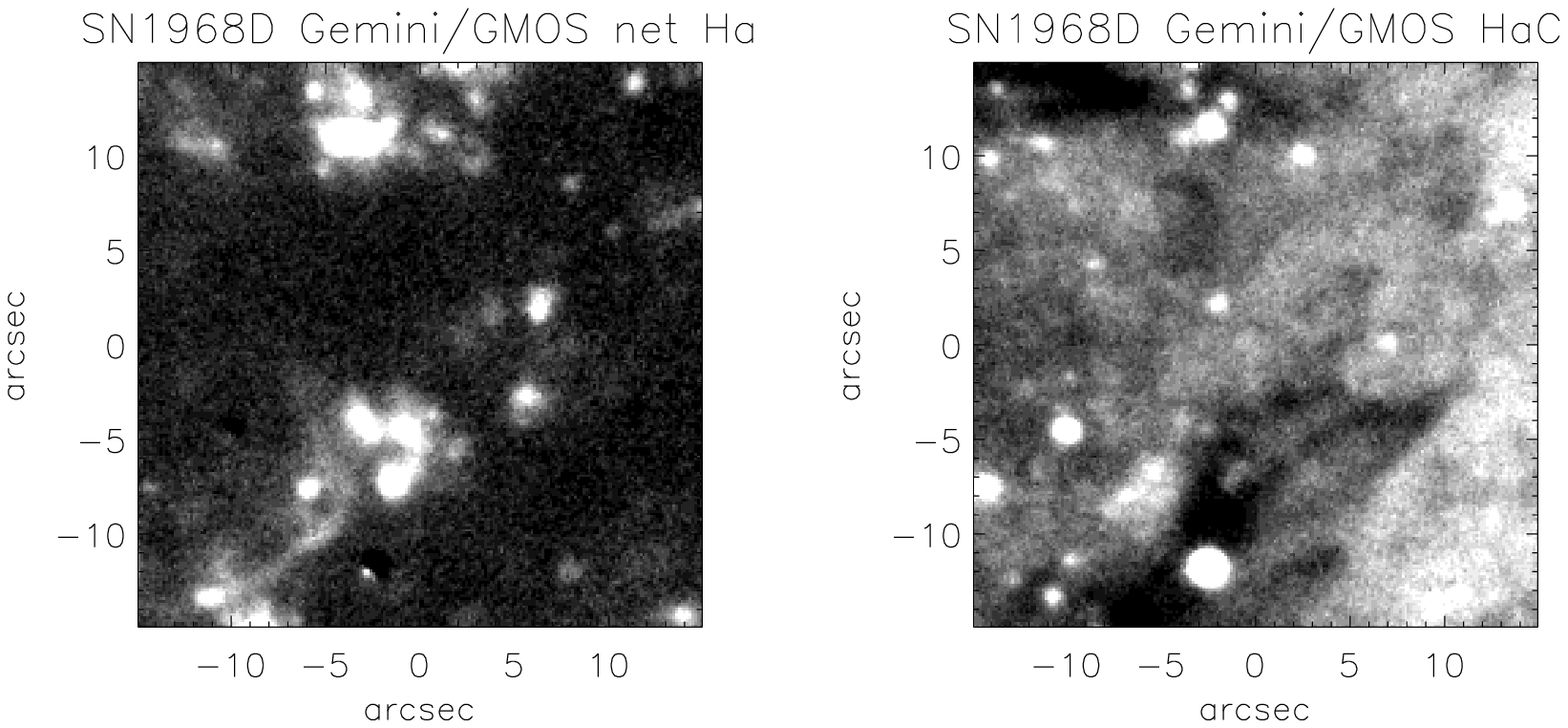}
\caption{(left) Gemini/GMOS net H$\alpha$  image (from GN-2009B-Q-4) 
showing the nebular environment of SN 1968D (at centre of image,  
Class 2). 
The 30$\times$30 arcsec$^{2}$ field of view projects to 1$\times$1 
kpc$^{2}$ at the 7.0 Mpc distance of NGC 6946; (right) 
Continuum image ($\lambda_{c}$ = 6620\AA).}
\label{sn1968d}
\end{figure*}

SN 1968D (II) was discovered in Feb 1968, in a region of NGC 6946 lacking 
nebular emission, 0.8$'$ (0.14 $R_{\rm 25}$)  NE of its centre (van Dyk et 
al. 1994). The low inclination of NGC 6946 implies projection effects are 
negligible, so 1$''$ = 35 pc for the adopted distance of 7 Mpc.
From our Gemini/GMOS imaging obtained in Aug 2009, the nearest major star
forming region is located SSE of the SN position (van Dyk et al. 1996), 
Figure~\ref{sn1968d} shows that this comprises diffuse  emission extending over 
6$''$ (200 pc) plus two  marginally extended  sources. Of these, the closest to 
the SN location  lies 4$''$.5 (150 pc)  away, and has a luminosity 
somewhat in 
excess of Orion, according to our  Gemini GMOS imaging. 

Additional isolated Orion-like H\,{\sc ii} regions 
are offset 6--7$''$ (200-230 pc) to the WNW and WSW, while brighter, 
extended  H\,{\sc ii} regions lie 12$''$ (400 pc) to the NNE and 17$''$ 
(550 pc) to  the NNW, with luminosities of 7$\times 10^{37}$ erg\,s$^{-1}$ 
and $1.4  \times 10^{38}$ erg\,s$^{-1}$, respectively. 

From inspection of 
HST/WFPC2 F547M and F656N imaging (GO 8591, PI Richstone) neither a star 
cluster
nor nebular emission are identified at the SN position. Isolated, 
moderately extended sources are responsible for the nearby  H$\alpha$ 
emission, while the brighter regions to the NNE and NNW are multiple.


\subsection{SN 1968L in M 83}

\begin{figure*}
\includegraphics[bb=20 10 500 235, width=0.75\textwidth]{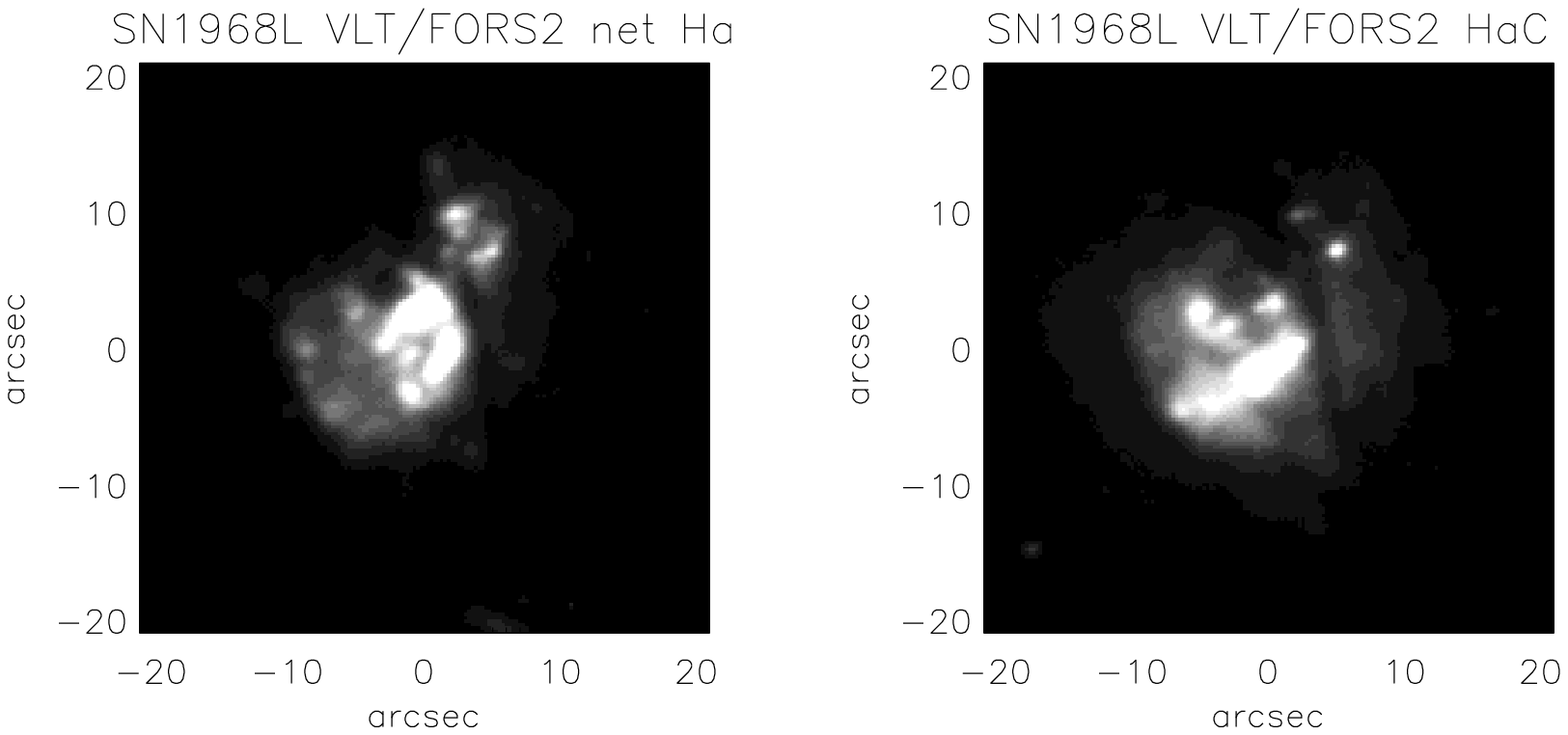}
\caption{(left) VLT/FORS2 net H$\alpha$  image 
(from Hadfield et al. 2005) 
showing the nebular environment of SN 1968L (at centre of image, 
Class 5). 
42$\times$42 arcsec$^{2}$ field of view projects to 1$\times$1 
kpc$^{2}$ at the 4.9 Mpc distance of M 83; (right) 
Continuum image ($\lambda_{c}$ = 6665\AA).}
\label{sn1968l}
\end{figure*}

This type II-P supernova was discovered in July 1968, and occurred within 
the nuclear starburst of M 83 (van Dyk et al. 1996). As for SN 1923A,
the low inclination of M~83 implies negligible projection effects, such 
that 1$''$ approximates to 25  pc at 4.9 Mpc. Figure~\ref{sn1968l} shows
that nebular emission is  extremely  strong at the position of the SN, albeit 
offset from the peak H$\alpha$ emission, which lies to the N and E. At the 
spatial resolution (FWHM $\sim$ 0$''$.9 or 20 pc) of our VLT/FORS2 imaging 
from Jun 2002, the closest H\,{\sc ii} region is located 1$''$ (25 pc) to 
the E, and 
has  a luminosity comparable to the Carina Nebula. Brighter giant H\,{\sc 
ii}  regions lie 1$''$.5 (40 pc)  to the SW and 2$''$ (50 pc) to the WNW, 
while  a  pair of exceptionally bright knots 4$''$ (100 pc) to the N, 
coincident 
with clusters \#23, \#24 and \#27 from Harris et al. (2001), have a 
combined luminosity of $1.2 \times 10^{40}$ erg\,s$^{-1}$ (similar to 30 
Doradus).  

From an inspection of the narrow-band continuum image, SN 1968L is not 
spatially coincident with a bright cluster. This is confirmed from 
inspection of HST WFPC2/F547M (GO 8234, PI D.Calzetti) and HST WCF3/F555W 
imaging (GO 11360, PI R.W. O'Connell), while WFC3/F657N imaging 
reveals solely diffuse emission at the SN position. The nebular 
emission to the E of SN 1968L is spatially extended, while a single
compact source is responsible for the giant H\,{\sc ii} region to the NW.
The nearest  continuum  sources to the SN location are clusters \#13 and 
\#15  from Harris et al. (2001) which are coincident with the giant 
H\,{\sc ii} region 1$''$.5 to the SW of the supernova position. 


\subsection{SN 1970G in M 101}

\begin{figure*}
\includegraphics[bb=20 10 500 235, width=0.75\textwidth]{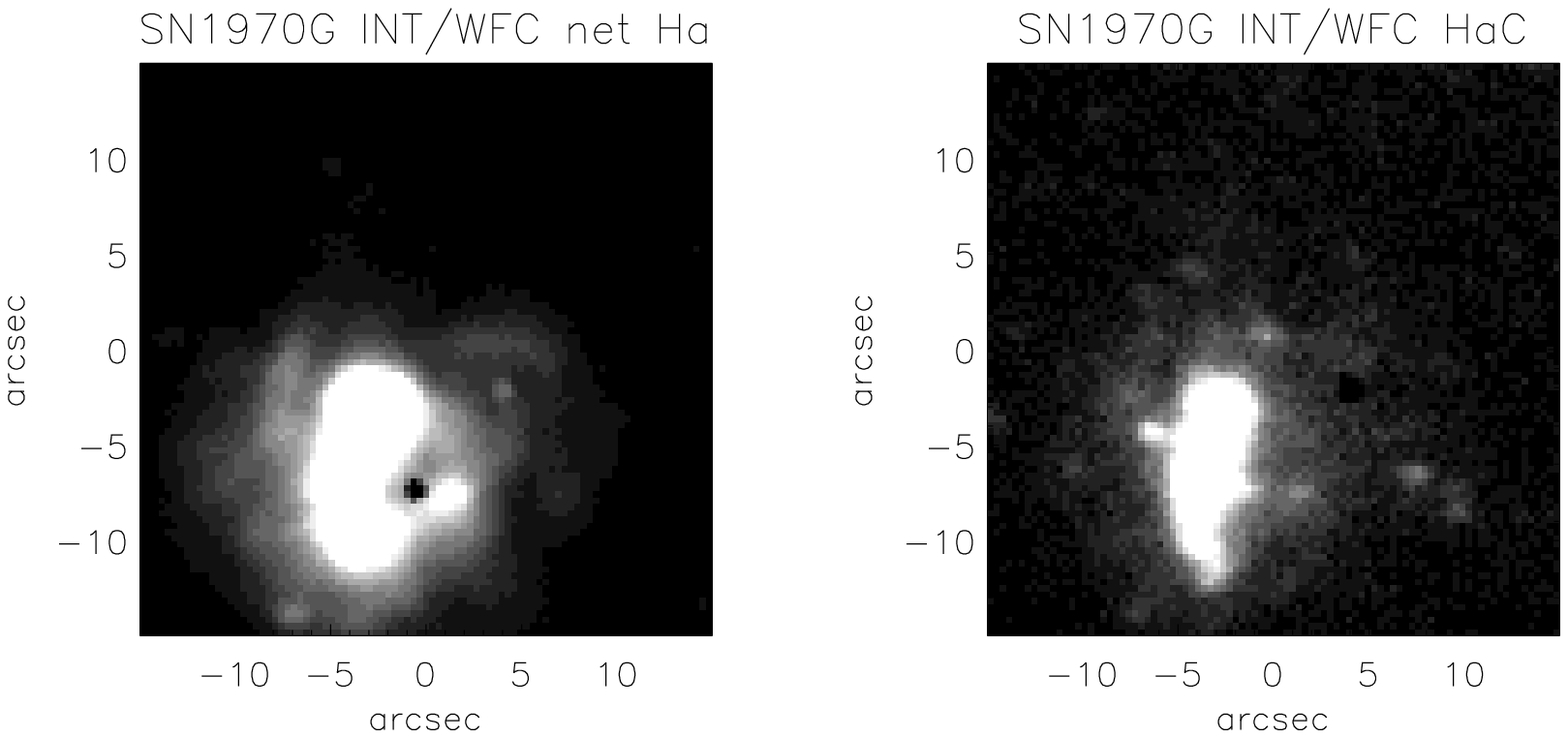}
\caption{(left) INT/WFC net H$\alpha$  image 
showing the nebular environment of SN 1970G (at centre of image,  
Class 5). 
The 30$\times$30 arcsec$^{2}$ field of view projects to 1$\times$1 
kpc$^{2}$ at the 6.96 Mpc distance of M 101; (right) 
Continuum image ($\lambda_{c}$ = 6657\AA).}
\label{sn1970g}
\end{figure*}

SN 1970G (II-L) was discovered in Aug 1970, and occurred  close to a very 
bright, spatially extended H\,{\sc ii} region NGC~5455 (Allen 
et al. 1976; Cowan et al. 1991), 6.6$'$ (0.46 $R_{\rm 
25}$) SW  of the nucleus of M 101  (NGC 5457). Owing to the low
inclination of M~101, there are no projection effects, so this
corresponds to a distance of 13.5 kpc for the adopted 6.9 Mpc
distance to M 101 (3$''$ corresponds to 100 pc). Due to 
the low resolution of  the  Hoopes 
et al.  (2001) imaging, we employ higher spatial resolution INT/WFC  
imaging from Jun 2006 (FWHM$\sim$1$''$.1) to assess the detailed nebular 
morphology.  From Fig.~\ref{sn1970g}, the SN position is located at the 
periphery of NGC~5455,  with the peak  H$\alpha$ and continuum emission 
(\# 416  from Hodge et al. 1990) located 5$''$ (170 pc)  to the SE of the 
SN  location. The  H$\alpha$-derived luminosity of this   region is 
2$\times  
10^{40}$  erg\,s$^{-1}$, comparable to 30 Doradus,  based on the Hoopes et  
al. (2001) imaging. The INT/WFC imaging reveals a faint continuum source 
within $\sim$1$''$ of the position of SN 1970G (right panel of 
Fig.~\ref{sn1970g}. Higher resolution 
HST WFPC2 imaging using the F606W filter (GO 6713, PI W.B. Sparks)  
suggests that the SN  is not coincident with nebular emission, with 
the nearest bright continuum source offset E by 1$''$.6.


\subsection{SN 1980K in NGC 6946}

\begin{figure*}
\includegraphics[bb=20 10 500 235, width=0.75\textwidth]{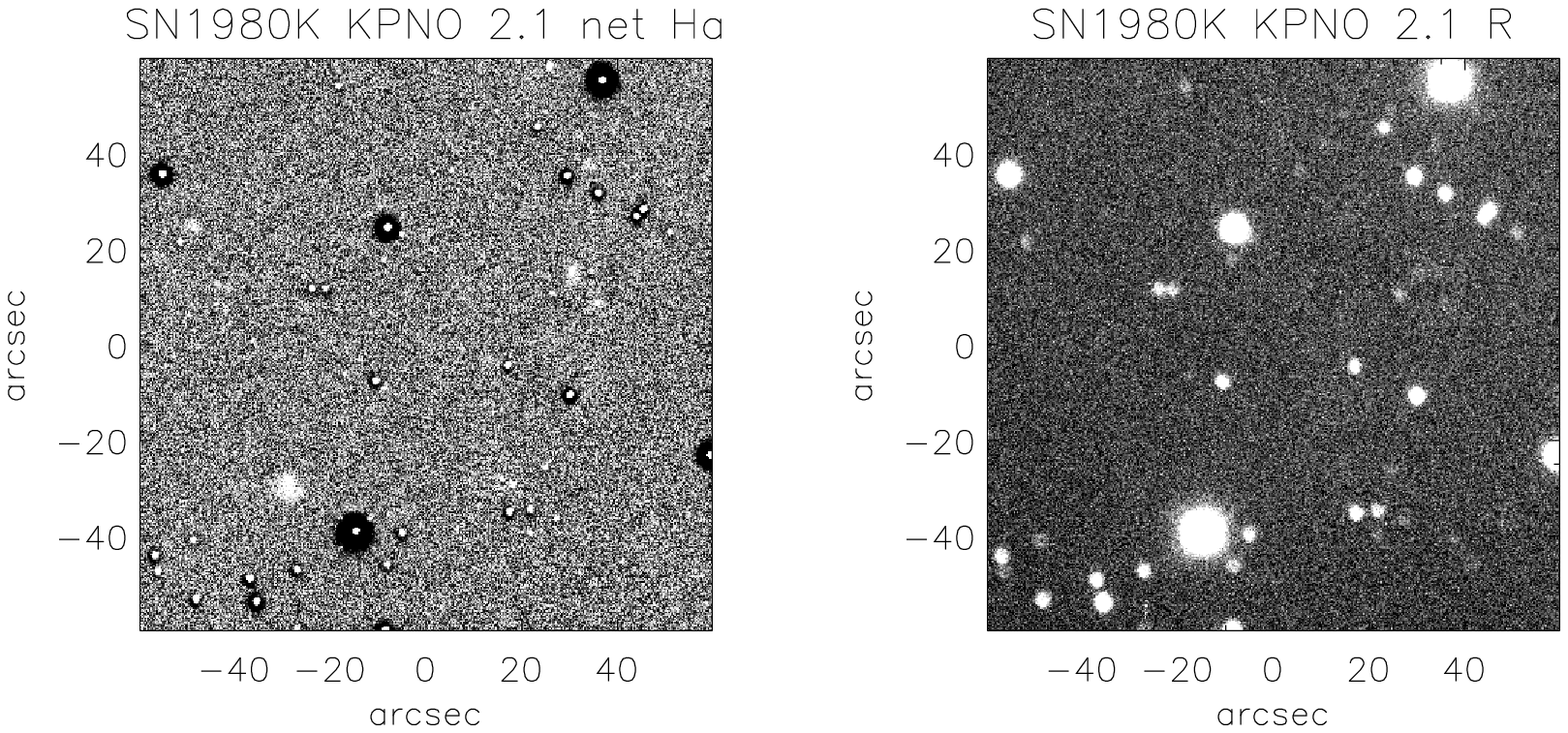}
\caption{(left) KPNO 2.1m net H$\alpha$  image (from Kennicutt et al.
2003) 
showing the nebular environment of SN 1980K (at centre of 
image, Class 2). Bright field stars are poorly subtracted. 
The 120$\times$120 arcsec$^{2}$ 
field of view projects to {\bf 4$\times$4} kpc$^{2}$ at the 7 Mpc distance 
of NGC 6946; 
(right) R-band image.}
\label{sn1980k}
\end{figure*}

This type II-L supernova, discovered in Oct 1980, occurred in the outer
disk of NGC~6946,  5$'$.45 to the SE of the nucleus, which corresponds
to a deprojected distance of 5$'$.7 (0.99 $R_{\rm 25}$) or 11.5kpc
for the 7 Mpc distance to NGC 6946.
Based on the radio SN position from Weiler et al. (1992), it
fell beyond the field-of-view of our Gemini GMOS imaging, so instead we
employ the KPNO 2.1m imaging from Mar 2001 (Kennicutt et al. 2003). 
Fig.~\ref{sn1980k} shows that there is no  
significant H$\alpha$ emission close to the position of the SN.  
Extended H\,{\sc ii} regions lie 50$''$ (1.75 kpc) W, 60$''$ (2.1 kpc) WNW and 
55--65$''$ (1.9--2.3 kpc) NE, with luminosities of $2 \times 10^{37}$  
erg\,s$^{-1}$, $4 \times 10^{37}$  erg\,s$^{-1}$ and $5  \times 10^{37}$ 
erg\,s$^{-1}$, respectively. SN 1980K is seen as a point source in 
HST/WFPC2 F606W imaging (GO 11229, PI M. Meixner), with a second, 
visually fainter point source 0$''$.5 (20 pc) to its E (Sugerman et al. 
2012).


\subsection{SN 1983N in M 83}

\begin{figure*}
\includegraphics[bb=20 10 500 235, width=0.75\textwidth]{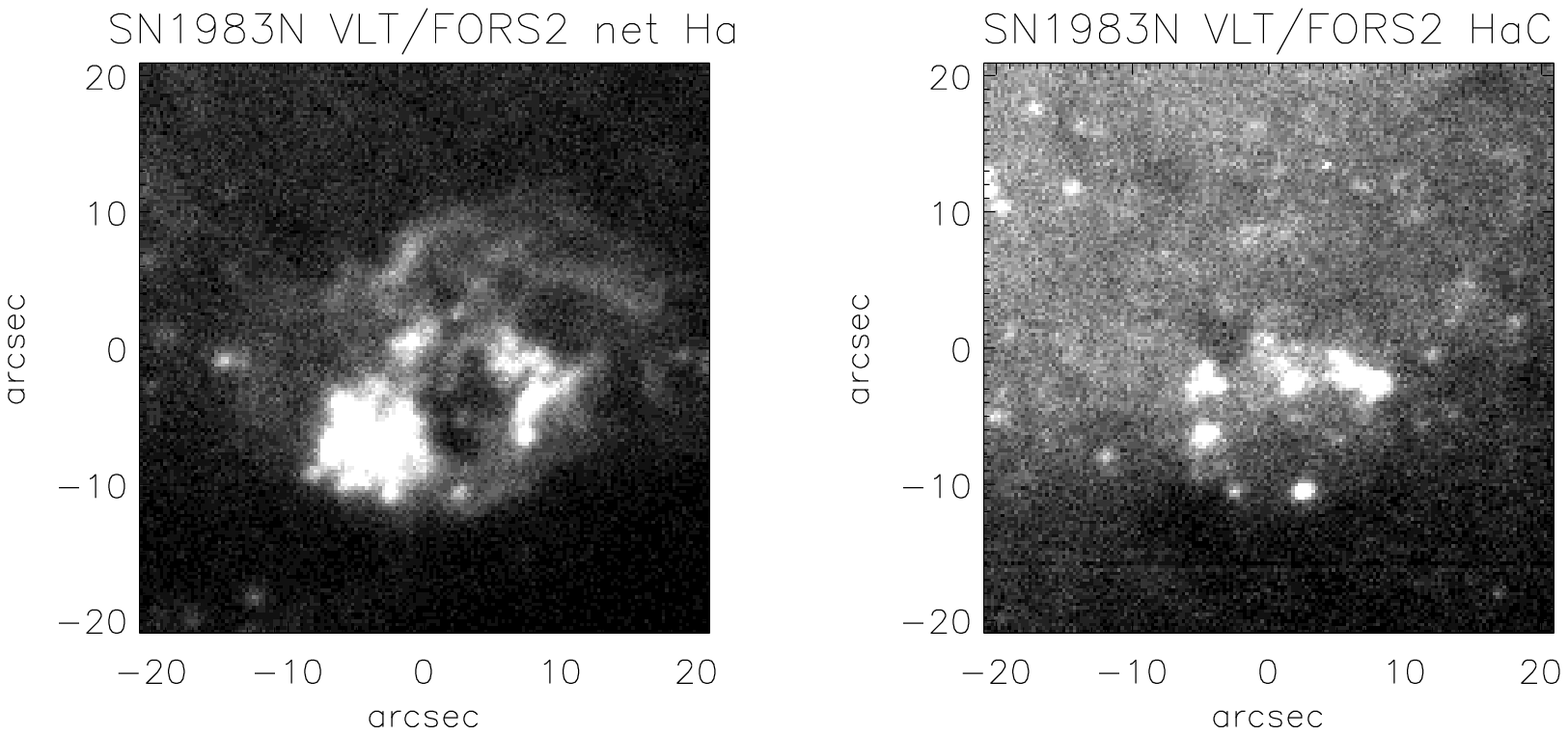}
\caption{(left) VLT/FORS2 net H$\alpha$  image (from Hadfield et 
al. 2005) showing the nebular environment of SN 1983N (at centre of 
image, Class 4). The 
42$\times$42 arcsec$^{2}$ field of view projects to 1$\times$1 
kpc$^{2}$ at the 4.9 Mpc distance of M 83; (right) 
Continuum image ($\lambda_{c}$ = 6665\AA).}
\label{sn1983n}
\end{figure*}

The type Ib SN 1983N was discovered in July 1983 (Wamsteker 1983), and 
occurred 3$'$ SW of the nucleus of M 83 (NGC 5236), corresponding to a
deprojected distance of 3$'$.1 (0.48 $R_{\rm 25}$) or 4.4 kpc for the
adopted distance of 4.9 Mpc to M 83. Based upon the radio 
position of Sramek et al. (1983) this is close to a complex of star 
formation to the E, SW 
and very bright  nebular emission to the SE. Clocchiatti et al. (1996)
enable relative astrometry from our Jun 2002 VLT/FORS2 imaging. 
Figure~\ref{sn1983n} illustrates that an extended H\,{\sc ii} region, 
\#222 from the catalogue of Rumstay \&  Kaufman (1983), lies  1$''$ (25 pc) to the 
east of the SN, which has a  luminosity intermediate between the Orion and 
Rosette 
nebulae. An arc of  nebular emission lies 9$''$ (220 pc) to the SW of SN 1983N, 
alias \# 234 from Rumstay \& Kaufman (1983). This has an integrated  luminosity of 
$1.6 
\times 10^{38}$ erg\,s$^{-1}$ (3$''$  radius) while a giant H\,{\sc ii} 
region, \#220 from Rumstay \& Kaufman  (1983), peaks  9$''$ (220 pc) to 
the SE and and has a luminosity comparable to N66 (SMC). Broad 
band HST/STIS  imaging (GO 9148, PI P. Garnavich) reveals several faint 
sources  consistent with the SN error box, with brighter clusters 
coincident peak 
emission from the H\,{\sc ii} regions to the E, SW and SE.



\subsection{SN 1985F in NGC 4618}

\begin{figure*}
\includegraphics[bb=20 10 500 235, width=0.75\textwidth]{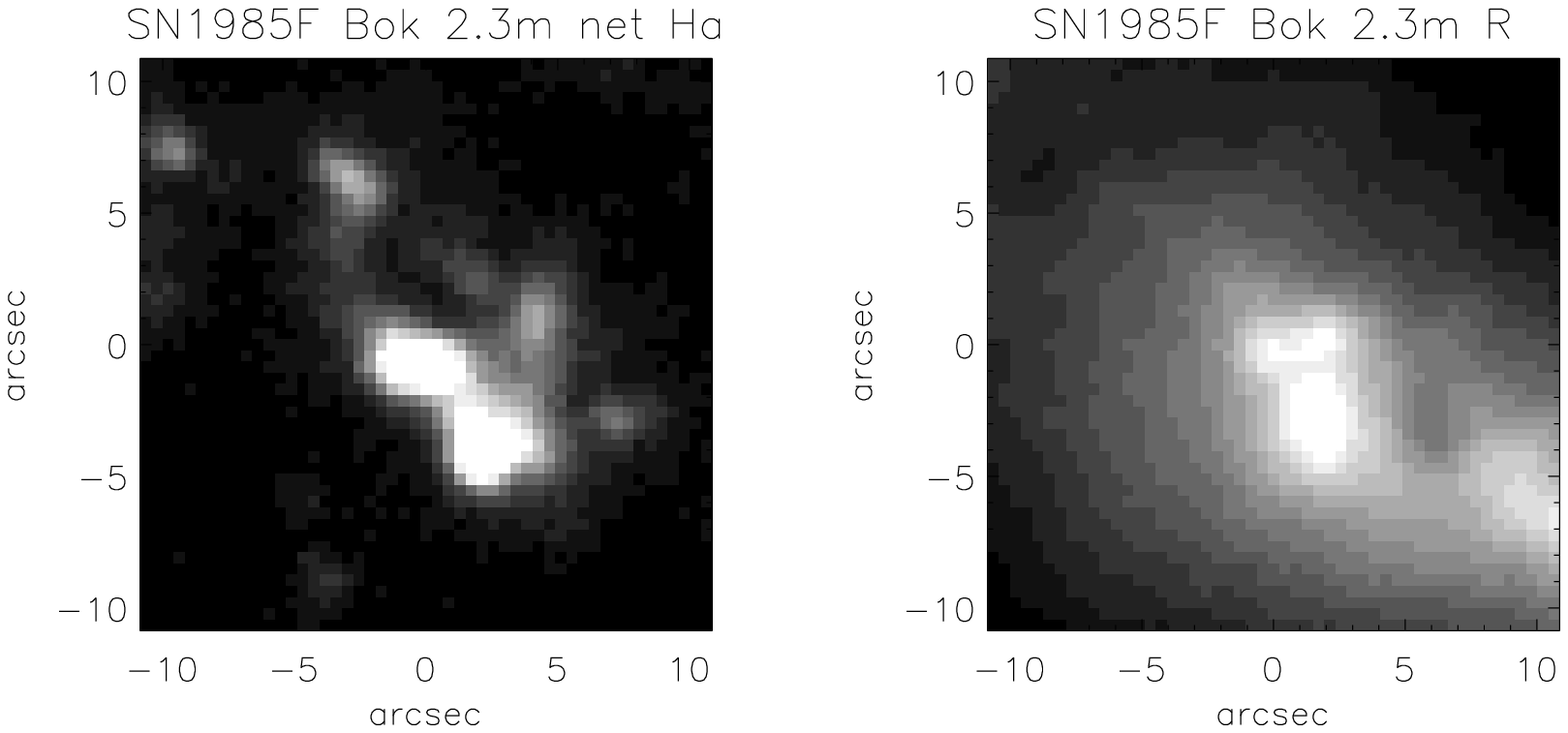}
\caption{(left) Bok 2.3m net H$\alpha$  image (from Kennicutt et al. 
2008) 
showing the nebular environment of SN 1985F (at 
centre of image, Class 5). The 22$\times$22 arcsec$^{2}$ field of 
view projects to 
1$\times$1 kpc$^{2}$ at the 9.2 Mpc distance of NGC 4618; (right) 
R-band image.}
\label{sn1985f}
\end{figure*}

This type Ib/c supernova was discovered in Feb 1985 (Filippenko \& 
Sargent 1985, 1986, Modjaz, priv. comm.), and is spatially coincident  
with a bright H\,{\sc ii} 
region and cluster, 10$''$ away from the centre of NGC~4618 (Filippenko et 
al. 1986), as shown in Fig.~\ref{sn1985f}. Based upon the HyperLeda 
inclination and PA of the major axis of NGC 
4618 this corresponds to a deprojected offset of 12$''$ (0.1 $R_{\rm 25}$)
or 0.5 kpc for the adopted distance of 9.2 Mpc to NGC 4618. Using 
the SN position of Filippenko \& Sargent (1986), the
2.3m Bok imaging of NGC~4618 from Apr 2001 (Kennicutt et al.
2008) reveals that the H$\alpha$-derived luminosity of this (extended)
region is N66-like, and is comparable to that of a neighbouring H\,{\sc 
ii} region, 4$''$ to the SW of SN 1985F, a deprojected distance of 220 pc 
away. HST/WFPC2 imaging using the F606W filter (GO 5446, PI Illingworth) 
confirms that the SN is coincident with  an extended source, while each of 
the ground-based continuum sources to the W and SW 
in spatially resolved into two primary extended clusters.





\subsection{SN 1986L in NGC 1559}

\begin{figure*}
\includegraphics[bb=20 10 500 235, width=0.75\textwidth]{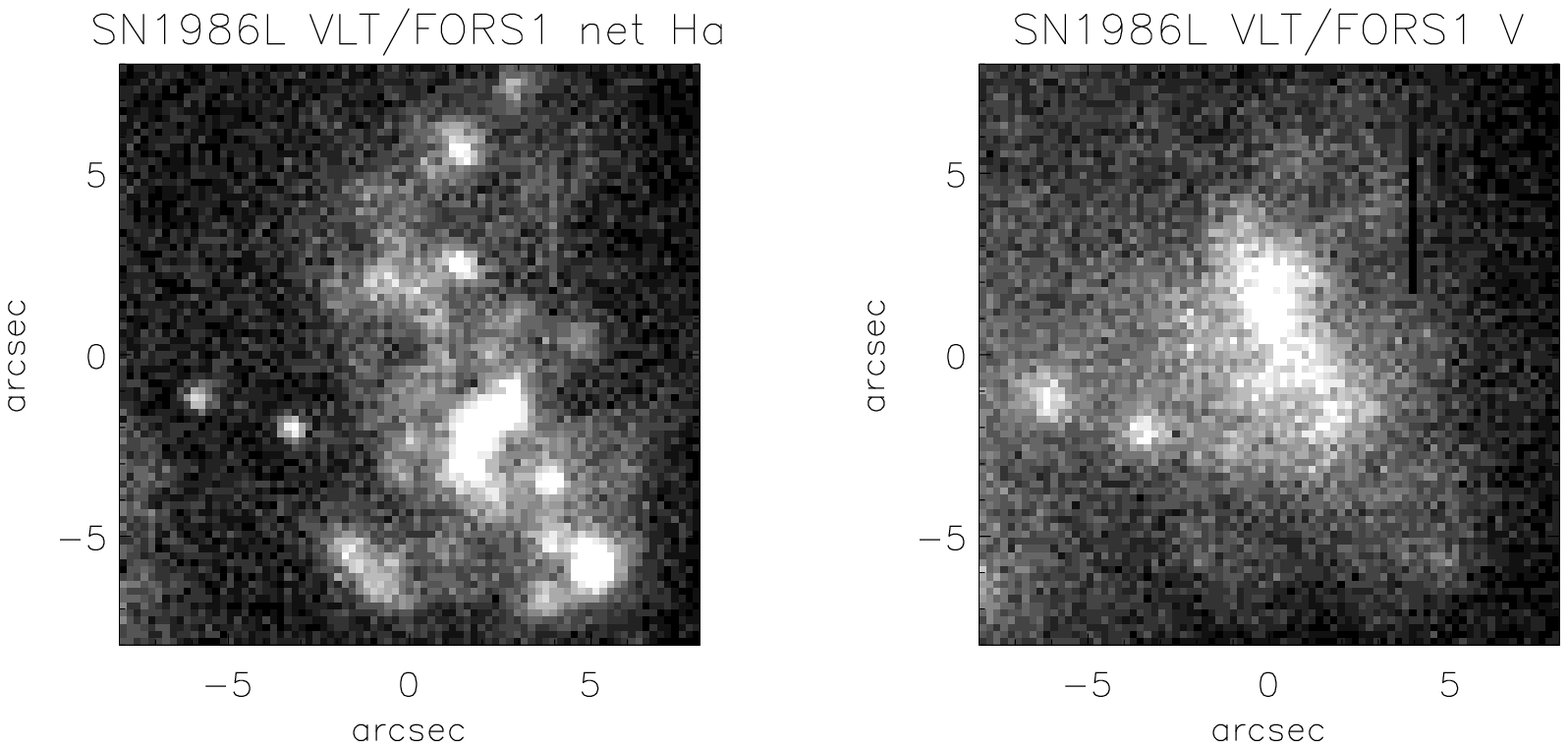}
\caption{(left) VLT/FORS1 net H$\alpha$  image (from
075.D-0213(A)) showing the nebular environment of SN 1986L (at 
centre of image, Class 3). The 16$\times$16 arcsec$^{2}$ field of 
view projects to 
1$\times$1 kpc$^{2}$ at the 12.6 Mpc distance of NGC 1559; (right) 
V-band image.}
\label{sn1986l}
\end{figure*}

SN 1986L (II-L) was discovered in NGC~1559 in Oct 1986 (Evans et al.
1986). Figure~\ref{sn1986l} presents net H$\alpha$ VLT/FORS1 imaging 
obtained in Aug 2005 and reveals that the SN lies in  an  extended region 
of nebular emission, which extends north from a giant H\,{\sc ii} region 
several arcsec to the SW. This  complex is 45$''$ west of the centre of 
NGC 1559, and corresponds to
a deprojected offset of 1$'$ (0.55 $R_{\rm 25}$) based on the
HyperLeda inclination and major axis PA, equivalent to 3.5 kpc for the
12.6 Mpc distance to NGC 1559 (100 pc corresponds to 1$''$.3). 
The nebular flux at the SN 
position is relatively faint, based upon the astrometry of McNaught \& 
Waldron (1986). H\,{\sc ii} 
regions are located 2$''$ (150 pc) to the NNE and  3$''$ (230 pc) to the 
NW with luminosities of $8 \times 10^{37}$  erg\,s$^{-1}$ and $1.2 \times 
10^{38}$ erg\,s$^{-1}$, respectively. In  addition, a giant H\,{\sc ii} 
region $\sim$3$''$ (230 pc) to the SW of SN  1986L has a luminosity 
comparable to the Carina Nebula. Unfortunately, SN 1986L occurred outside
the field of view of HST/WFPC2  imaging (GO 9042, PI S.J. Smartt).



\subsection{SN 1987A in LMC}

\begin{figure*}
\includegraphics[bb=20 10 500 235, width=0.75\textwidth]{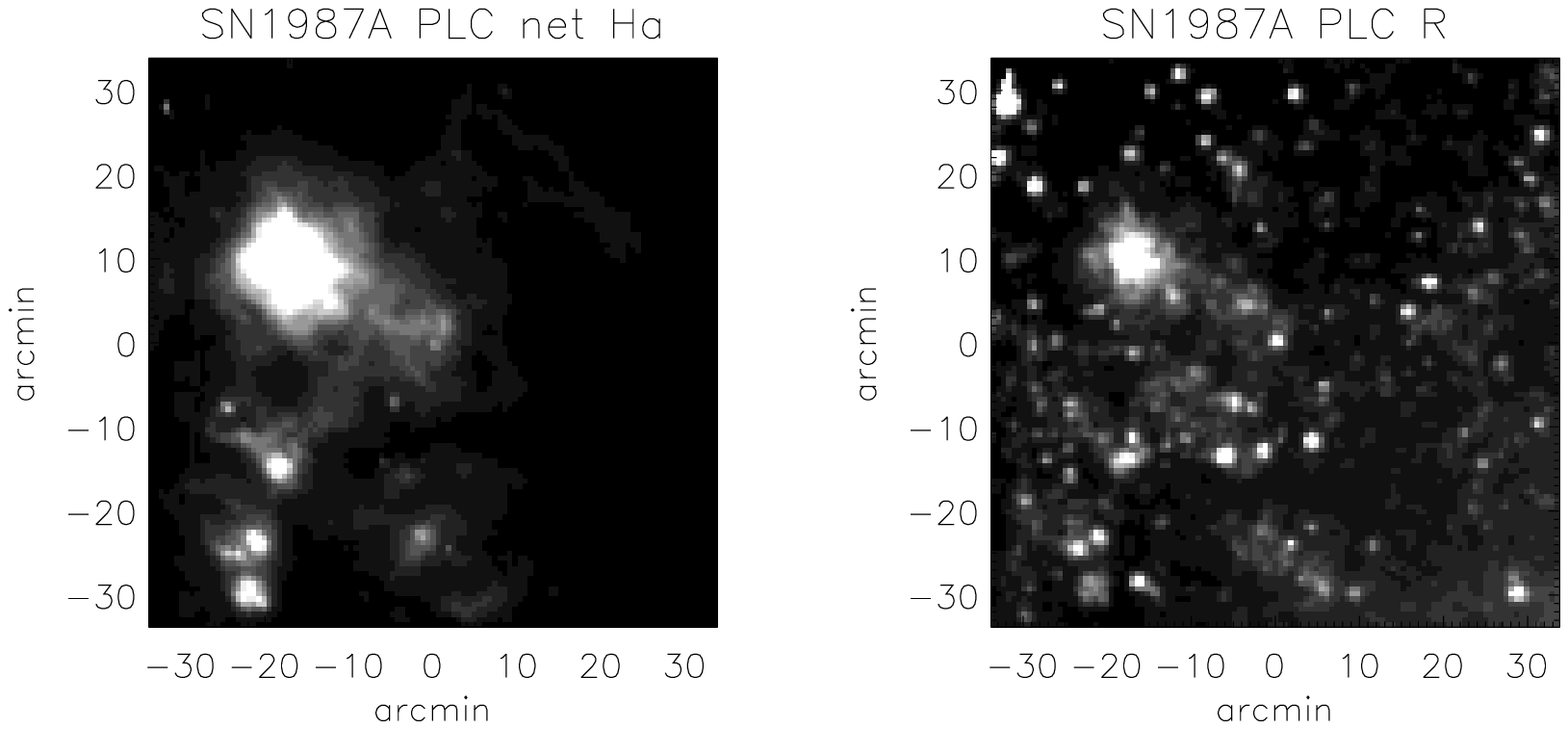}
\caption{(left) Parking Lot Camera net H$\alpha$  image (from Kennicutt et 
al. 1995) showing the nebular environment of SN 1987A (at centre of 
image, Class 4). The 68$\times$68 arcmin$^{2}$ field of view 
projects 
to  1$\times$1 kpc$^{2}$ at the 50 kpc distance of the LMC; (right) R-band 
image (from Bothun \& Thompson 1988).}
\label{sn1987a}
\end{figure*}

SN 1987A, the best studied supernova of the modern era, lies in the 
periphery of the 30 Doradus (Tarantula Nebula) region of the LMC. 
H$\alpha$ imaging from the Nikon Survey Camera (M.S. Bessell, priv. comm.) 
and MCELS (Smith et al. 2000) indicates faint nebular emission at the 
position of the SN. In Figure~\ref{sn1987a}, we present the lower spatial 
resolution Parking Lot Camera images (Kennicutt et al. 1995), 70$''$ or 17 
pc at the 50 kpc LMC distance. Nebular emission is present at the SN site, 
although it is relatively faint, and would not necessarily be detected in 
ground-based  imaging of other ccSNe in our survey. Still, SN 1987A is 
is close proximity to a pair of bright knots of nebular 
emission 2--3$'$ ($\sim$40 pc) to the NW, with an 
integrated luminosity of $6 \times 10^{37}$ erg\,s$^{-1}$ (2$'$ radius 
aperture), as measured from Nikon continuum-subtracted imaging, calibrated 
via Kennicutt et al. (1995). The centre of 30 Doradus lies 20$'$ (300 pc) 
to the NE, with an integrated luminosity of $6 \times 10^{39}$ 
erg\,s$^{-1}$ (10.5$'$ radius aperture). Panagia et al. (2000) discuss the 
faint cluster coincident with SN 1987A, but again this would not be 
detected via ground-based imaging of SN beyond the Local Group.




\subsection{SN 1992ba in NGC 2082}


This type II supernova was discovered by R. Evans in NGC~2082 in late Sep 
1992 (Evans \& Phillips 1992). SN 1992ba occurred
26$''$ to the W of the nucleus of NGC~2082, which deprojects to 29$''$ 
(0.5 $R_{\rm 25}$) based on the low inclination from HyperLeda, plus an 
adopted PA=0.This corresponds to a galactocentric distance of 1.9 kpc for 
the 13.1 Mpc distance to NGC 2082. We do not have access to calibrated 
H$\alpha$ imaging, so we have inspected archival R-band imaging from Oct 
1992 (AAT Prime Focus) and Apr 2000 (3.5m NTT/EMMI). The latter are
reveal continuum sources 2$''$ NW and 2$''$.5 SE, with a brighter 
source 3$''$.5 SE of SN 1992ba.  Schmidt et al.  (1994) note that this 
supernova occurred in, or near, a bright H\,{\sc ii} region.



\subsection{SN 1993J in M 81}

\begin{figure*}
\includegraphics[bb=20 10 500 235, width=0.75\textwidth]{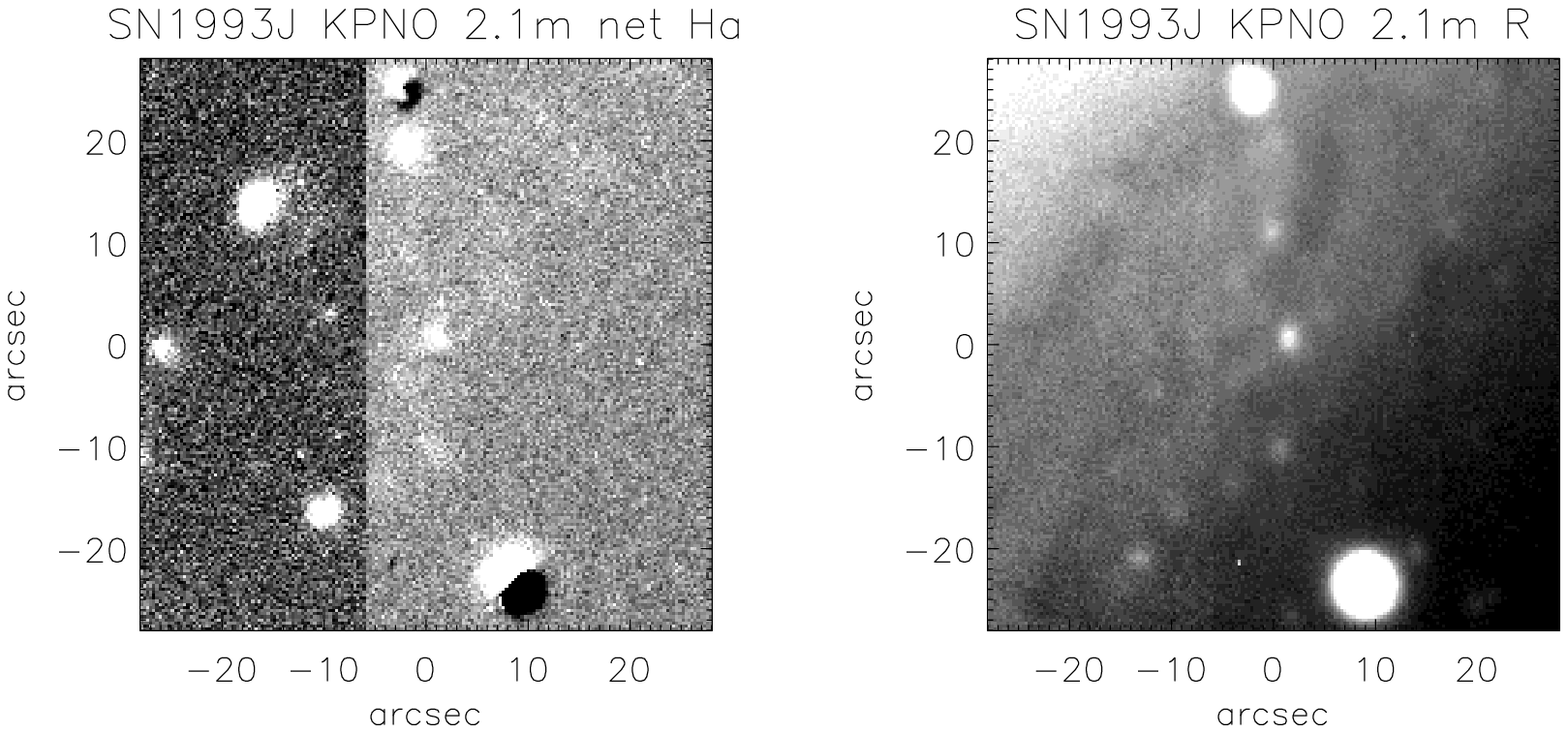}
\caption{(left) KPNO 2.1m net H$\alpha$  image (from Kennicutt et al. 
2003)
showing nebular emission close to the position of SN 1993J (at 
centre of image, Class 2). The 56$\times$56 arcsec$^{2}$ field of 
view projects to 
1$\times$1 kpc$^{2}$ at the 3.6 Mpc distance of M~81. The apparent change in
sky background arises from the mosaicing of several pointings of M~81; 
(right) R-band image.}
\label{sn1993j}
\end{figure*}

This well-studied type IIb supernova was discovered in March 1993 
(Ripero et al. 1993), and occurred 2$'$.8  SSW of the nucleus of M 81 (NGC 
3031). The HyperLeda inclination of M~81 is high (62.7$^{\circ}$ so
this position corresponds to a deprojected offset of 4$'$.4 
(0.32 $R_{\rm 25}$), equivalent to a distance of 4.6 kpc for the
3.65 Mpc distance to M 81. From the KPNO 2.1m imaging from Mar 
2001 (Kennicutt et al. 2003), shown in Fig.~\ref{sn1993j},
SN 1993J is spatially coincident with a  faint  nebular emission,  likely arising 
from the SNR itself. A  brighter,  Orion-like region  lies 21$''$ to the NE, 
or a deprojected distance of 580 pc. An extended 
H\,{\sc ii}  region, with a H$\alpha$+[N\,{\sc ii}] flux  of $4 \times 
10^{-14}$ erg\,s$^{-1}$\,cm$^{-2}$ lies 38$''$ (1.0 kpc deprojected) to the NW. 
The blue supergiant companion to the red supergiant progenitor is detected
in late-time spectroscopy (Maund et al. 2004) and spectroscopy
(Maund \& Smartt 2009), the latter based upon extensive HST ACS and WFPC2 
imaging. 



\subsection{SN 1994I in M 51a}

\begin{figure*}
\includegraphics[bb=20 10 500 235, width=0.75\textwidth]{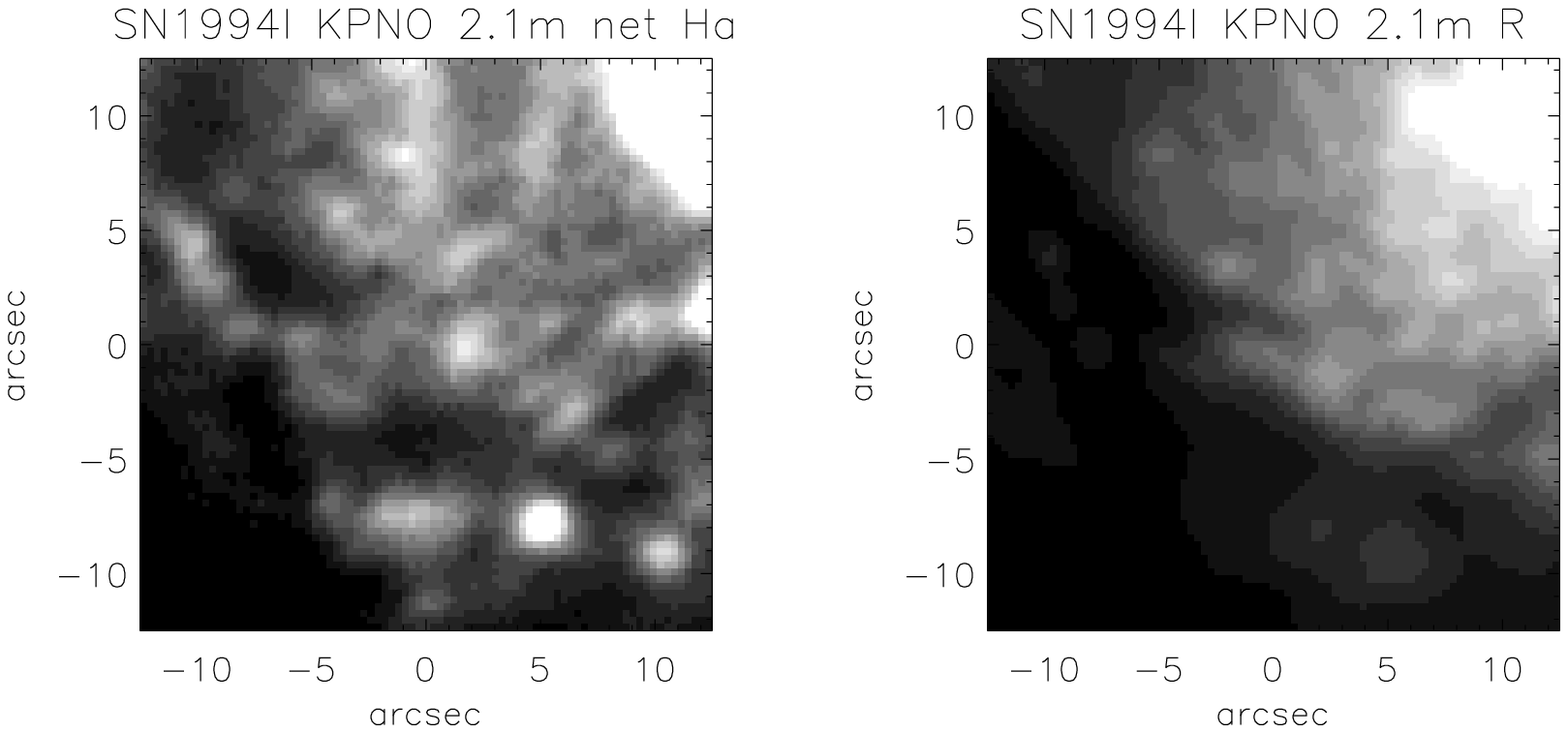}
\caption{(left) KPNO 2.1m net H$\alpha$  image (from Kennicutt et al. 
2003)
showing nebular emission close to the position of SN 1994I (at 
centre of image, Class 5). The 25$\times$25 arcsec$^{2}$ field of 
view projects to 
1$\times$1 kpc$^{2}$ at the 8.4 Mpc distance of M~51a; (right) R-band image.}
\label{sn1994i}
\end{figure*}

This type Ic supernova occurred close to the nucleus of M 51a in Apr 1994 
(Puckett et al. 1994), 18$''$ to its SE. The low inclination of 
32.6$^{\circ}$ for M 51a implies negligible projection effects, so this 
corresponds to a distance of 0.8 kpc (0.06 $R_{\rm 25}$) for the adopted 
distance of 8.39 Mpc. Based on radio-derived coordinates from Rupen et al. 
(1994), Fig.~\ref{sn1994i} illustrates that SN 1994I lies within a large 
region of diffuse nebular emission. The closest identifiable source is a 
giant H\,{\sc ii} region 2$''$ (80 pc) to the W in the KPNO 2.1m imaging 
from Mar 2001 (Kennicutt et al. 2003). Brighter sources lie 9$''$ (0.38 
kpc) SW and 13$''$ (0.55 kpc) W. SN 2004I does not appear to coincide with 
a bright star cluster in the KPNO 2.1m R-band imaging, although 
identification is severely hindered by diffuse emission and the moderate 
spatial resolution of the ground-based images. We have therefore inspected 
archival Hubble Space Telescope ACS/WFC (GO 10452, PI S. Beckwith) images 
obtained with the F555W and F658N (H$\alpha$+[N\,{\sc ii}]) filters (e.g.  
Chandar et al. 2011), which confirm no bright cluster is spatially 
coincident with SN 1994I.





\subsection{SN 1995V in NGC 1087}

\begin{figure*}
\includegraphics[bb=20 10 500 235, width=0.75\textwidth]{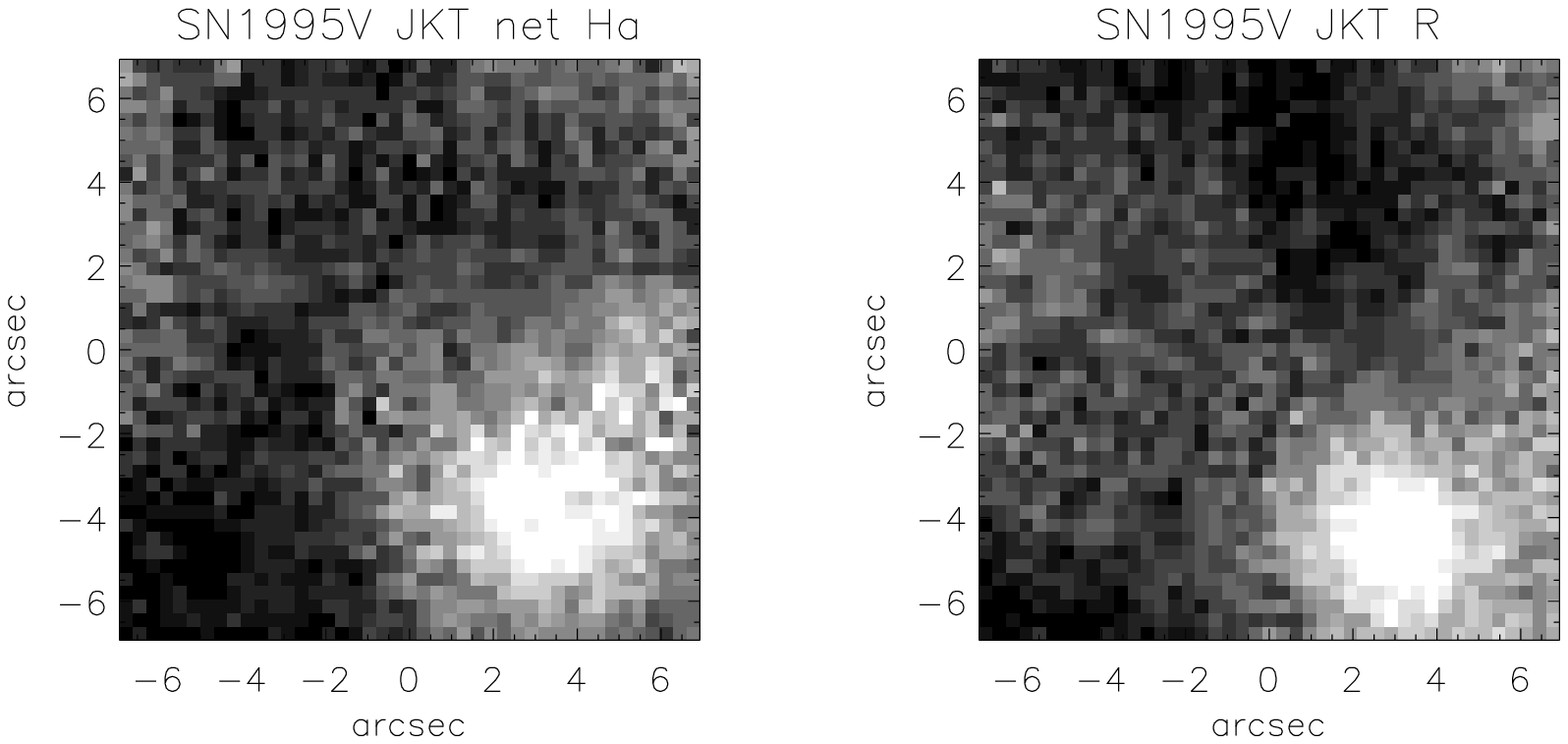}
\caption{(left) JKT net H$\alpha$  image (from James et al. 2004)
showing the nebular environment of SN 1995V (at 
centre of image, Class 2). The 14$\times$14 arcsec$^{2}$ field of 
view projects to 
1$\times$1 kpc$^{2}$ at the 14.4 Mpc distance of NGC~1087; (right) R-band image.}
\label{sn1995v}
\end{figure*}

This type II-P supernova was discovered in Aug 1995 (Evans et al. 1995), 
and occurred 24$''$ E of the  nucleus of NGC 1087 (Balam 1995). This
corresponds to a deprojected offset of 40$''$ (0.36 $R_{\rm 25}$) or
2.8 kpc for the adopted distance of 14.4 Mpc to NGC 1087. Fig.~\ref{sn1995v}
shows JKT H$\alpha$ imaging from Jan 2000 (James et al. 2004). Although
seeing conditions were poor (FWHM$\sim$3$''$.5), SN 1995V does not 
appear to be coincident with nebular emission, with the nearest identifiable 
H\,{\sc ii} region (and bright cluster) located 5$''$ (0.6 kpc) to the SW. 





\subsection{SN 1995X in UGC 12160}


The type II supernova SN 1995X was discovered in UGC 12160 in Aug 1995 
(Mueller et al. 1995). We do not have access to calibrated H$\alpha$ 
imaging  of UGC 12160, although Mueller et al. (1995) report narrow 
H$\alpha$ emission from a H\,{\sc ii} region superimposed upon an early SN 
spectrum. The SN occurred 22$''$  NW of the centre of UGC 
12160 (Sicoli et al.  1995), which corresponds to a deprojected distance
of 26$''$ (0.42 $R_{\rm 25}$) or 1.8 kpc for the adopted distance of 14.4 Mpc to
UGC 12160. At this location, Anderson et al. (2012) 
report a NCR pixel value of 0.903 from Liverpool Telescope  RATCam 
imaging, indicating it occurred close to the peak of H$\alpha$ emission from
the host.


\subsection{SN 1996cr in Circinus}

\begin{figure*}
\includegraphics[bb=20 10 500 235, width=0.75\textwidth]{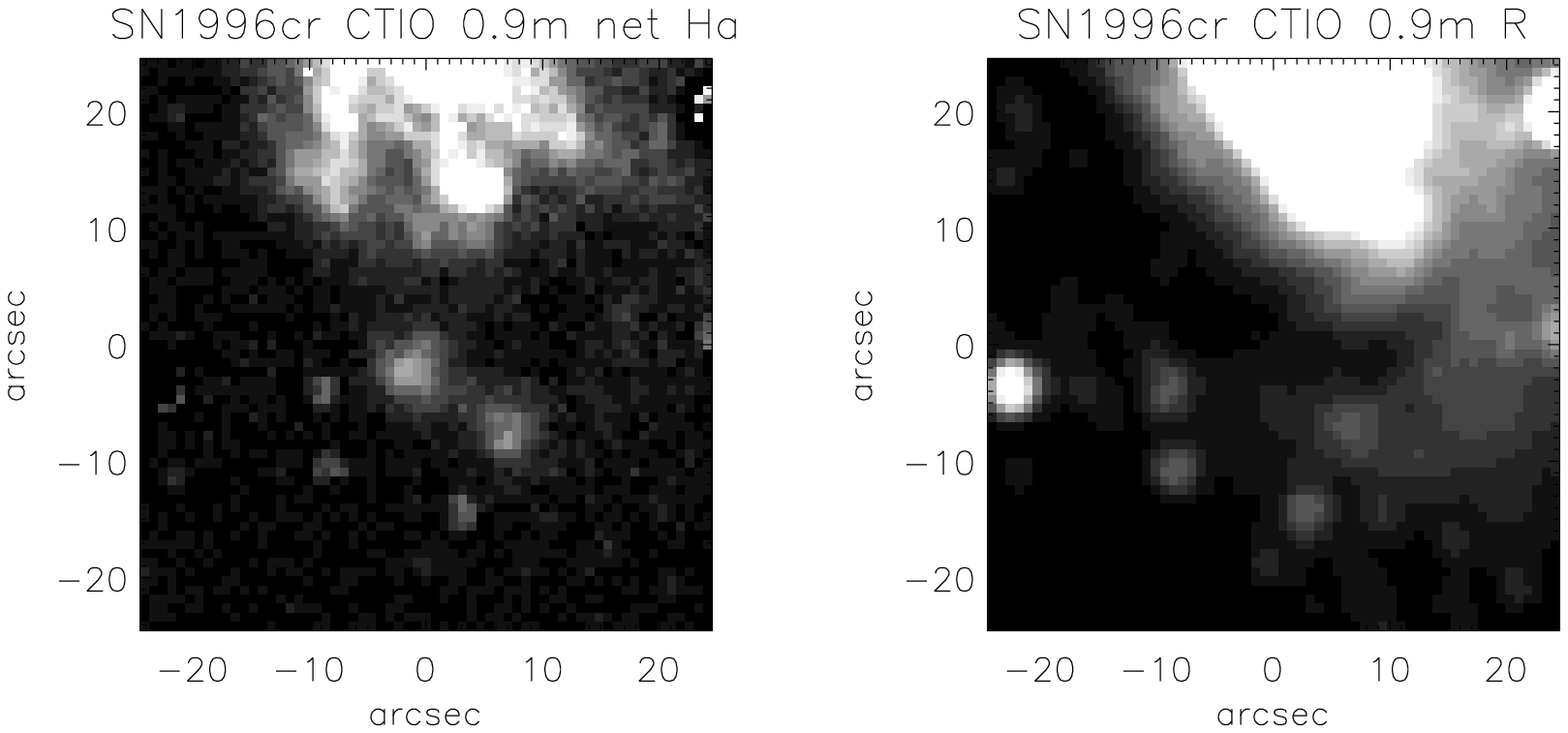}
\caption{(left) CTIO 0.9m net H$\alpha$  image (from Kennicutt et al. 2008)
showing the nebular environment of SN 1996cr (at 
centre of image, Class 5). The 50$\times$50 arcsec$^{2}$ field of 
view projects to 
1$\times$1 kpc$^{2}$ at the 4.21 Mpc distance of Circinus; (right) R-band 
image.}
\label{sn1996cr}
\end{figure*}

This type IIn supernova was discovered in Mar 1996,
although was originally identified as an X-ray source (Bauer 2007). It 
occurred 24$''$ to the S of the nucleus of Circinus (ESO 097-G13), 
corresponding to a deprojected offset 37$''$ (0.18 $R_{\rm  25}$) or 0.75
kpc for a 4.21 Mpc distance to Circinus. This galaxy
suffers extensive foreground extinction ($A_{\rm B} \sim$ 2 mag) 
due to  its low galactic latitude ($b$ = --3.8$^{\circ}$). H$\alpha$ 
imaging from Apr 2002 using the CTIO 0.9m (Kennicutt et al. 2008)
is presented in 
Fig.~\ref{sn1996cr}. This reveals faint nebular emission at the 
position of SN 1996cr, as discssed by Bauer (2007), which connects to a 
bright H\,{\sc ii} region, 3$''$ to its SE, a 
deprojected distance of 100 pc away. A 
starburst ring of H\,{\sc ii} regions surrounds the (type 2 Seyfert) 
nucleus of  Circinus (Marconi et al. 1994), the most southerly component 
of which lies 15$''$ (0.5 kpc) NNW of SN 1996cr. No cluster is apparent at 
the location of the SN, although the CTIO imaging was obtained during 
poor seeing conditions (FWHM$\sim$3$''$.5). 
We have therefore
inspected HST WFPC2  F547M and F656N imaging (GO 7273) of Circinus (Wilson et al. 
2000). No  cluster is detected at the  location of SN 1996cr, although compact 
H$\alpha$ emission is confirmed, potentially resulting from the
SN itself, in part.



\subsection{SN 1997X in NGC 4691}

\begin{figure*}
\includegraphics[bb=20 10 500 235, width=0.75\textwidth]{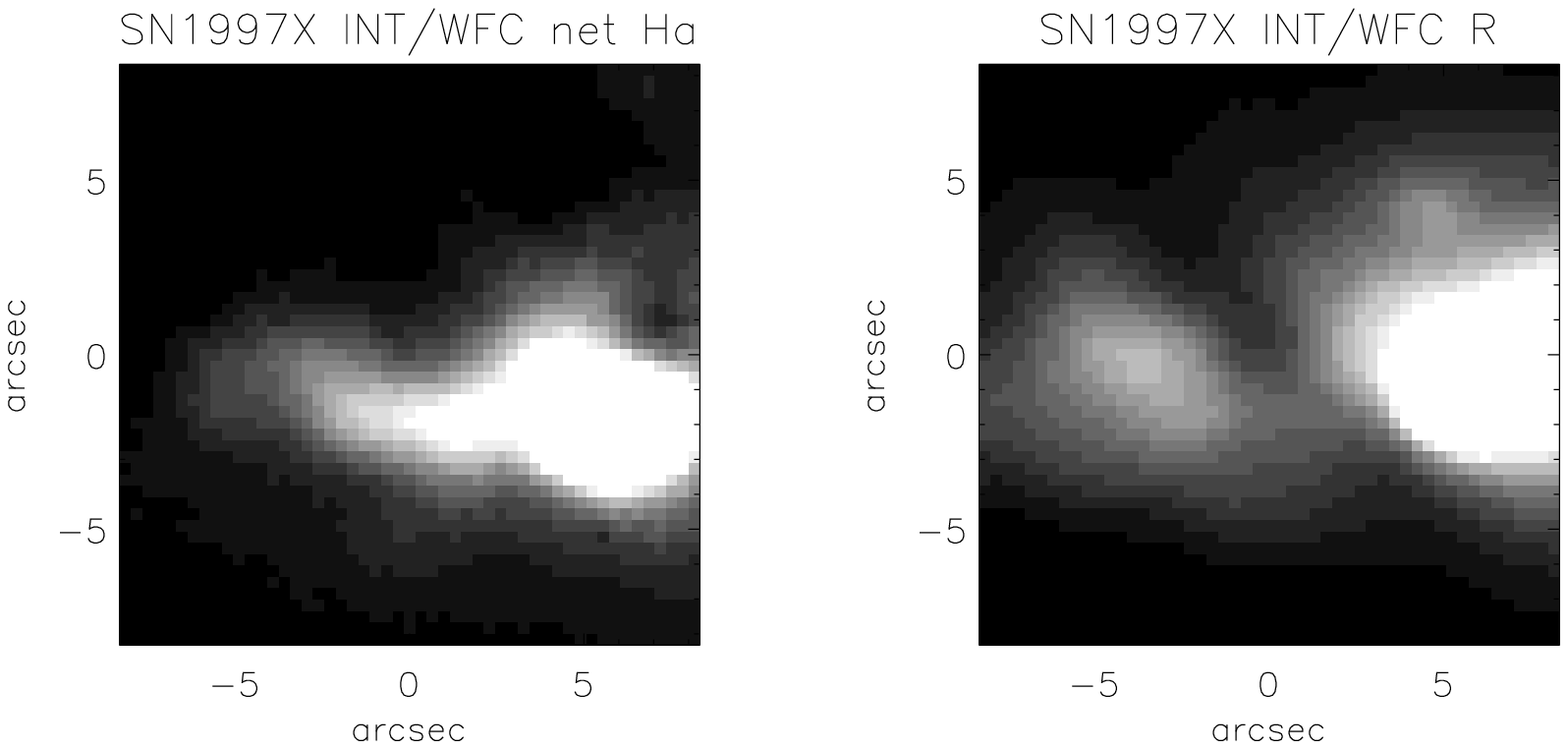}
\caption{(left) INT/WFC net H$\alpha$  image (from Anderson \& James 2008)
showing the nebular environment of SN 1997X (at  centre of image,  
Class 3). The 
17$\times$17 arcsec$^{2}$ field of view projects to 1$\times$1 kpc$^{2}$ at the 
12 Mpc distance of NGC~4691; (right) R-band image.}
\label{sn1997x}
\end{figure*}

SN 1997X was discovered in Feb 1997, approximately 9$''$ E of the 
centre of NGC 4691 (Nakano et al. 1997). Using the inclination and
major axis PA from HyperLeda this corresponds to a deprojected offset
of 11$''$ (0.14 $R_{\rm 25}$) or 0.7 kpc for a 12 Mpc distance to NGC 
4691. Although we do not have access to  calibrated H$\alpha$ imaging of 
NGC~4691, Anderson \& James (2008) discuss INT/WFC imaging, from which
a normalized  cumulative rank (NCR)  pixel  value of 0.323 is obtained
for SN 1997X.  Munari et al. (1998)  report nebular H$\alpha$+[N\,{\sc ii}]  
emission  from  a 
H\,{\sc ii} region superimposed upon an early spectrum of this 
type Ib supernova (Modjaz et al., in prep.). Figure~\ref{sn1997x} 
presents the INT H$\alpha$ imaging
from Mar 2007,  from which we note diffuse nebular  emission centred upon the   
nucleus of NGC 4691, extending $\pm$15$''$  east-west, Superimposed upon the  
diffuse nebular emission are several  spatially  extended knots, the 
brightest of which lies 6$''$ (0.4 kpc) WSW of the SN location.









\subsection{SN 1998dn in NGC 337A}

\begin{figure*}
\includegraphics[bb=20 10 500 235, width=0.75\textwidth]{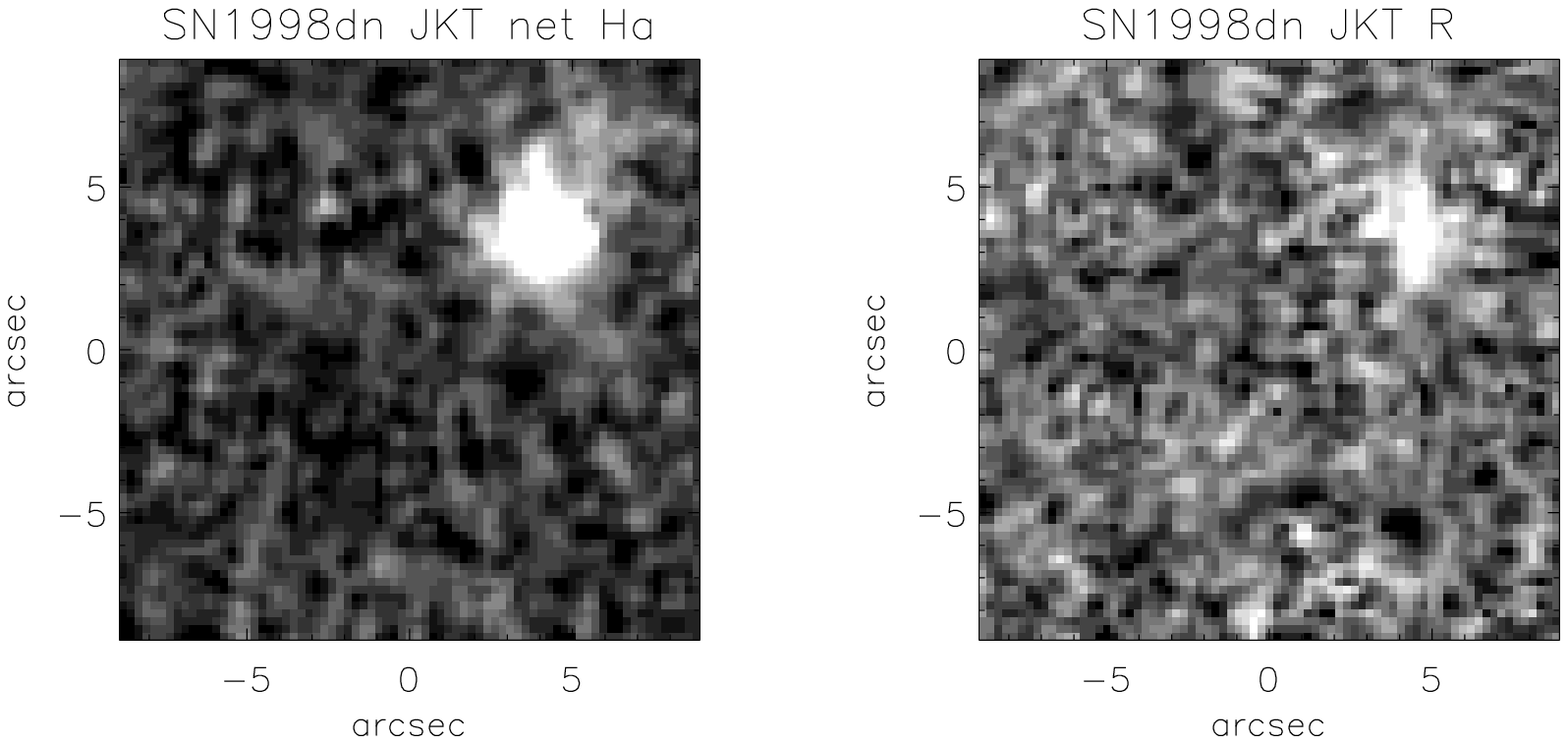}
\caption{(left) JKT net H$\alpha$  image (from Knapen et al. 2004)
showing the nebular environment of SN 1998dn (at  centre of image,  
Class 2). 
The 
18$\times$18 arcsec$^{2}$ field of view projects to 1$\times$1 kpc$^{2}$ at the 
12 Mpc distance of NGC~337A; (right) R-band image.}
\label{sn1998dn}
\end{figure*}

SN 1998dn was discovered in Aug 1998 in NGC 337A, 2.1$'$  SW of its 
nucleus (Cao 1998), corresponding to a deprojected offset of
3.5$'$ (1.2 $R_{\rm 25}$) or 11.6 kpc based on a 11.4 Mpc distance to NGC 
337A. JHK H$\alpha$ imaging from Apr 2002 
(Knapen et  al. 2004) is presented in Fig.~\ref{sn1998dn} and shows 
that this type II SN lies in a region 
devoid of nebular emission, with the nearest bright H\,{\sc ii} region 
5$''$.5 to the NW, which corresponds to a deprojected distance of 0.5
kpc for the HyperLeda inclination and PA.




\subsection{SN 1999em in NGC 1637}

\begin{figure*}
\includegraphics[bb=20 10 500 235, width=0.75\textwidth]{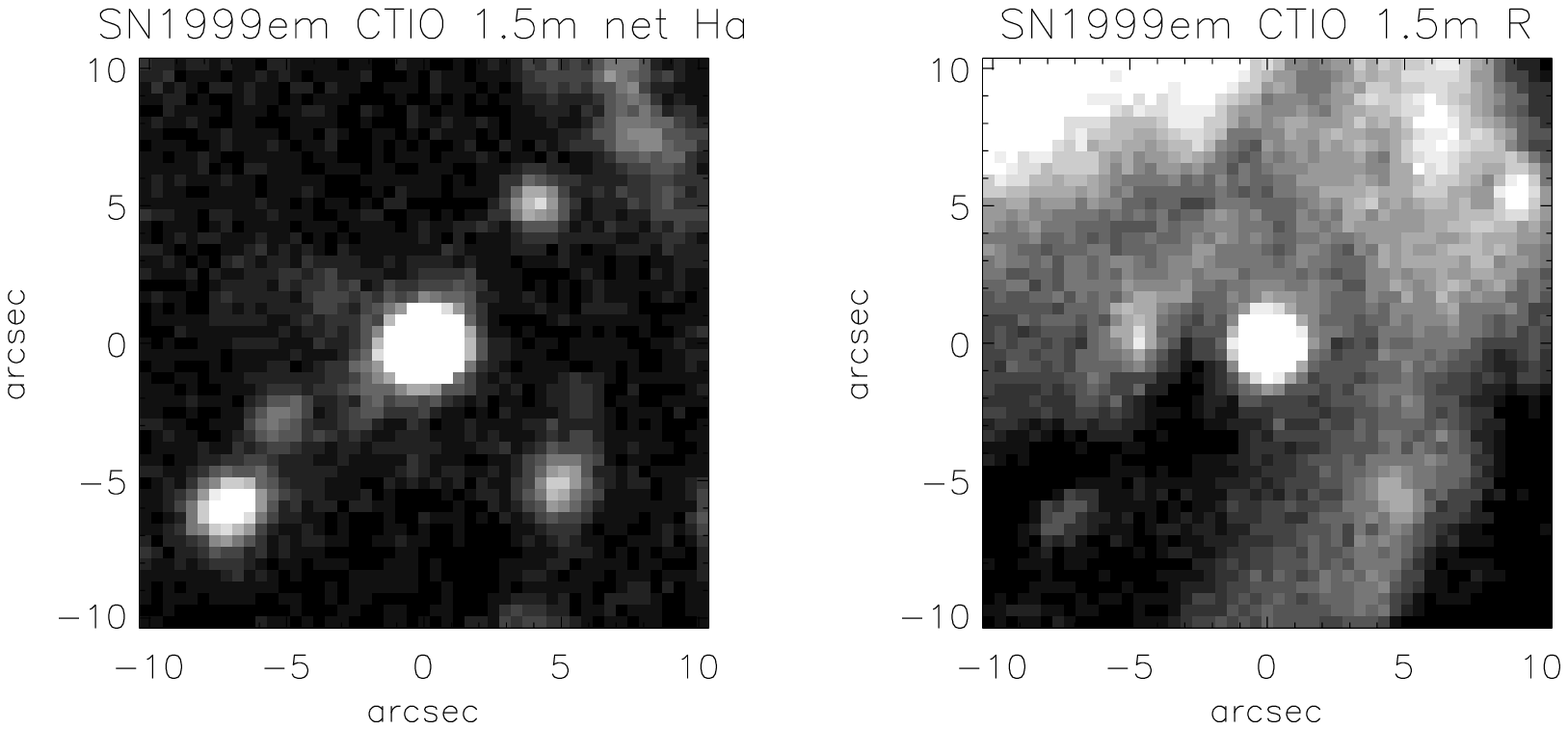}
\caption{(left) CTIO 1.5m net H$\alpha$ image from 26 Oct 2000 (Meurer et al. 2006)
showing the nebular environment of SN 1999em (bright source at centre of image,
Class 2). The 
21$\times$21 arcsec$^{2}$ field of view projects to 1$\times$1 kpc$^{2}$ at the 
9.77 Mpc distance of NGC~1637; (right) R-band image.}
\label{sn1999em}
\end{figure*}

This type II-P supernova was discovered in Oct 1999, 24$''$ 
SW of the nucleus of NGC 1637 (Li 1999; Jha et al. 1999), corresponding
to a deprojected offset of 27$''$ (0.22 $R_{\rm  25}$) which
is equivalent to a galactocentric distance of 1.3 kpc for a 9.77 Mpc 
distance to NGC 1637. As shown in Fig.~\ref{sn1999em}, SN 1999em was 
still very bright in Oct 2000 when  the CTIO  1.5m H$\alpha$ imaging 
of Meurer et al. (2006)  was obtained. We have therefore inspected pre-SN 
H$\alpha$  imaging from Ryder \& Dopita (1993) obtained with the Siding 
Spring 40$''$  telescope which indicates negligible nebular emission at 
the SN site. The closest H\,{\sc ii} region to the SN location lies 6$''$ 
(0.3 kpc) SE and has a luminosity comparable to the Orion nebula. Other 
H\,{\sc ii} regions to the NW and SW each lie 7$''$ (375 pc) away, with 
H$\alpha$+[N\,{\sc ii}] fluxes of $1.7 \times 10^{-15}$ 
erg\,s$^{-1}$\,cm$^{-2}$ and 
$3.1 \times 10^{-15}$ erg\,s$^{-1}$\,cm$^{-2}$, respectively.
A brighter source lies 9$''$.5 (0.5 kpc) 
to the SE, with a luminosity comparable to the Rosette nebula. Smartt et 
al. (2002) analysed Jan 1992 broad-band imaging from 3.6m CFHT in which 
the 
progenitor was undetected, i.e. there is no evidence for an host cluster.
Note that the SN is detected in HST WFPC2 F555W imaging (GO 9155, PI D.C. 
Leonard) obtained in Sep 2001.




\subsection{SN 1999eu in NGC 1097}

\begin{figure*}
\includegraphics[bb=20 10 500 235, width=0.75\textwidth]{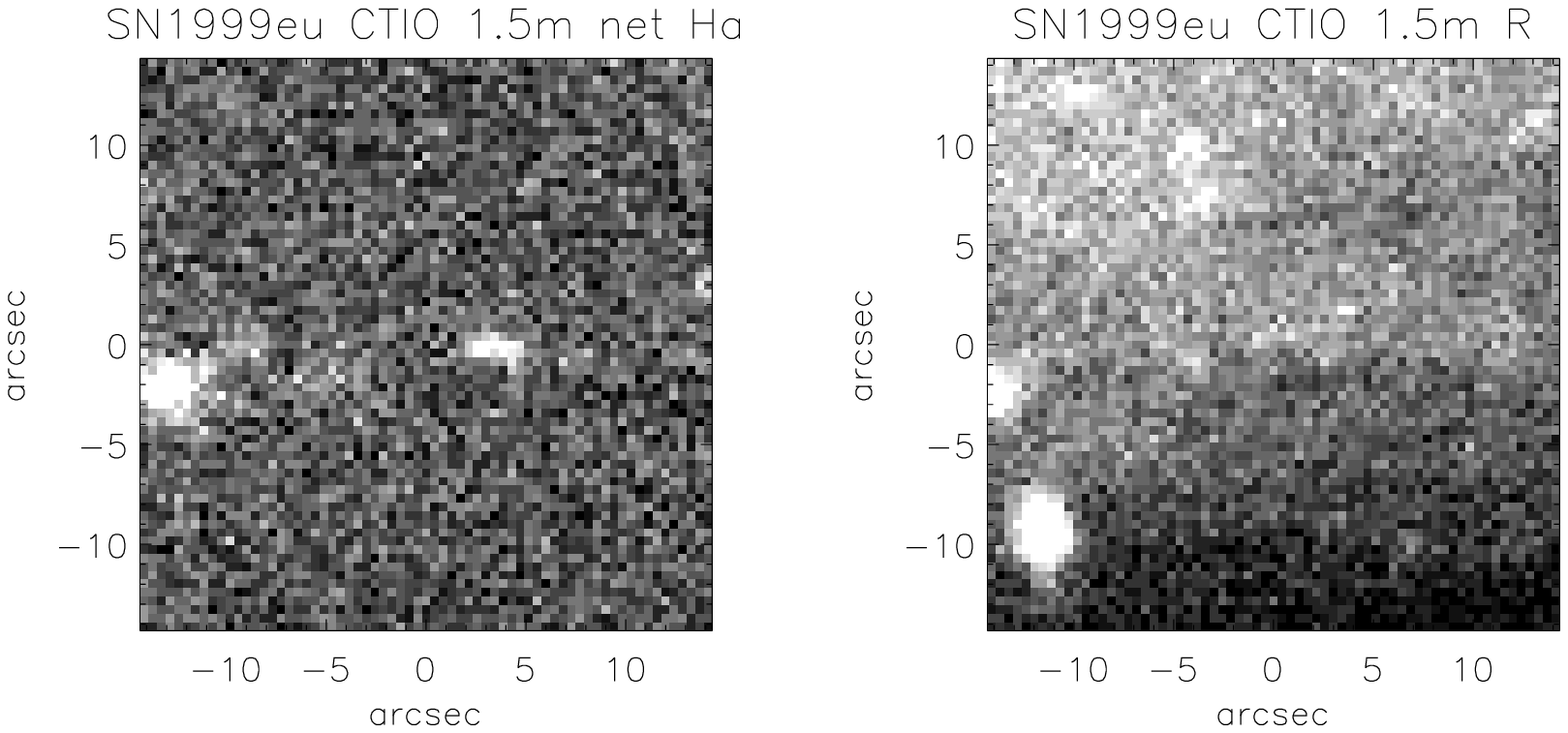}
\caption{(left) CTIO 1.5m net H$\alpha$ image (Kennicutt et al. 2003)
showing the nebular environment of SN 1999eu (at centre of image, 
Class 2). The 
29$\times$29 arcsec$^{2}$ field of view projects to {\bf 2$\times$2} 
kpc$^{2}$ at the 
14.2 Mpc distance of NGC~1097; (right) R-band image.}
\label{sn1999eu}
\end{figure*}

This type II-P supernova was discovered in Nov 1999 (Nakano et al. 
1999a) within a spiral arm of NGC 1097, 2.6$'$ to the 
SSE of its nucleus. This corresponding to a de-projected galactocentric 
distance of 3.9$'$ (0.83 $R_{\rm 25}$) or 16 kpc for the HyperLeda 
inclination and major axis PA plus an EDD distance of 14.2 Mpc.
Fig.~\ref{sn1999eu} presents 
CTIO 1.5m H$\alpha$ imaging from Oct 2001 (Kennicutt 
et al.  2003), revealing no nebular emission at
the position of SN 1999eu. The closest H\,{\sc ii} region lies 3$''$.75 $\pm$
to the W, at a deprojected distance of 375 pc, and is spatially extended 
EW, while a brighter complex with a H$\alpha$+[N\,{\sc ii}] flux of 
2.5$\times 10^{-15}$  erg\,s$^{-1}$\,cm$^{-2}$ (3 arcsec aperture radius) 
lies 13$''$ (1.3 kpc deprojected) to the E.



\subsection{SN 1999gi in NGC 3184}

\begin{figure*}
\includegraphics[bb=20 10 500 235, width=0.75\textwidth]{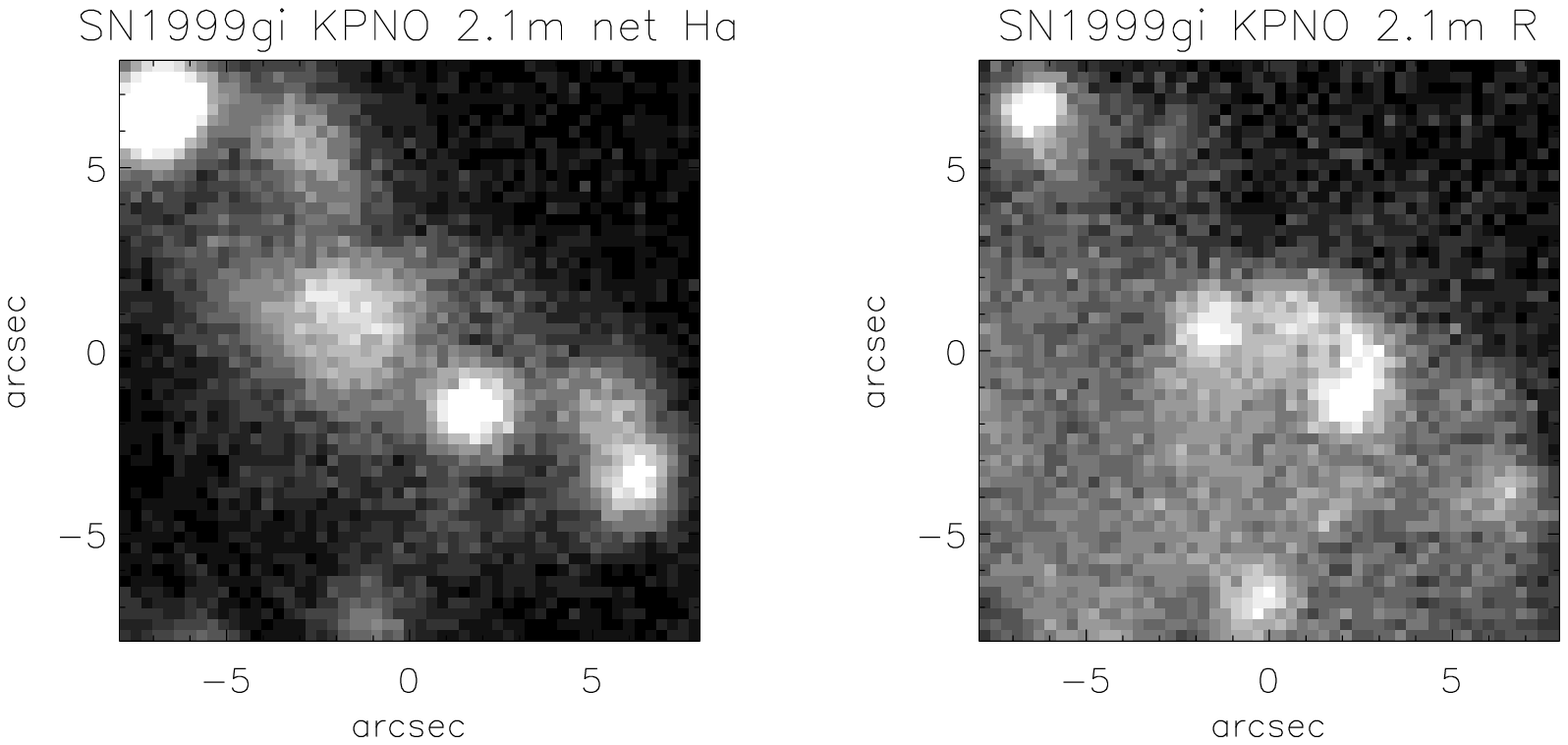}
\caption{(left) KPNO 2.1m net H$\alpha$ image (Kennicutt et al. 2003)
showing the nebular environment of SN 1999gi (at centre of image,
Class 5). The 
16$\times$16 arcsec$^{2}$ field of view projects to 1$\times$1 kpc$^{2}$ at the 
13 Mpc distance of NGC~3184; (right) R-band image.}
\label{sn1999gi}
\end{figure*}

SN 1999gi was discovered in Dec 1999, 1$'$ north of the nucleus of NGC 
3184 (Nakano et al. 1999b), whose low inclination of 14$^{\circ}$ 
implies negligible projection effects (0.28 $R_{\rm 25}$), such
that the galactocentric distance is 3.9 kpc based on a 13 Mpc distance to 
NGC 3184. Fig.~\ref{sn1999gi} shows KPNO 2.1m
H$\alpha$ imaging from Apr 2002 (Kennicutt et al. 2003) which
reveals diffuse emission at the position of the SN, intermediate between 
extended H\,{\sc 
ii} regions 2$''$ (125 pc) to  the SW and NE. The former has a H$\alpha$ 
luminosity comparable to the  Rosette nebula, while the latter is more 
extended and has a luminosity a  factor of $\sim$3 times higher. Within 
the larger star forming complex, additional knots lie 6$''$.9 (430 pc) to 
the SW and  10$''$ (630 pc) to the NE, with  H$\alpha$+[N\,{\sc ii}] 
fluxes of  4$\times 10^{-15}$  erg\,s$^{-1}$\,cm$^{-2}$ and  1$\times 
10^{-14}$  erg\,s$^{-1}$\,cm$^{-2}$, respectively.  Pre-SN WFPC2/WF images 
obtained wth the F606W filter in Jun 1994 (GO 5446, PI G.D. Illingworth), 
identify nearby OB stars but not the progenitor of the type II-P 
supernova,
while the SN is detected in WFPC2/PC F555W images (GO 8602, PI A.V. 
Filippenko) from Jan 2001 (Smartt et al. 2001; Leonard et al. 2002; Hendry 2006).


\subsection{SN 2001X in NGC 5921}

\begin{figure*}
\includegraphics[bb=20 10 500 235, width=0.75\textwidth]{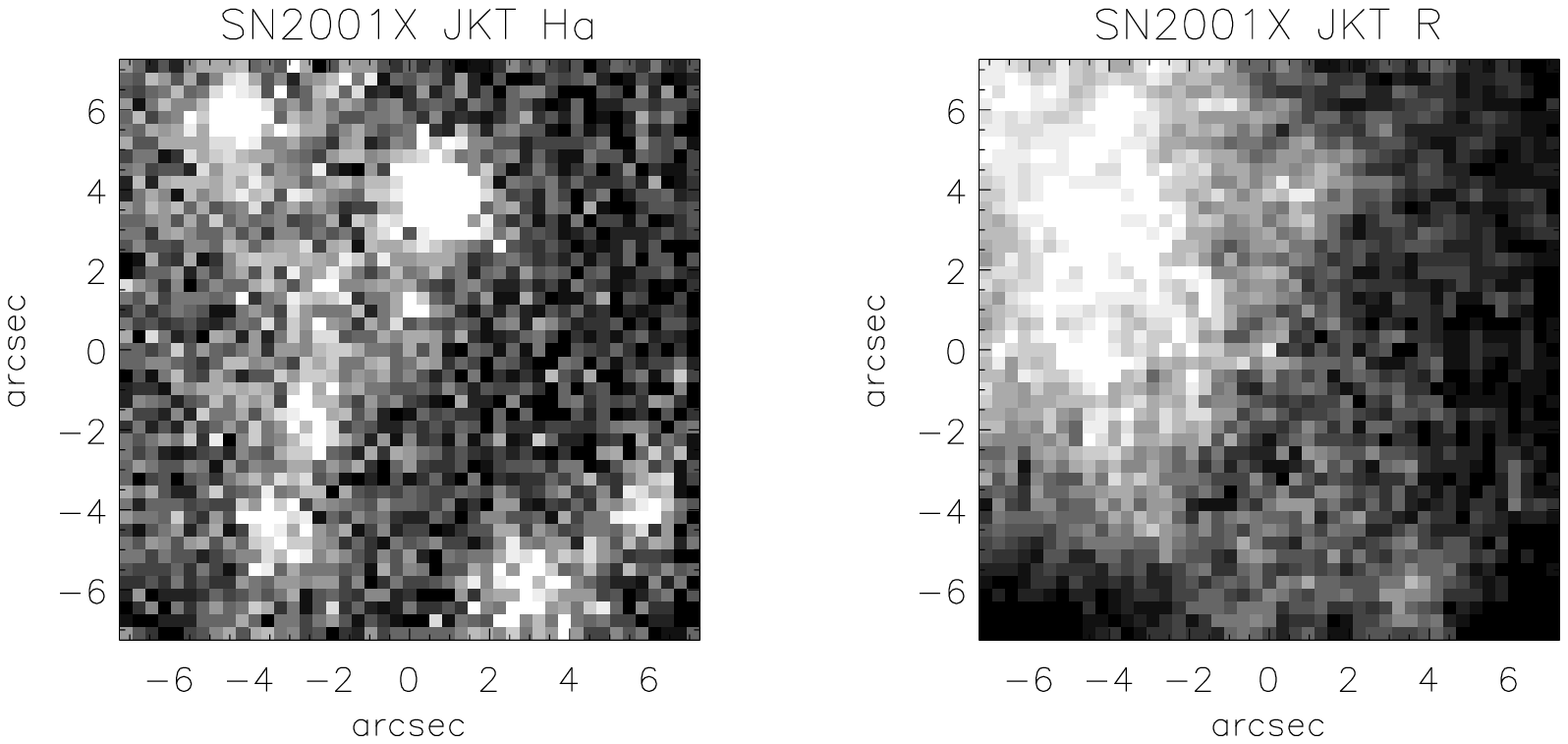}
\caption{(left) JKT H$\alpha$ image from Mar 1999 showing the nebular environment of 
SN 2001X (source at centre of image, Class 3). The 
14.7$\times$14.7 arcsec$^{2}$ field of view projects to 1$\times$1 kpc$^{2}$ at the 
14 Mpc distance of NGC~5921; (right) R-band image from Mar 2003.}
\label{sn2001x}
\end{figure*}

This type II-P supernova was discovered in Feb 2001, 33$''$ 
SSW of the nucleus of NGC 5921 (Li et al. 2001). For the HyperLeda
inclination and major axis PA, this deprojects to 49$''$ (0.34 $R_{\rm 
25}$) or 3.4 kpc, with a scale of 1$''$ = 100 pc for the 14 Mpc 
distance to NGC 5921. From inspection of the 
net H$\alpha$ imaging from Apr 2001 (James et al. 2004),
SN 2001X was exceptionally bright. Fig.~\ref{sn2001x} shows the
JKT H$\alpha$ imaging from Mar 1999, plus R-band imaging from 
Mar 2003 to assess the nebular environment. SN 2001X is coincident with faint 
nebulosity, while extended emission lies 3$''$ (0.3 kpc) SE of SN 2001X, comparable 
in  luminosity to the Rosette nebula. A brighter compact source 
lies 4$''$ to the N (0.4 kpc), although SN 2001X is not strictly
associated with either. Anderson et al. (2012) quote a NCR pixel value of 
0.698 for SN 2001X based on Liverpool Telescope imaging from Apr 2009, 
obtained archival  
indicating a potential contribution from the SN remnant.

\subsection{SN 2001ig in NGC 7424}

\begin{figure*}
\includegraphics[bb=20 10 500 235, width=0.75\textwidth]{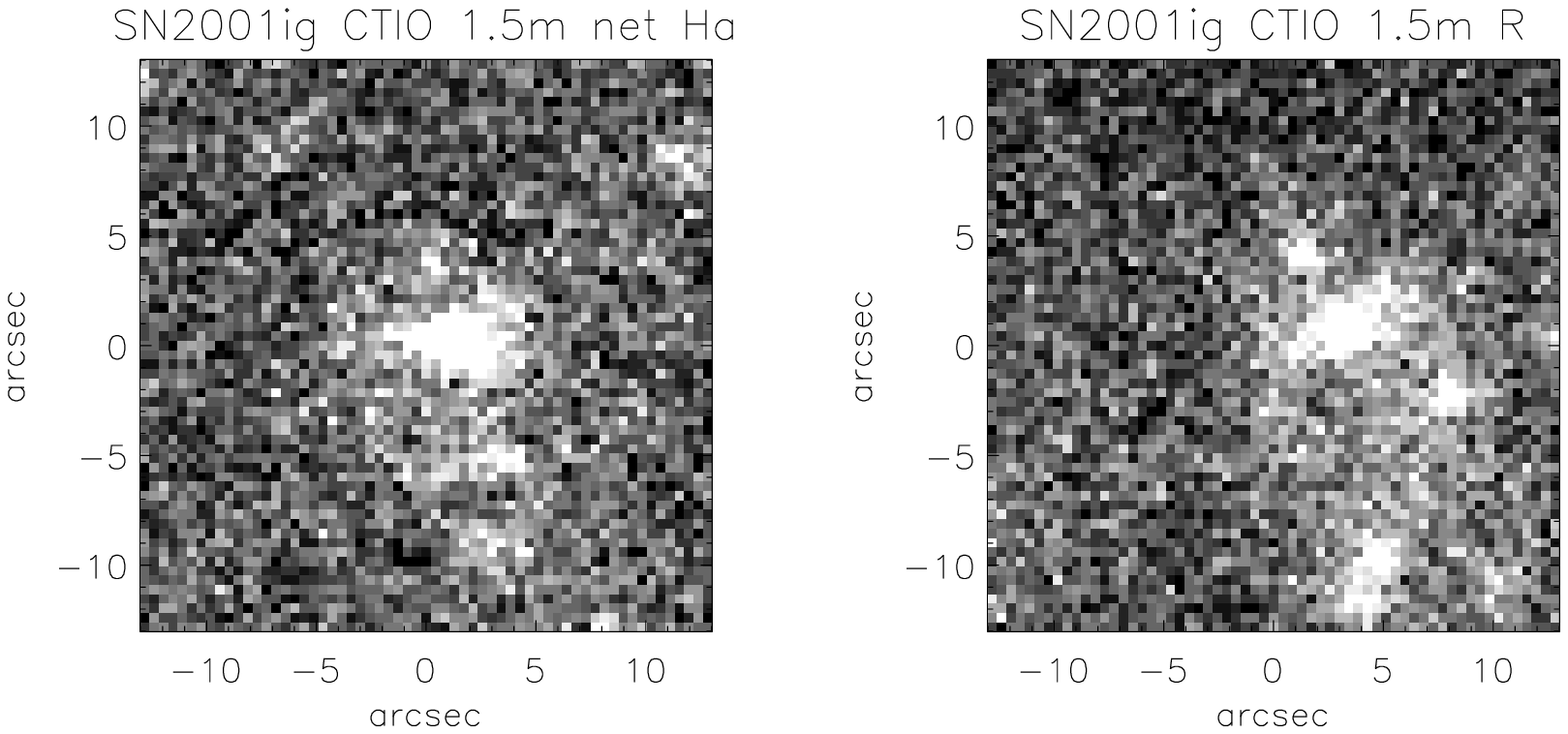}
\caption{(left) CTIO 1.5m H$\alpha$ imaging (Meurer et al. 2006) showing 
the nebular environment of SN 2001ig (source at centre of image,  
Class 4). The 
16$\times$16 arcsec$^{2}$ field of view projects to 1$\times$1 kpc$^{2}$ 
at the 7.94 Mpc distance of NGC~7424; (right) R-band imaging.}
\label{sn2001ig}
\end{figure*}

SN 2001ig (IIb) was discovered by R. Evans in Dec 2001, 2.95$'$  
NE  of the nucleus of NGC 7424 (Evans et al. 2002). Based on
the HyperLeda inclination of NGC 7424 and an adopted PA of the
major axis of 0, the deprojected distance is 4$'$.9 (1.0
$R_{\rm 25}$) or 11.2 kpc for the adopted distance of 7.94 Mpc.
Fig.~\ref{sn2001ig} presents pre-explosion CTIO 1.5m H$\alpha$  imaging 
from Sep 2000 (Meurer et al. 2006). The figure shows that SN 2001ig
occurred at the periphery of a modest luminosity  (Orion-like), extended 
H\,{\sc ii} region. Faint diffuse emission is also detected 6--10$''$ to 
the S of  the supernova, at a deprojected distance of 0.4--0.6 kpc.  A 
brighter H\,{\sc ii} region, with F(H$\alpha$+[N\,{\sc 
ii}]) = 2.5$\times 10^{-15}$   erg\,s$^{-1}$\,cm$^{-2}$ (2$''$.5 radius 
aperture), is located 17$''$ to the SW, corresponding to a deprojected
distance of 1.1 kpc. Astrometry was verified
from VLT/FORS2 R-band imaging of SN 2001ig obtained in Jun 2002 
(069.D-0453, PI E.Cappellaro). High spatial resolution $u', g', r'$ 
Gemini GMOS imaging from Sep 2004 is discussed by Ryder et al. (2006), 
who remark upon arcs of diffuse nebulosity from their deep $u'$ imaging.



\subsection{SN 2002ap in M 74}

\begin{figure*}
\includegraphics[bb=20 10 500 235, width=0.75\textwidth]{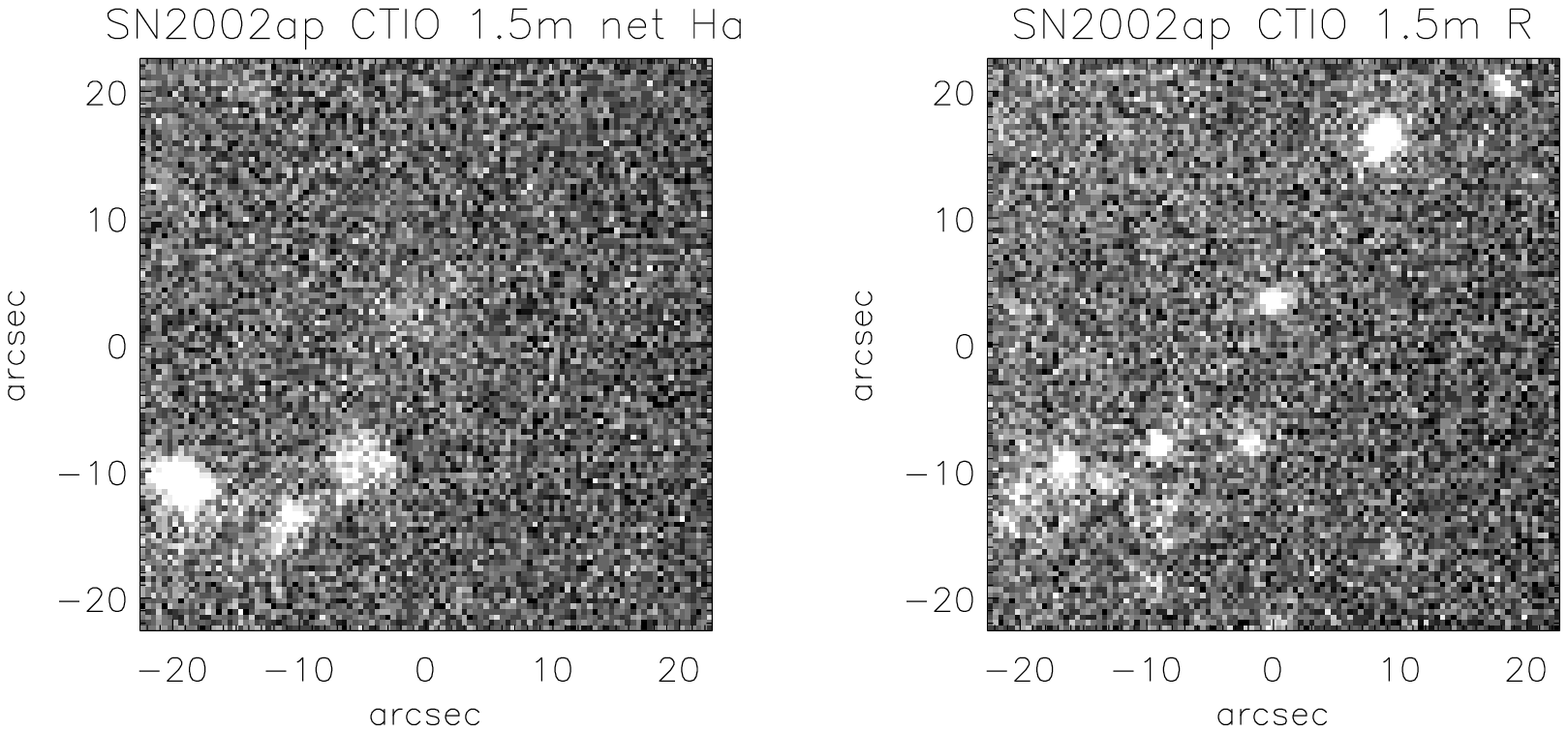}
\caption{(left) CTIO 1.5m H$\alpha$ imaging (Kennicutt et al. 2003) 
showing the nebular environment of SN 2002ap (source at centre of image, 
Class 2). The 46$\times$46 arcsec$^{2}$ field of view projects to 
{\bf 2$\times$2}
kpc$^{2}$ at the 9 Mpc distance of M~74; (right) R-band imaging.}
\label{sn2002ap}
\end{figure*}

This well studied type Ic supernova was discovered by Y. Hirose in Jan 
2002 in the  outer disk  of M 74, 4.7$'$ (0.89 $R_{\rm 25}$) 
SW of the nucleus (Nakano et al. 2002). The low inclination of M 74 
(Kamphuis \& Briggs 1992) implies that deprojection effects are 
negligible, such that the galactocentric distance is 12.2 kpc for
a 9 Mpc distance to M 74. Its location was beyond the  field-of-view 
of VLT/FORS1 and  VATT 1.8m imaging, so we present CTIO 1.5m H$\alpha$ 
imaging from Oct 2001 (Kennicutt et al. 2003) in Fig.~\ref{sn2002ap}. The 
closest nebular emission to SN 2002ap is an extended, Orion-like, H\,{\sc 
ii}  region 10$''$ (0.4 kpc) to  the SSE. Additional sources lie 17$''$ 
(0.75 kpc) and 22$''$ (1 kpc) to  the  SE, the latter with a 
H$\alpha$+[N\,{\sc ii}] flux of 4$\times 
10^{-15}$  erg\,s$^{-1}$\,cm$^{-2}$ (4$''$ radius aperture). 

Crockett et 
al. (2007) present deep, 3.6m CFHT broad-band imaging from Oct 1999, 
together 
with post-SN
HST ACS/HRC imaging from Jan 2003--Aug 2004 from which the progenitor 
star could not be identified. We have also inspected CFHT/CFH12K 
H$\alpha$ images of the site of SN2002ap from Jun 1999 (PI J.-C. 
Cuillandre) which  confirm  the CTIO  results, while  nebular emission is 
not detected in  shallow HST ACS/HRC F658N images from Jan 2003 (GO 9144, 
PI R.P. Kirschner).






\subsection{SN 2002hh in NGC 6946}

\begin{figure*}
\includegraphics[bb=20 10 500 235, width=0.75\textwidth]{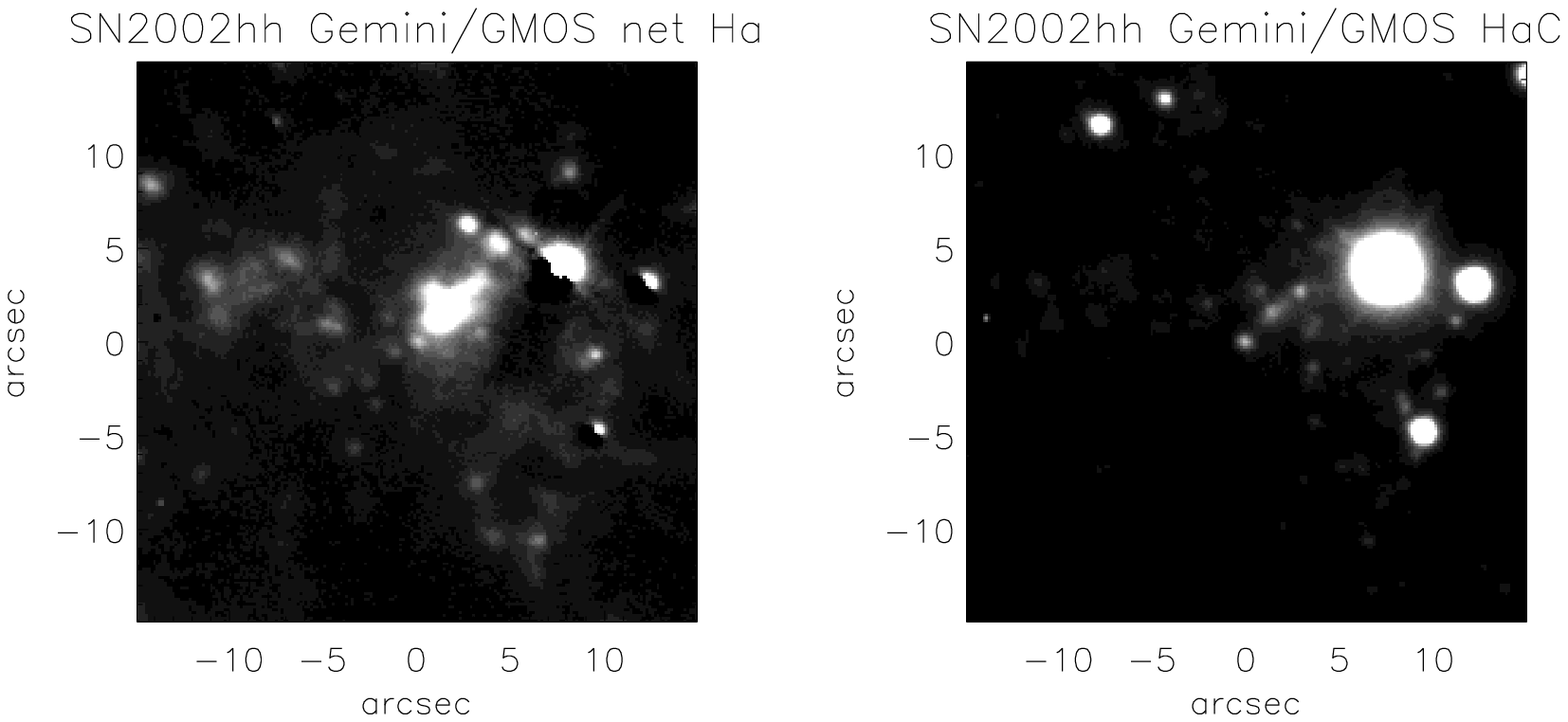}
\caption{(left) Gemini/GMOS net H$\alpha$  image (from GN-2009B-Q-4) 
showing 
nebular emission  close to the position of SN 2002hh (at centre of image, 
Class 5). 
The 30$\times$30 arcsec$^{2}$ field of view projects to 1$\times$1 
kpc$^{2}$ at the 7.0 Mpc distance of NGC 6946; (right) 
Continuum image ($\lambda_{c}$ = 6620\AA).}
\label{sn2002hh}
\end{figure*}

This type II-P supernova was discovered in Oct 2002,  2.2$'$ (0.4 $R_{\rm 
25}$) SW of the nucleus of NGC 6946 (Li 2002), corresponding to 4.5 kpc
for the adopted distance of 7 Mpc to NGC 6946 (3$''$ approximates to 100 pc). 
Fig.~\ref{sn2002hh} presents our Oct 2009 Gemini GMOS H$\alpha$ 
imaging. SN 2002hh  occurred close to the periphery of an extended  giant 
H\,{\sc ii} region  1$''$.5--3$''$.5 (50--120 pc) to its NW, whose radius 
is 
$\sim4''$. In addition, a  faint  knot of nebular emission is observed at 
the position of SN 2002hh, likely arising from the SNR, with a  
H$\alpha$+[N\,{\sc ii}] flux of  4$\times 10^{-16}$  
erg\,s$^{-1}$\,cm$^{-2}$ (1$''$ radius 
aperture), corresponding to a luminosity several times lower than Orion.
As discussed by e.g. Otsuka et al. (2012), SN 2002hh is detected in F606W 
filter observations with HST ACS/HRC from Sep 2005 (GO 10607, PI. 
B.Sugerman) and WFPC2/PC images from July 2007 (GO 11229, PI M. Meixner). 
In  addition, the extended nebular emission to 
the NW is spatially resolved in ACS/HRC F658N imaging from Sep 2005, with  
the brightest knot 2$''$.3 (80 pc) NW of SN 2002hh.






\subsection{SN 2003B in NGC 1097}

\begin{figure*}
\includegraphics[bb=20 10 500 235, width=0.75\textwidth]{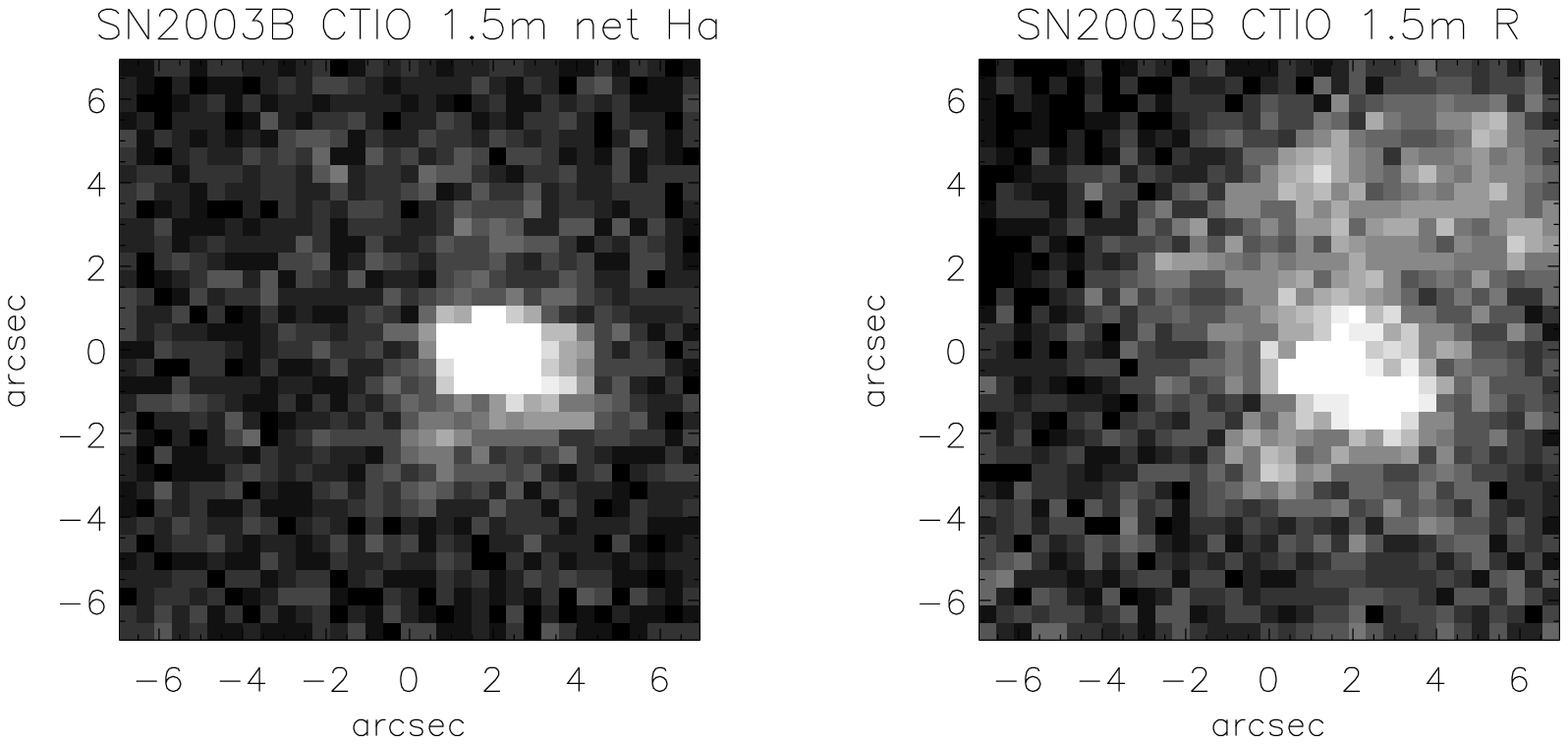}
\caption{(left) CTIO 1.5m net H$\alpha$ image (Kennicutt et al. 2003)
showing the nebular environment of SN 2003B (at centre of image,  
Class 5). The 
14.5$\times$14.5 arcsec$^{2}$ field of view projects to 1$\times$1 
kpc$^{2}$ at the  14.2 Mpc distance of NGC~1097; (right) R-band image.}
\label{sn2003b}
\end{figure*}

SN 2003B (II-P) was discovered by R. Evans in Jan 2003, 3$'$ 
NW of the nucleus of NGC 1097 (Evans \& Quirk 2003), equivalent
to a galactocentric distance of 4.9$'$ (1.04 $R_{\rm 25}$) or 20 kpc
for the 14.2 Mpc distance to NGC 1097. Fig.~\ref{sn2003b} presents
CTIO 1.5m H$\alpha$ imaging from Oct 2001 (Kennicutt et al. 2003).
SN 2003B lies at the periphery of a spatially extended 
H\,{\sc ii}  region, cited in early spectroscopy of SN 2003B by
Kirshner \& Silverman (2003). The luminosity of this giant H\,{\sc ii}
region is comparable to N66 in the SMC. 




\subsection{SN 2003gd in M 74}

\begin{figure*}
\includegraphics[bb=20 10 500 235, width=0.75\textwidth]{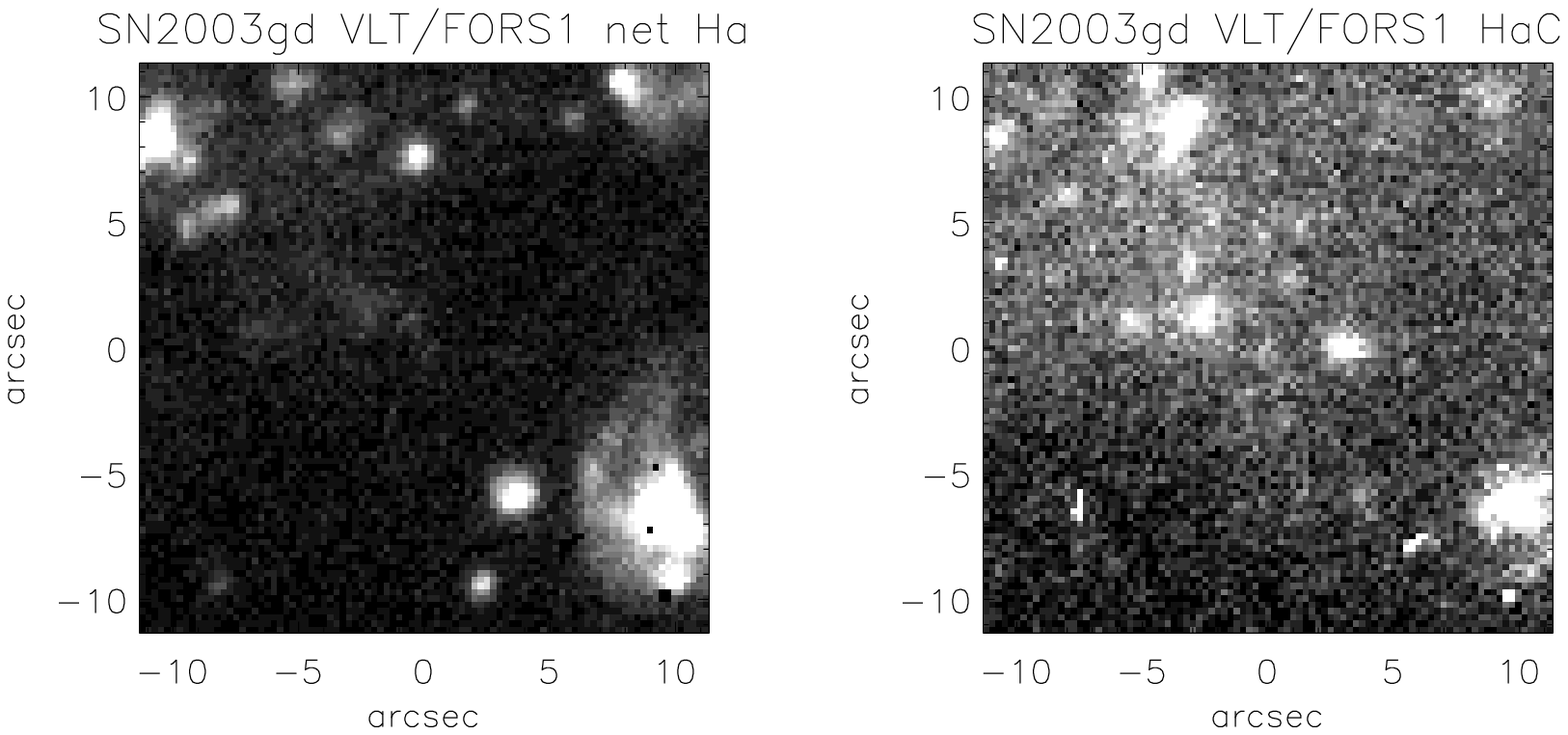}
\caption{(left) VLT/FORS1 net H$\alpha$ image 
showing the nebular environment of SN 2003gd (at centre of image,  
Class 2). The 
23$\times$23 arcsec$^{2}$ field of view projects to 1$\times$1 
kpc$^{2}$ at the  9.0 Mpc distance of M~74; (right) -band image.}
\label{sn2003gd}
\end{figure*}

This type II-P supernova was also discovered by R. Evans in Jun 
2003, 2.7$'$ (0.51 $R_{\rm 25}$) SSE of the nucleus of M 74 (Evans \& 
McNaught 2003). As for SN 2002ap, deprojection effects are negligible
owing to the low inclination of M 74 (Kamphuis \& Briggs 1992), so
the galactocentric distance is 7.0 kpc based on our 9 Mpc distance. The 
location of the SN is intermediate between three 
large, star forming complexes to the SW, NE and NW (Hodge 1976). 
Fig.~\ref{sn2003gd} presents the Oct 2007 VLT/FORS1 imaging, for which the
precise  SN location is identified from differential astrometry using
post-explosion HST ACS/HRC F555W observations from Aug 2003 (GO 9733, PI 
S.J. Smartt). Diffuse  nebulosity is observed several arcsec
to the NE of SN 2003gd, whose integrated H$\alpha$+[N\,{\sc   ii}] flux  
of 4$\times 10^{-16}$   erg\,s$^{-1}$\,cm$^{-2}$ (1$''$.5  radius 
aperture) implies a luminosity significantly inferior to the Orion nebula.  

The closest H\,{\sc ii} regions to SN 2003gd are compact sources to the N 
and SW, each 7$''$  (0.3 kpc) away, catalogued as \#641 
and \#640 from Hodge (1976), respectively. Each has a luminosity 
comparable  to the Orion nebula. The main complex of  the SW nebulosity  
(\#639 from Hodge 1976) peaks 12$''$ (0.5 kpc) from  SN 2003gd, and has a 
H$\alpha$+[N\,{\sc ii}] flux an order of magnitude larger. The luminosity 
of another complex, \#636 and \#637 from Hodge (1976), which peaks 18$''$ 
(0.8 kpc) to the NW of the SN, is comparable to N66 in the SMC, while
the complex 15$''$ to the NE (\#649--651 from Hodge 1976) has
an intermediate luminosity. Hendry et al. (2005) refer to the NW complex
in their study of SN 2003gd via \#72--73 from Belley \& Roy (1992).

HST  ACS/HRC F625W imaging (GO 10272, P.I. 
A.V. Filippenko) also reveals very faint nebulosity several arcsec W of 
the SN position, while no cluster is seen at the site of the SN. Pre-SN 
HST WFPC2 and Gemini GMOS  imaging from May-Aug 2002  enabled Smartt et 
al. (2004) to identify a red  supergiant as  the progenitor of SN 2003gd 
(see also Maund \& Smartt 2009).






\subsection{SN 2003jg in NGC 2997}

\begin{figure*}
\includegraphics[bb=20 10 500 235, 
width=0.75\textwidth]{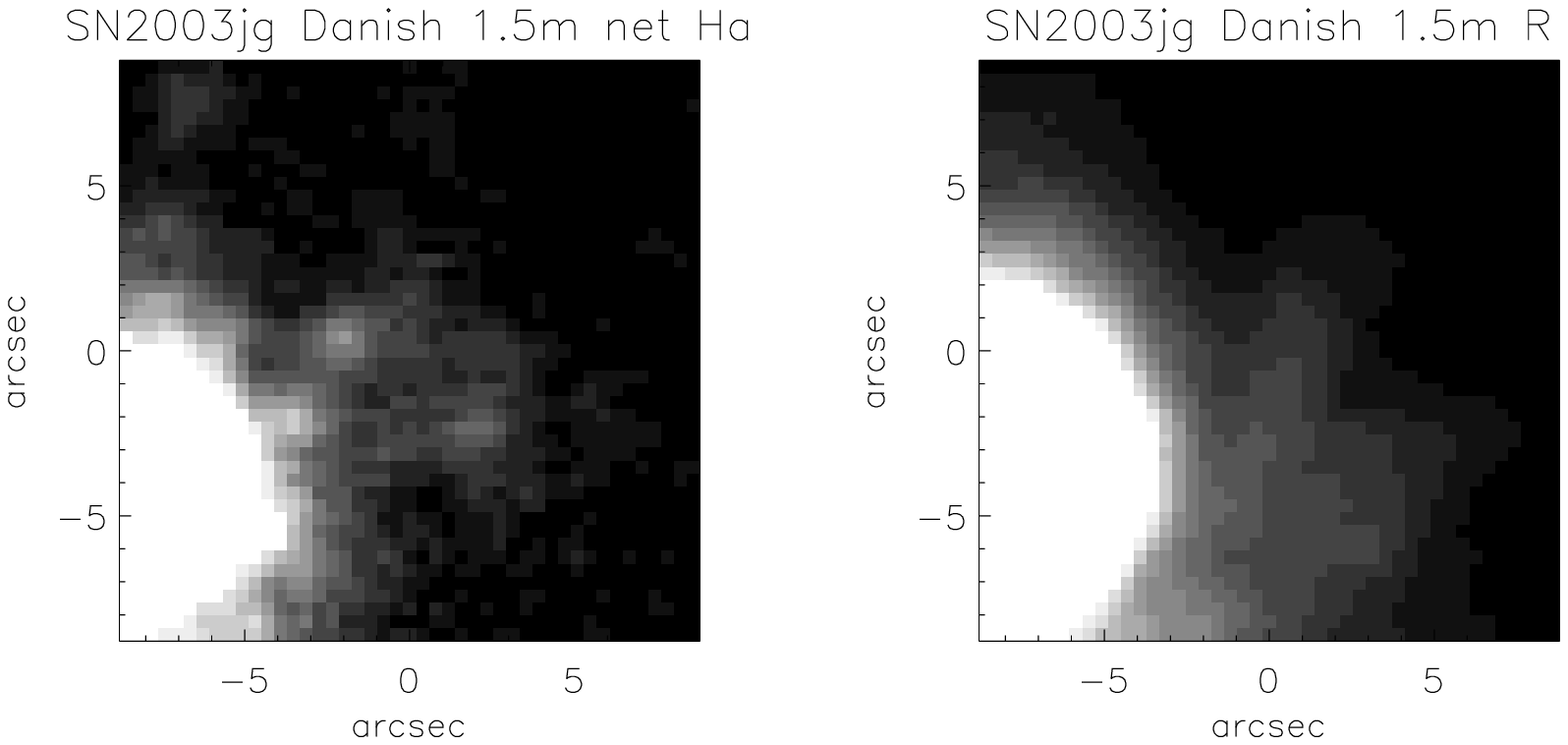}
\caption{(left) Danish 1.5m net H$\alpha$ image (from Larsen \& Richtler
1999) showing the nebular environment of SN 2003jg (at centre of image, 
Class 2). 
The 18$\times$18 arcsec$^{2}$ field of view projects to 1$\times$1 
kpc$^{2}$ at the  11.7 Mpc distance of NGC~2997; (right) R-band image.}
\label{sn2003jg}
\end{figure*}

This type Ib/c supernova was discovered by R. Martin in Oct 2003,
13$''$ NW of the nucleus of NGC 2997 (Martin \& Biggs 2003), which 
deprojects to 18$''$ (0.07 $R_{\rm 25}$) or a galactocentric
distance of 1.0 kpc using a 11.7 Mpc distance to NGC 2997. This
galaxy is undergoing an intense, ring-like star 
formation episode in its nucleus, with a luminosity comparable to 30 
Doradus in the LMC. Fig.~\ref{sn2003jg} presents Danish 1.5m 
H$\alpha$ images  (Larsen \&  Richtler 1999), revealing that SN 2003jg 
lies 2$''$ (140 pc) W of a modest, Orion-like H\,{\sc ii}  region, 
apparently associated with this intense activity. 
It is unclear whether this region 
is extended from the Danish 1.5m imaging, so we have examined high 
quality VLT/FORS1  R-band imaging from Mar 1999 (60.A-9203), which 
indicate that the source is spatially extended NS. No source is 
coincident with the SN progenitor based on HST WFPC2 imaging from Aug 
2001 using the  F450W filter (GO 9042, PI S.J. Smartt).

%

\subsection{SN 2004dj in NGC 2403}

\begin{figure*}
\includegraphics[bb=20 10 500 235, width=0.75\textwidth]{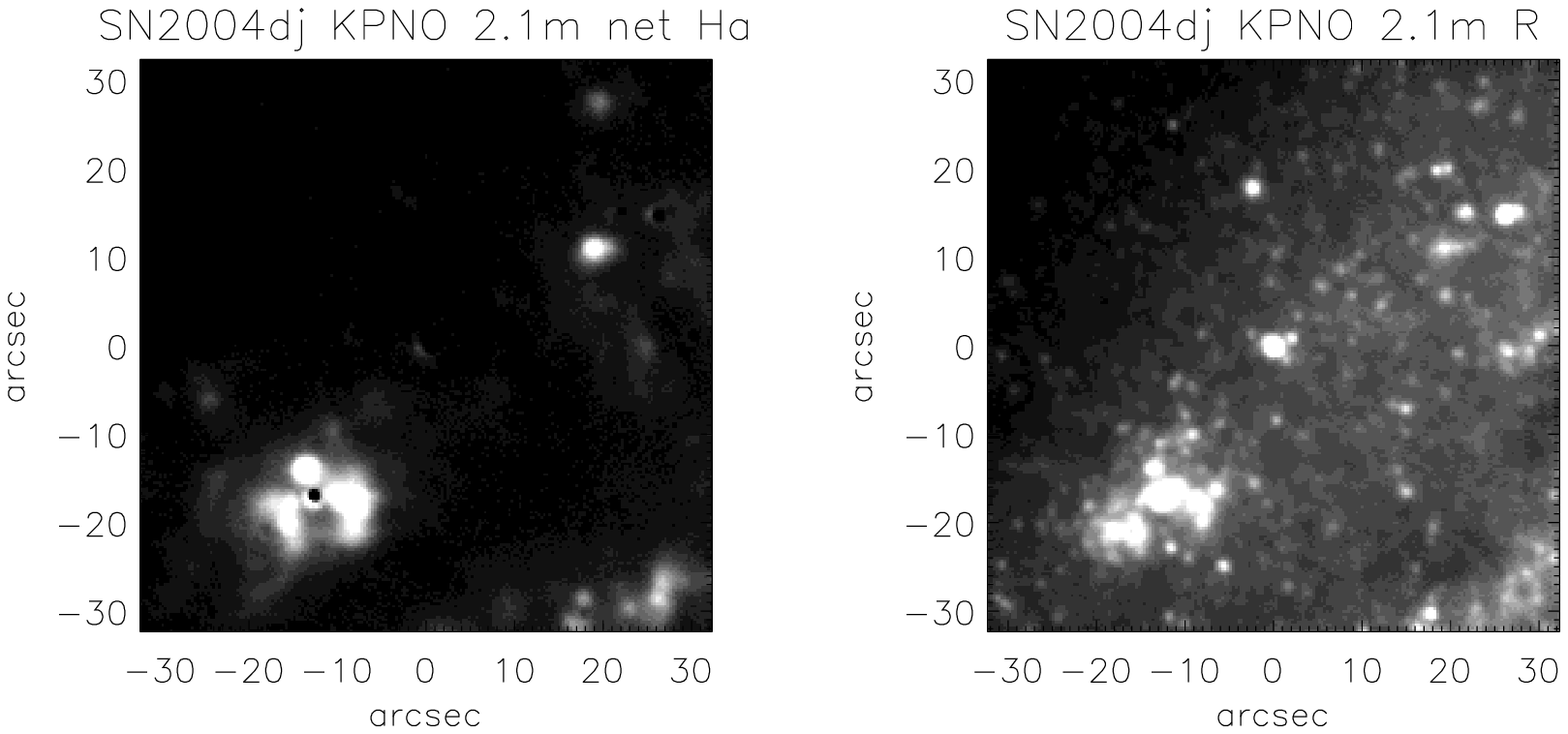}
\caption{(left) 2.1m KPNO H$\alpha$ image (from Kennicutt et al.
2003) showing the nebular environment of SN 2004dj (at centre of image,
Class 1). 
The 65$\times$65 arcsec$^{2}$ field of view projects to 1$\times$1 
kpc$^{2}$ at the  3.16 Mpc distance of NGC~2403; (right) R-band image.}
\label{sn2004dj}
\end{figure*}

SN 2004dj (II-P) was discovered by K. Itagaki in July 2004, 2.7$'$ 
east of the nucleus of NGC 2403 (Nakano et al. 2004). Based on the
inclination and PA of the major axis of NGC 2403 from HyperLeda, this
corresponds to a deprojected distance of 3.7$'$ (0.34 $R_{\rm 25}$)
or 3.4 kpc for a 3.16 Mpc distance to NGC 2403, the second closest
ccSNe in our sample. Figure.~\ref{sn2004dj} shows  KPNO 2.1m 
H$\alpha$ observations  of NGC~2403 from Nov 2001 (Kennicutt et  al. 
2003), while we have also inspected higher spatial resolution H$\alpha$ 
observations of the central region of  NGC~2403 from NOT/ALFOSC (Larsen 
\& Richtler 1999).


Ma\'{i}z-Apell\'{a}niz et al. (2004) identified SN 2004dj with
a young, compact star cluster, \#96 from Sandage (1984).
From Fig.~\ref{sn2004dj}, this is not a source of H$\alpha$ emission, with
the faint, diffuse emission $\sim$8$''$ to the SE, a deprojected 
distance of 170  pc away. The nearest prominent star forming regions each 
lie  21$''$ (450 pc) NW and SE. The former  H\,{\sc  ii} region is spatially 
extended, albeit relatively compact, with a luminosity  similar to the 
Rosette nebula, while the latter contains multiple  knots, separated by 
several arcsec and has an integrated luminosity  comparable to N66.

Vink\'{o} et al. (2006, 2009) have studied Sandage 96, from which both 
`young' (10--16 Myr, 15--20 $M_{\odot}$) and  `old' (30--100 Myr, $< 10 
M_{\odot}$) solutions were obtained. From post-explosion 2.3m Bok 
H$\alpha$ imaging,  they also note that Sandage 96 lacks extended 
H$\alpha$ emission and  attribute the bulk of the compact H$\alpha$ 
emission that they detected to SN  2004dj itself.  Diffuse emission is not 
detected in HST ACS/HRC F658N  imaging  from Aug 2005 (GO 10607, PI B.E. 
Sugerman), while the compact nature of the NW source is confirmed from HST 
WFPC2 F606W imaging from Apr 2008 (GO  11229, PI M. Meixner).







\subsection{SN 2004et in NGC 6946}

\begin{figure*}
\includegraphics[bb=20 10 500 235, width=0.75\textwidth]{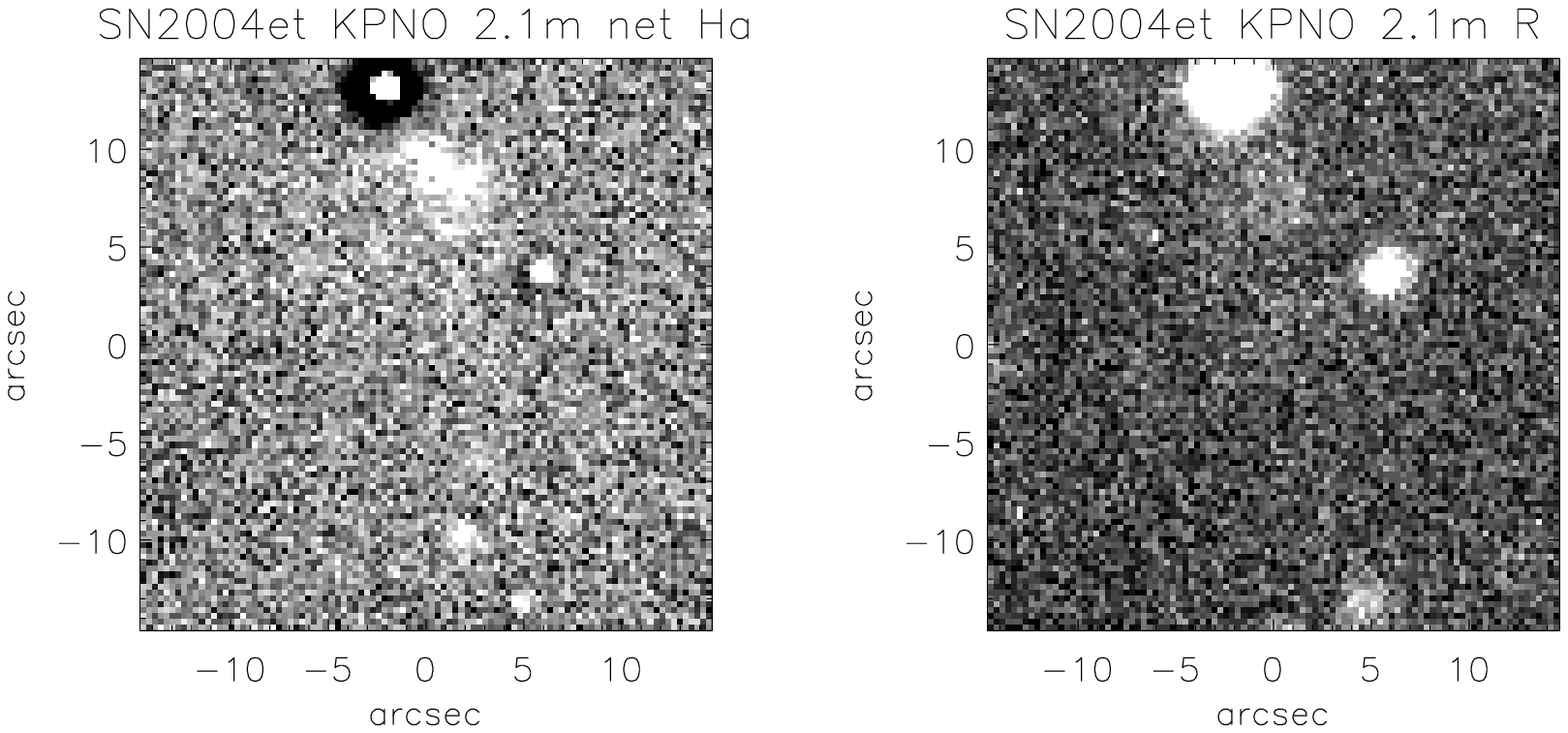}
\caption{(left) 2.1m KPNO H$\alpha$ image (from Kennicutt et al.
2003) showing the nebular environment of SN 2004et (at centre of image,
Class 2). 
The 30$\times$30 arcsec$^{2}$ field of view projects to 1$\times$1 
kpc$^{2}$ at the  7.0 Mpc distance of NGC~6946; (right) R-band image.}
\label{sn2004et}
\end{figure*}

This type II-P supernova was discovered by S. Moretti in Sep 2004,
4.5$'$ (0.8 $R_{\rm 25}$) SE of the nucleus of NGC 6946 (Zwitter
et al. 2004) corresponding to a galactocentric distance of 9.6 kpc based 
on a 7.0 Mpc distance to NGC 6946. This region, far from any large star 
forming regions,
falls beyond the field-of-view of the Gemini GMOS imaging, so 
Fig.~\ref{sn2004et} shows
KPNO 2.1m H$\alpha$ imaging from Mar 2001 (Kennicutt et al. 2003). 
The closest nebular emission lies 9$''$ (300 pc) N of the position of SN 
2004et, extends NE--SW over several arcsec, and has an integrated 
luminosity comparable to the Orion nebula. Crockett et al. (2011) 
discuss constraints upon the progenitor of SN 2004et from various 
facilities, including pre-SN CFHT/CFH12K R-band imaging and post-SN
HST WFPC2/PC (GO 11229, P.I. M. Meixner) F606W  imaging from Jan 2008, 
with a second (point) source detected 0$''$.25 E of the SN position.
The extended nebular emission N of SN 2004et is apparent in both datasets.




\subsection{SN 2005at in NGC 6744}

\begin{figure*}
\includegraphics[bb=20 10 500 235, width=0.75\textwidth]{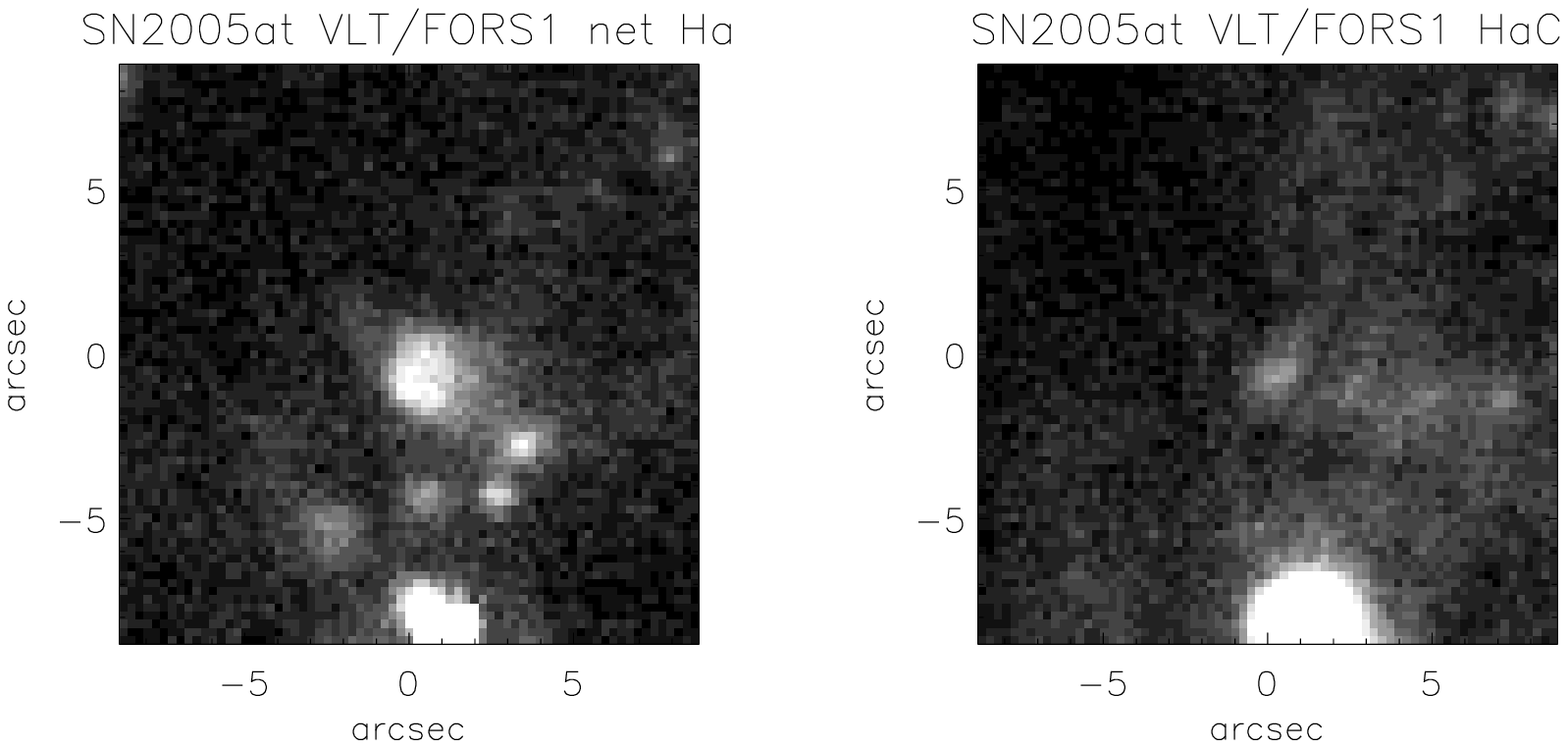}
\caption{(left) VLT/FORS1 H$\alpha$ image (081.B-0289(C), PI 
P.A.~Crowther)  showing the nebular environment  of SN 2005at (at centre 
of image, Class 4). The 18$\times$18 arcsec$^{2}$ field of  view 
projects to 
1$\times$1   kpc$^{2}$ at the  11.6 Mpc distance of NGC~6744; (right) 
Continuum ($\lambda_{c}$ =  6665\AA) image.} \label{sn2005at}
\end{figure*}

This type Ic supernova was jointly discovered by R. Martin and L. 
Monard in Mar 2005, 2$'$.3  NNE of the 
nucleus of NGC 6744 (Martin et al. 2005), corresponding to a
galactocentric distance of 3$'$.0  (0.30 $R_{\rm 25}$) or 10.0 kpc for
a 11.6 Mpc distance to NGC 6744. In Fig.~\ref{sn2005at} we present
our VLT/FORS1  continuum-subtracted  H$\alpha$ imaging from Jun 2008. 
The SN occurred   within an extended H\,{\sc ii} region whose luminosity is 
comparable to the Rosette  nebula. This nebula is also detected in pre-SN 
Danish  1.5m imaging from Feb 1998 (Larsen \& Richtler 1999). 
The faint continuum source at the position of SN 2005at in 
Fig.~\ref{sn2005at} is likely due to the SNR itself. A bright continuum 
source  lies 8$''$.5 S of SN  2005at, at a deprojected distance of 0.6 
kpc, which is poorly subtracted in the VLT/FORS1 net H$\alpha$ image. The 
SN has  also been imaged with HST  WFPC2/PC using the F555W 
filter (GO 10877, PI W. Li) in Apr 2007.




\subsection{SN 2005cs in M 51a}

\begin{figure*}
\includegraphics[bb=20 10 500 235, width=0.75\textwidth]{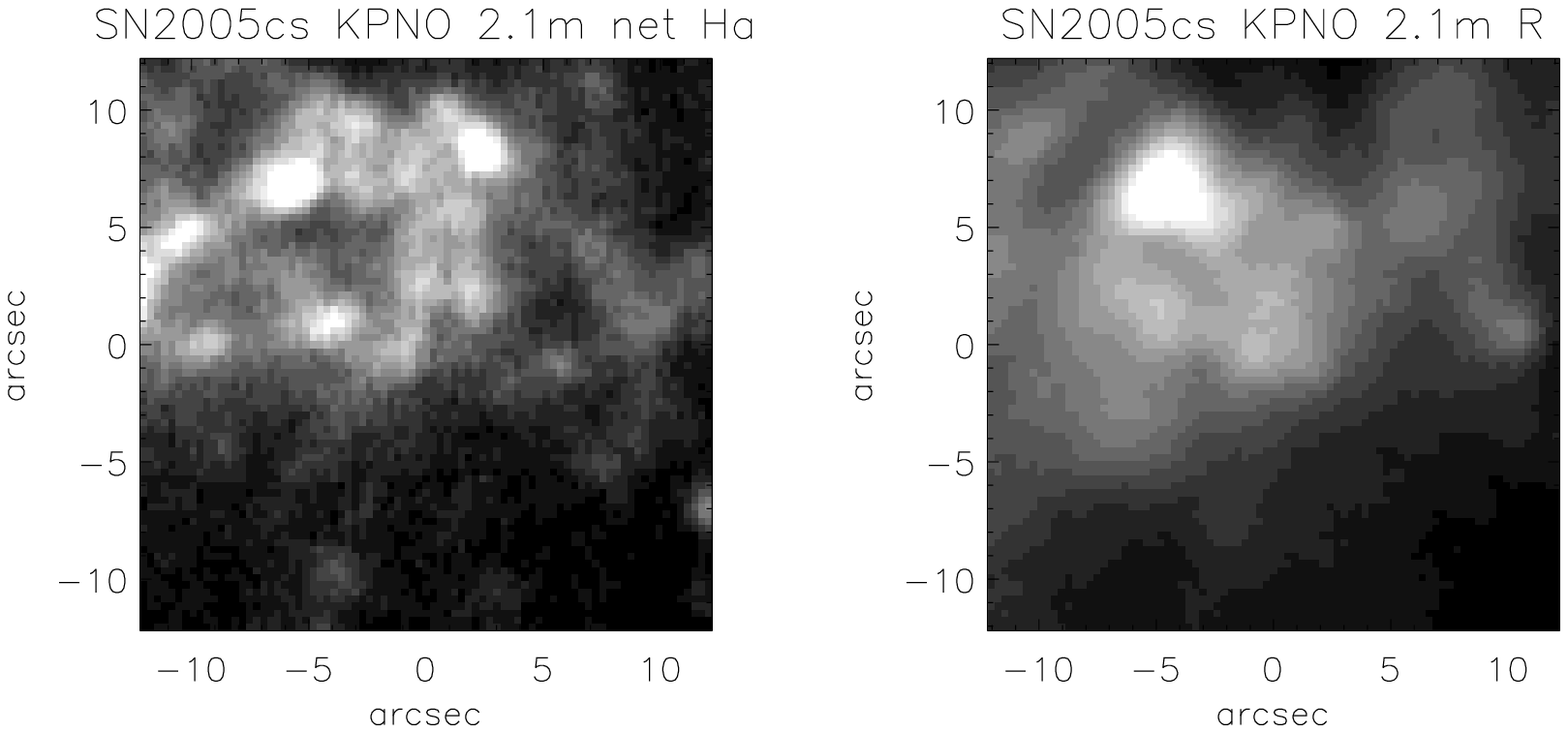}
\caption{(left) KPNO 2.1m net H$\alpha$ image (from Kennicutt et al.
2003) showing the nebular environment of SN 2005cs (at centre of image,
Class 4).
The 25$\times$25 arcsec$^{2}$ field of view projects to 1$\times$1
kpc$^{2}$ at the  8.4 Mpc distance of M~51a; (right) R-band image.}
\label{sn2005cs}
\end{figure*}

This type II-P supernova was discovered by W. Kloehr in June 2005,
within a complex star forming region, 1.1$'$ (0.2 $R_{\rm 25}$) SSW of the
nucleus of M~51a (Kloehr et al. 2005), with negligible projection effects,
such that the galactocentric distance is 2.7 kpc for a 8.39 Mpc distance 
to M 51a. Fig.~\ref{sn2005cs} presents KPNO 2.1m imaging from Mar 2001
(Kennicutt et al. 2003), revealing that SN 2005cs coincides with a region 
of nebular emission extending N-S. Orion-like H\,{\sc ii} regions are
located 1$''$ (40 pc) E, 2$''$.7 (110 pc) N and 4$''$.3 (175 pc) E, 
while brighter star forming knots are located $>$9$''$ ($>$0.35 kpc) to 
the N, NE  and  E,  the brightest of which is spatially extended N-S, lies 
13$''$  (0.55 kpc) E of the SN and has a luminosity comparable to the 
Rosette  nebula. These sources are poorly resolved in ground-based R-band 
imaging,  except that by far the brightest continuum source is that 9$''$ 
NE of the SN. 

We have inspected HST ACS/WFC (GO 10452, PI S. Beckwith) images 
obtained with the F555W and F658N (H$\alpha$+[N\,{\sc ii}]) filters in 
Jan 2005. As shown in Fig.~\ref{sn2005cs_2}, SN 2005cs  is not associated 
with nebular emission, although compact H$\alpha$ sources lie 1$''$ (40 
pc) to the SE  and NE, plus there is a compact star cluster $\sim$0$''$.2 
(8 pc) to the SW (see Li et al. 2006). 




\subsection{SN 2005kl in NGC 4369}

\begin{figure*}
\includegraphics[bb=20 10 500 235, width=0.75\textwidth]{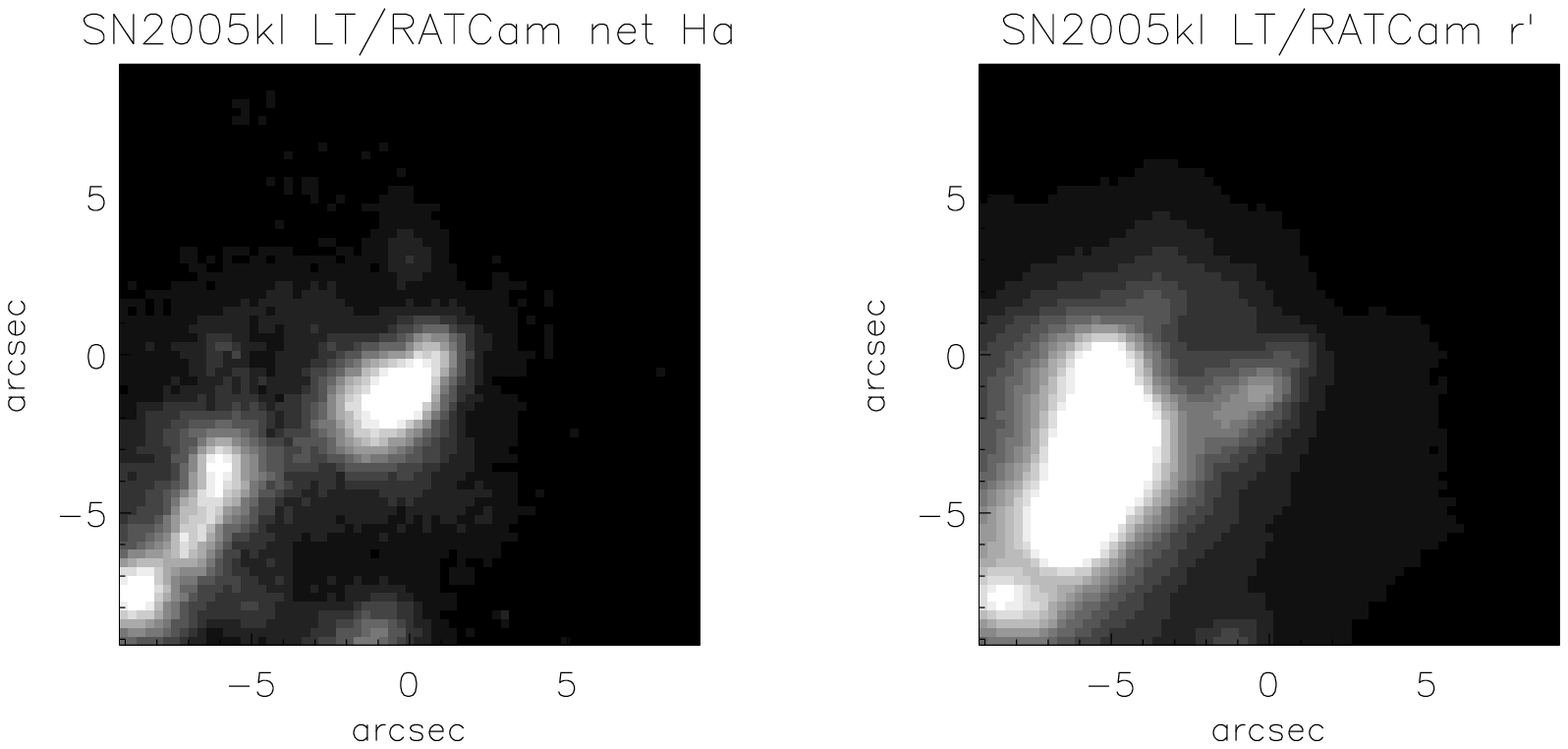}
\caption{left) Liverpool Telescope RATCam net H$\alpha$ image (from Anderson
\& James 2008) showing the nebular environment of SN 2005kl (at centre of image,
Class 5).
The 18$\times$18 arcsec$^{2}$ field of view projects to 1$\times$1
kpc$^{2}$ at the 11.2 Mpc distance of NGC~4369; (right) Sloan r$'$-band image.}
\label{sn2005kl}
\end{figure*}

SN 2005kl was discovered by M. Migliardi in Nov 2005, 10$''$ 
(0.17 $R_{\rm 25}$) NW of the 
centre of NGC 4369 (Migliardi et al. 2005). In view of the low inclination 
of this galaxy (18.9$^{\circ}$), there are negligible projection effects,
so the galactocentric distance is 0.6 kpc for  a 11.2 Mpc distance to 
NGC4369. Although we do not have access to  calibrated H$\alpha$ 
imaging of NGC~4369, Fig.~\ref{sn2005kl} presents Liverpool Telescope RATCam 
H$\alpha$ and  (Sloan) r$'$-band imaging of NGC 4369 from Feb 2007 (Anderson \& 
James  2008). This type Ic supernova occurred at the periphery of a  bright, 
spatially  extended H\,{\sc ii} region, the peak of which lies 1$''$.5 (80 
pc) to 
the SE of the supernova, with the main (continuum) body of the galaxy further to 
the E. Anderson \& James (2008) quote a normalized cumulative rank (NCR) pixel 
value of 0.570 from their RATCam imaging.




\subsection{SN 2007gr in NGC 1058}

\begin{figure*}
\includegraphics[bb=20 10 500 235, width=0.75\textwidth]{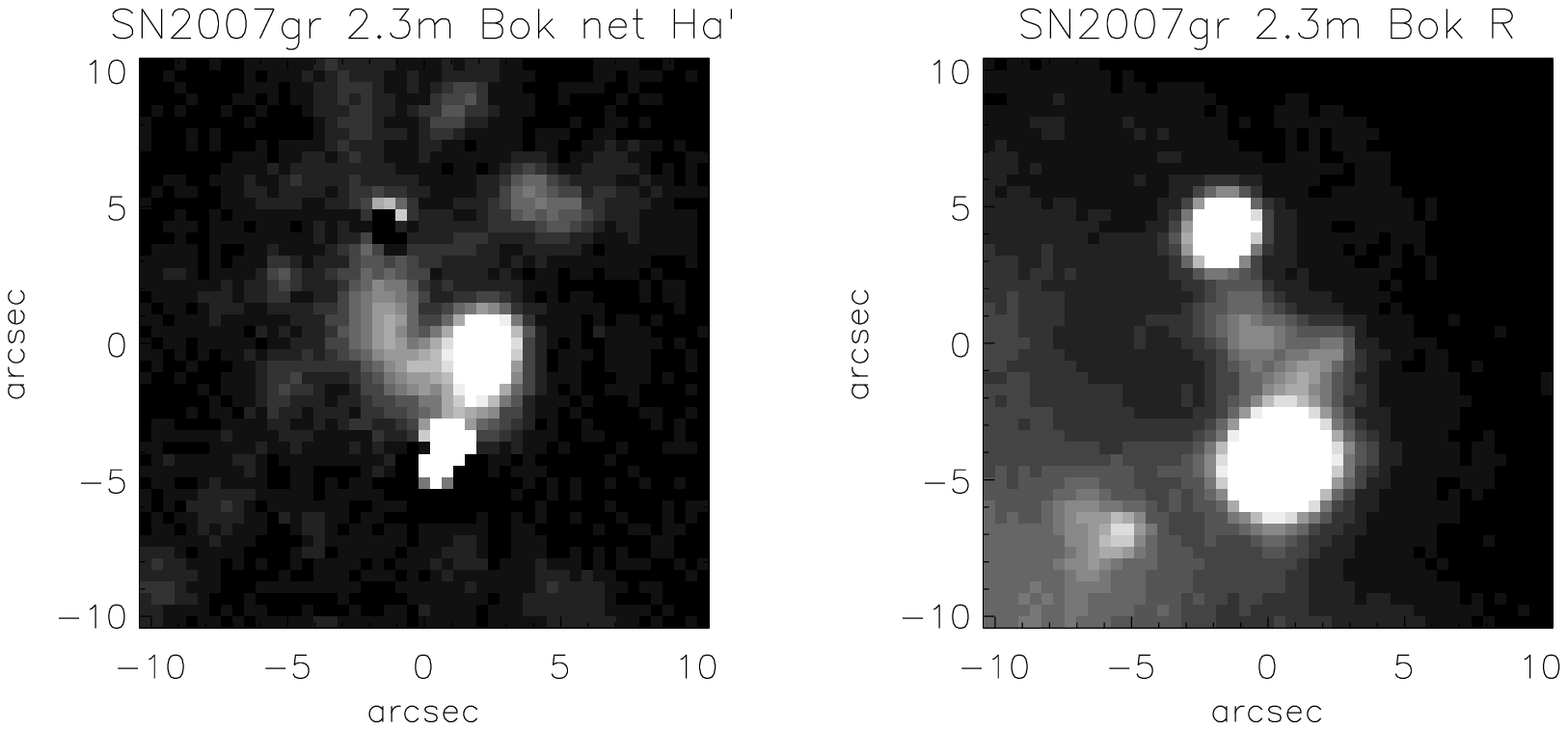}
\caption{left) Bok 2.3m net H$\alpha$ image (from Kennicutt et al. 
2008) showing the nebular environment of SN 2007gr (at centre of image,
Class 5).
The 21$\times$21 arcsec$^{2}$ field of view projects to 1$\times$1
kpc$^{2}$ at the 9.86 Mpc distance of NGC~1058; (right) R-band image.}
\label{sn2007gr}
\end{figure*}

This type Ic supernova was discovered by W. Li in Aug 2007, 0.5$'$ 
NW of the nucleus of NGC 1058 (Li et al. 2007), corresponding to a
deprojected offset of 0.7$'$ (0.45 $R_{\rm 25}$) for the HyperLeda
inclination and major axis PA, i.e. a 2.0 kpc galactocentric distance 
for the 9.86 Mpc distance to NGC 1058. This
SN occurred within a large star forming complex (Crockett et al. 2008). 
Fig.~\ref{sn2007gr} shows 2.3m  Bok imaging obtained in Nov 2003 
(Kennicutt et al. 
2008), in which the primary  source of H$\alpha$  emission within this complex is 
a giant H\,{\sc ii}  region, comparable in luminosity to the SMC's N66. This lies 
$\sim$2$''$ W of SN 2007gr, corresponding to a deprojected distance of 130  
pc, and itself is spatially  extended. 
Diffuse  nebular emission  is  found at the location SN 2007gr, while 
another  bright component is observed 1$''$.5 to the NW, with a luminosity 
similar  to the Rosette  nebula.  Anderson et al. (2012) quote a NCR pixel 
value of 0.157 for SN 2007gr based on JKT imaging. Crockett et al. (2008) have 
studied  pre- and post-SN images of NGC~1058 including archival INT imaging.
INT/WFC H$\alpha$ and $r'$-band imaging from Jan  2005 reveal bright, 
spatially extended emission 2$''$.7 (170 pc) W of the SN location,  plus 
faint nebulosity $1''$ to the S, and to the E. More recent HST 
WFPC2/F675W imaging from Nov 2008 (GO 10877, PI W. Li) and WFC3/F625W
imaging from Jan 2010 (GO 11675, PI J. Maund) indicate fainter nebulosity 
at the SN position.



\subsection{SN 2008bk in NGC 7793}

\begin{figure*}
\includegraphics[bb=20 10 500 235, width=0.75\textwidth]{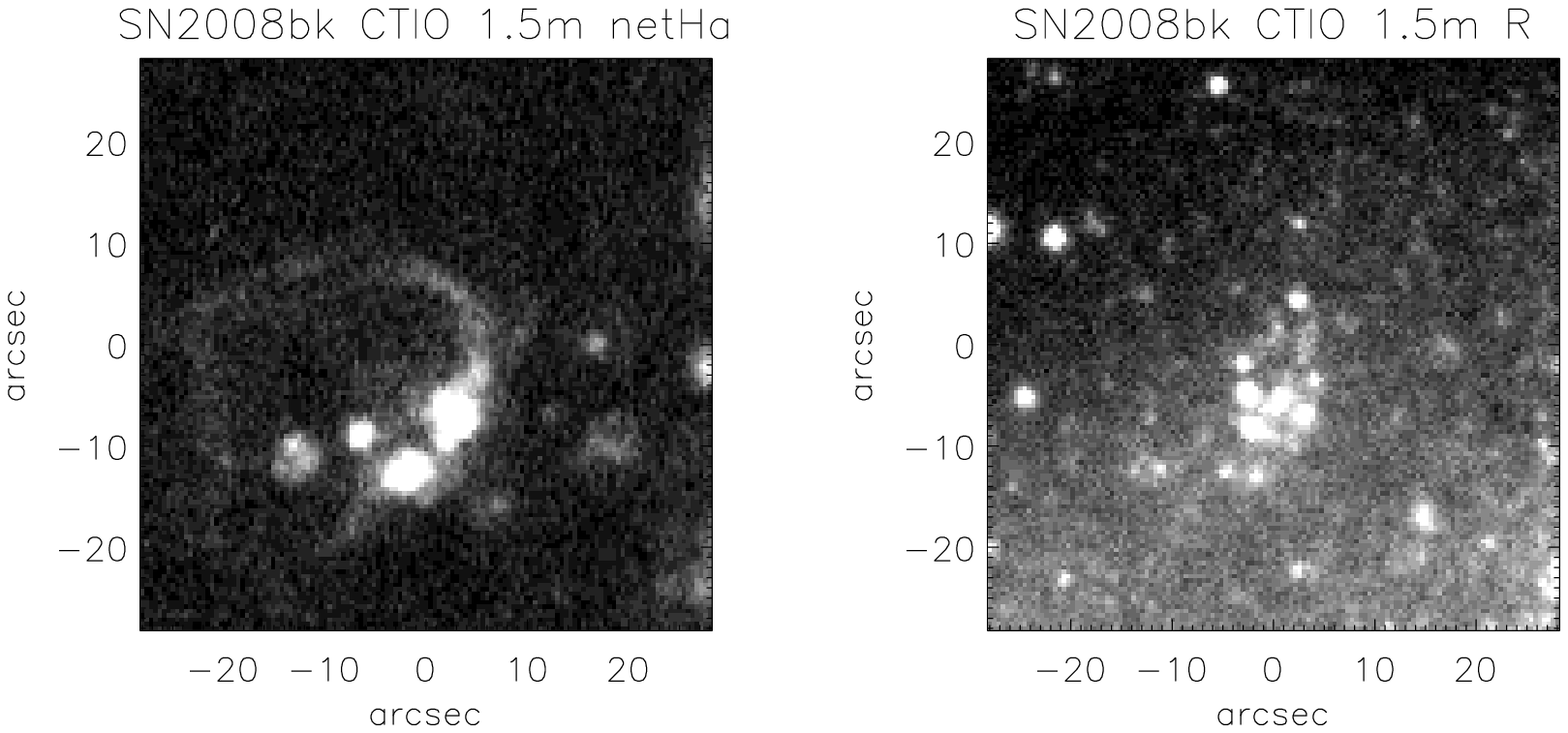}
\caption{left) CTIO 1.5m net H$\alpha$ image (from Kennicutt et al. 
2003) showing the nebular environment of SN 2008bk (at centre of image,
Class 2).
The 57$\times$57 arcsec$^{2}$ field of view projects to 1$\times$1
kpc$^{2}$ at the 3.61 Mpc distance of NGC~7793; (right) R-band image.}
\label{sn2008bk}
\end{figure*}

This type II-P supernova was discovered in Mar 2008, offset 2.1$'$ NNE 
of the nucleus of NGC 7793 (Monard 2008a). This corresponds to a 
deprojected radial offset of 3.5$'$ (0.75 $R_{\rm 25}$) for the adopted 
inclination and PA (Carignan \& Puche 1990), equivalent to 3.7 kpc
for a 3.61 Mpc distance to NGC 7793. Since this position 
lays beyond the  field-of-view of our VLT/FORS1 H$\alpha$ images, so
Fig.~\ref{sn2008bk} presents continuum
subtracted CTIO 1.5m H$\alpha$  images from Oct 2001 
(Kennicutt et al. 2003). 
SN 2008bk is neither spatially coincident with nebular emission
nor a bright continuum source. A H$\alpha$ arc extends to the N and W of
the SN position, $\sim$6$''$ away, corresponding to a deprojected 
distance of 175 pc, and connects to a spatially 
extended  H\,{\sc ii} region, 7$''$ (200 pc) to the SW. In addition to 
this  source, which is  somewhat  more luminous than the Orion nebula, 
other fainter star  formation knots lie 11--12$''$ to the S and SE of SN 
2008bk. Mattila et al. (2008) and van Dyk et al. (2012) confirm a red 
supergiant progenitor for SN 2008bk from, respectively, pre-explosion VLT/FORS1 
and Gemini/GMOS imaging. Archival VLT/FORS1 H$\alpha$ and R-band 
imaging from Sep 2001 (067.D-0006(A)), the former uncalibrated, 
confirm the nebular morphology of CTIO 1.5m imaging. Post-SN F814W 
imaging has also been obtained with HST WFC3 from Apr 2011 (GO 12262, PI 
J.R. Maund).




\subsection{SN 2008eh in NGC 2997}

\begin{figure*}
\includegraphics[bb=20 10 500 235, 
width=0.75\textwidth]{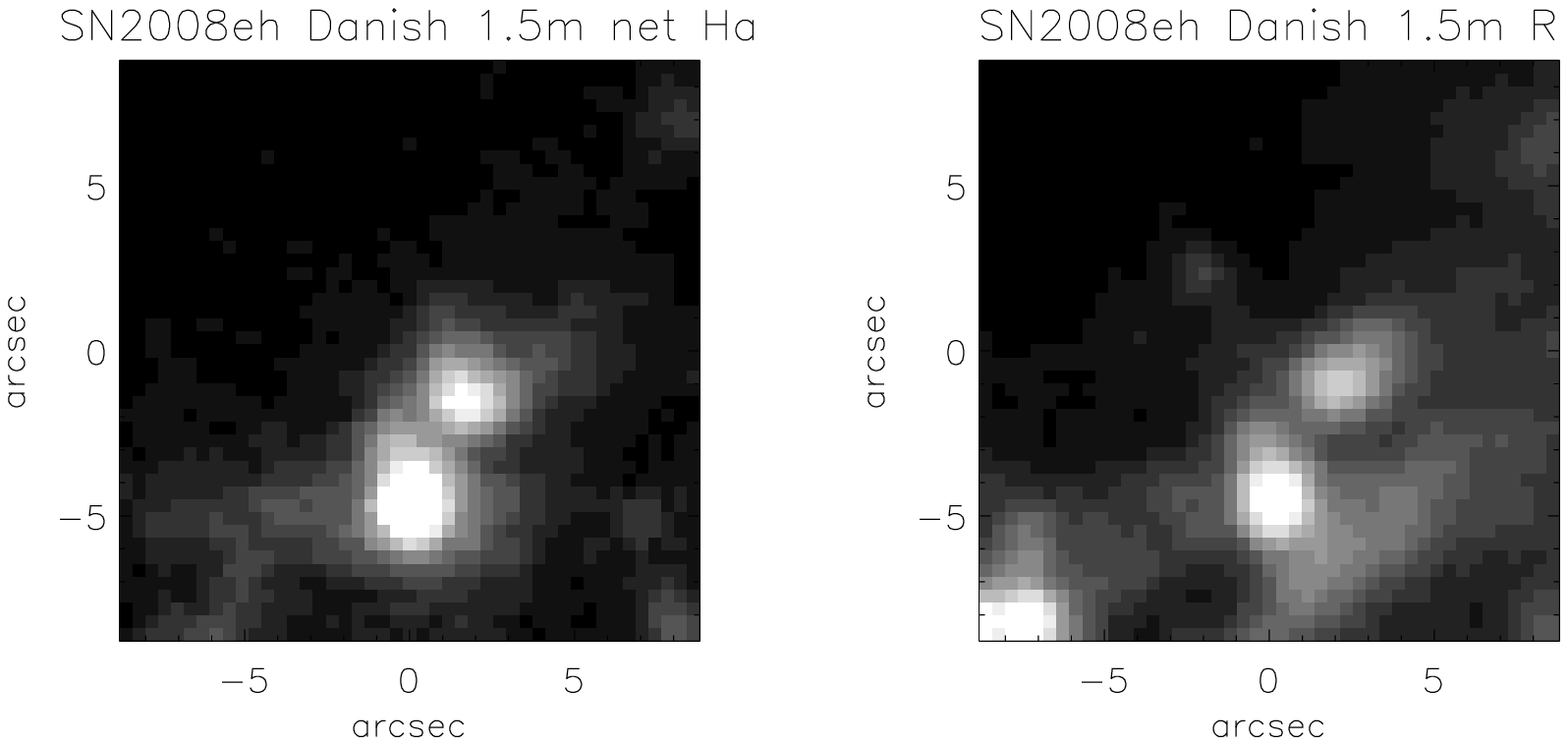}
\caption{(left) Danish 1.5m net H$\alpha$ image (from Larsen \& Richtler
1999) showing the nebular environment of SN 2008eh (at centre of image,
Class 5). 
The 18$\times$18 arcsec$^{2}$ field of view projects to 1$\times$1 
kpc$^{2}$ at the  11.7 Mpc distance of NGC~2997; (right) R-band image.}
\label{sn2008eh}
\end{figure*}

This probable type Ib/c supernova (Horiuchi et al. 2011) was discovered in 
July 2008,  
2.1$'$ ENE of the nucleus of NGC 2997 (Monard 2008b), close to its
major axis, such that the deprojected radial distance is 2$'$.2 (0.5 
$R_{\rm 25}$) or 7.4 kpc for a 11.3 Mpc distance to NGC 2997. 
Fig.~\ref{sn2008eh} shows the site of SN 2008eh at the periphery of 
a giant H\,{\sc ii} region, 2$''$ (110 pc) to the SW, based on
1.5m Danish H$\alpha$ imaging of Larsen \& Richtler (1999). A 
brighter, N66-like, giant H\,{\sc ii} region 4$''$.5 (250 pc) lies to the 
S. Both sources are prominent in R-band images.


\subsection{SN 2009N in NGC 4487}

\begin{figure*}
\includegraphics[bb=20 10 500 235, width=0.75\textwidth]{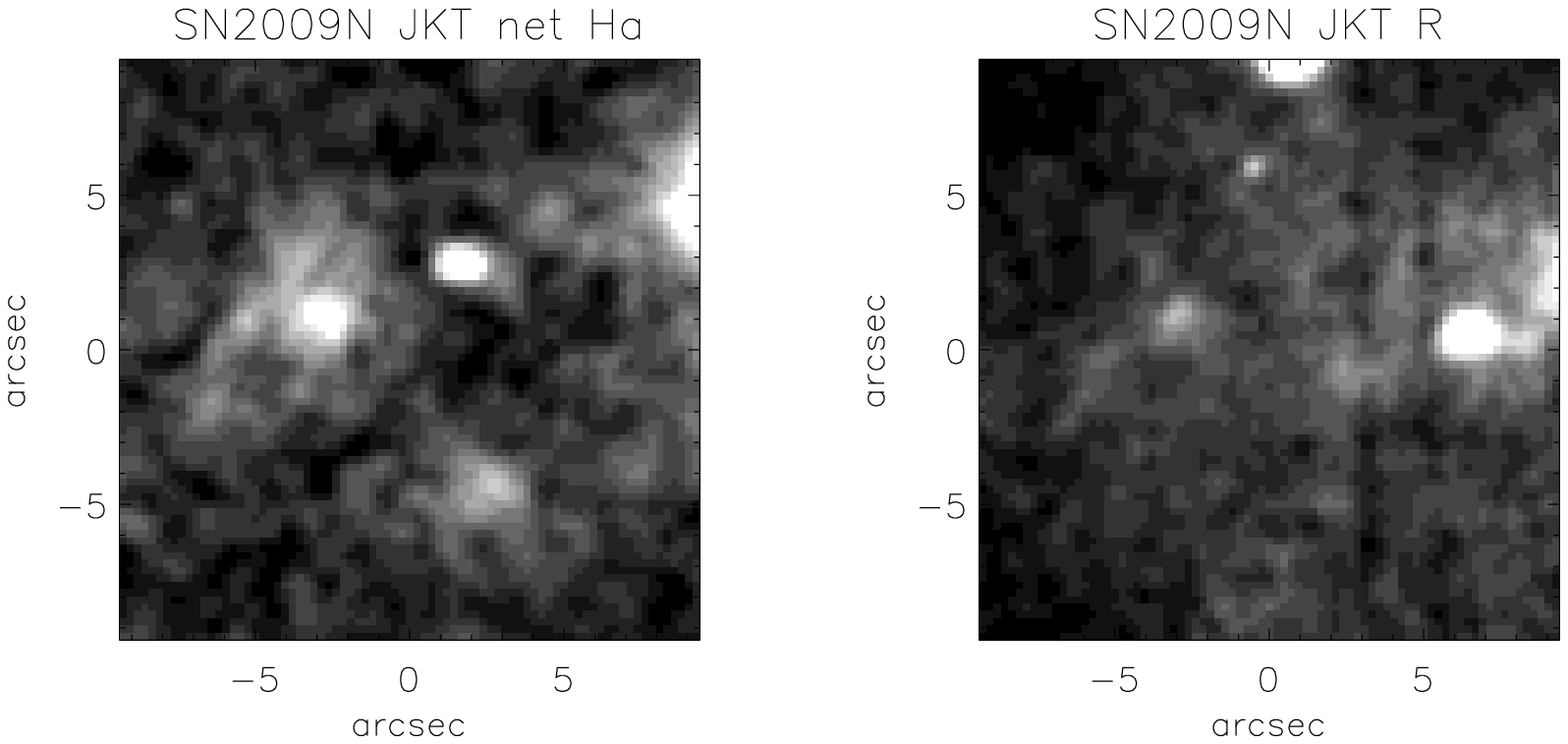}
\caption{left) JKT net H$\alpha$ image (from Knapen et al. 
2004) showing the nebular environment of SN 2009N (at centre of image,
Class 2).
The 19$\times$19 arcsec$^{2}$ field of view projects to 1$\times$1
kpc$^{2}$ at the 11.0 Mpc distance of NGC~4487; (right) R-band image.}
\label{sn2009n}
\end{figure*}

This type II-P supernova was discovered by K. Itagaki in Jan 2009, 
1.3$'$ ENE of the nucleus of NGC 4487 (Nakano et al. 2009), which
corresponds to a deprojected radial distance of 1.6$'$ (0.77 $R_{\rm 
25}$)  or 5.1 kpc for a 11 Mpc distance to NGC 4487. Figure~\ref{sn2009n}
shows continuum-subtracted JKT H$\alpha$ imaging 
(Knapen et al. 2004), from which we identify a compact H\,{\sc ii} region 
3$''$ 
to the NW, a deproject distance of 200 pc away, with a luminosity 
comparable to the Orion nebula. In 
addition, a more extended H\,{\sc ii} region lies 3$''$ (200 pc) to the 
NE, closer in luminosity to the Rosette nebula. SN 2009N is not associated
with either of these star forming regions.



\subsection{SN 2009ib in NGC 1559}

\begin{figure*}
\includegraphics[bb=20 10 500 235, width=0.75\textwidth]{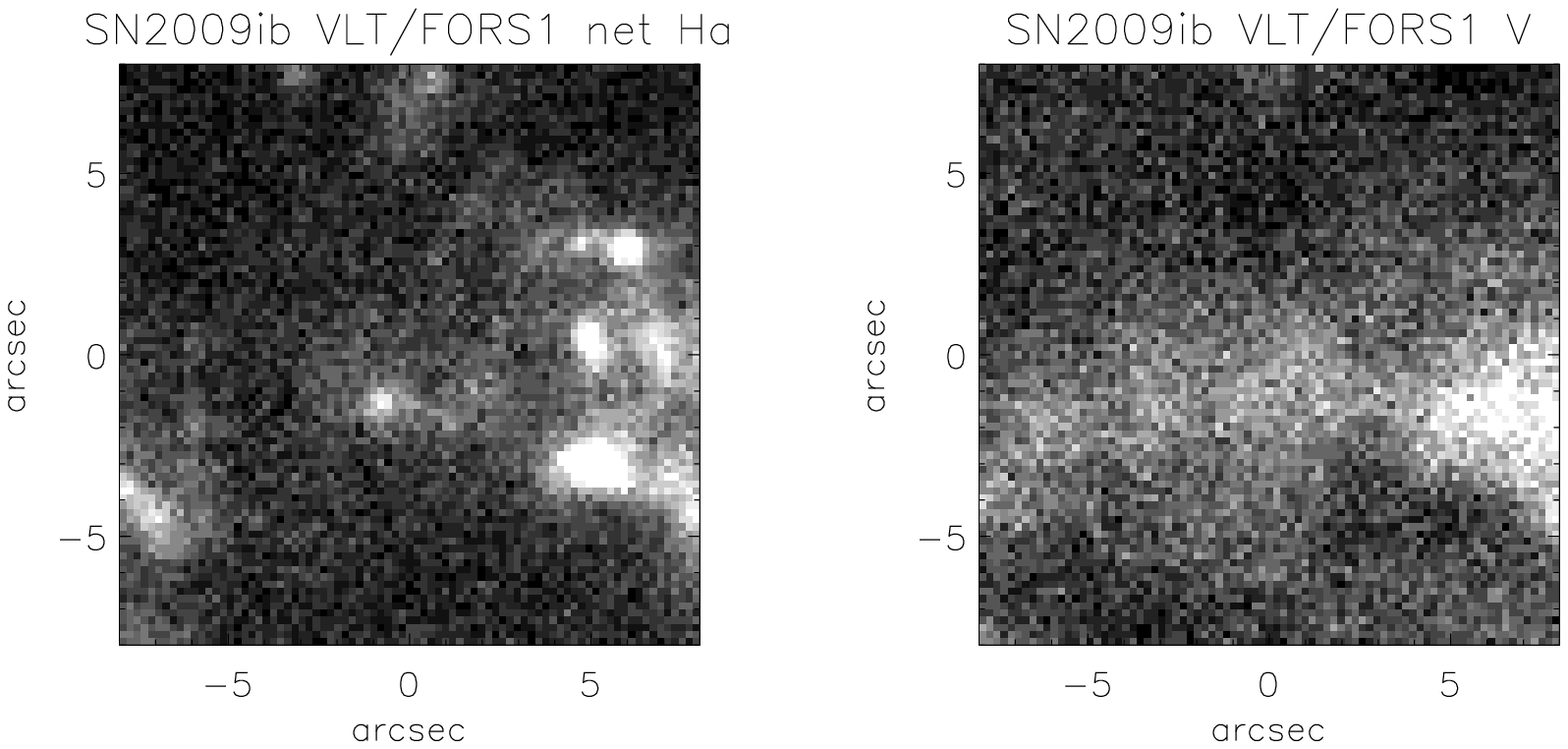}
\caption{left) VLT/FORS1 net H$\alpha$ image (from 075.D-0213(A))
showing the nebular environment of SN 2009ib (at centre of image,
Class 2).
The 16$\times$16 arcsec$^{2}$ field of view projects to 1$\times$1
kpc$^{2}$ at the 12.6 Mpc distance of NGC~1559; (right) V-band image.}
\label{sn2009ib}
\end{figure*}

SN 2009bi (II-P) was disovered in Aug 2009, 37$''$ NE 
of the centre of NGC 1559 (Pignata et al. 2009). This corresponds to
a radial distance of 70$''$ (0.67 $R_{\rm 25}$) or galactocentric distance 
of 4.2 kpc for a 12.6 Mpc distance to NGC 1559. Fig.~\ref{sn2009ib}
shows pre-SN VLT/FORS1 continuum subtracted H$\alpha$ imaging 
from Aug 2005. Faint nebulosity is observed at the site of the SN, 
with an extended H\,{\sc ii}  
region 1$''$.5 to the SE, a deprojected distance of 170 pc away. 
Extended, giant H\,{\sc ii} regions lie 
to the W of SN 2009bi, including a source 6$''$ (0.7 kpc) SW of the SN 
comparable in luminosity to N66,  plus a supergiant H\,{\sc ii} region 
13$''$ (1.5 kpc) to the ESE with a  H$\alpha$+[N\,{\sc  ii}] flux of 
1.5$\times  10^{-13}$  erg\,s$^{-1}$\,cm$^{-2}$ (2$''$.5 aperture radius). 
Pre-SN HST imaging with WFPC2 using the F606W filter (GO 9042, PI S.J. 
Smartt) reveals a potential host cluster within the error circle of the SN 
location, plus diffuse nebular emission.


\subsection{SN 2011dh in M 51a}

\begin{figure*}
\includegraphics[bb=20 10 500 235, width=0.75\textwidth]{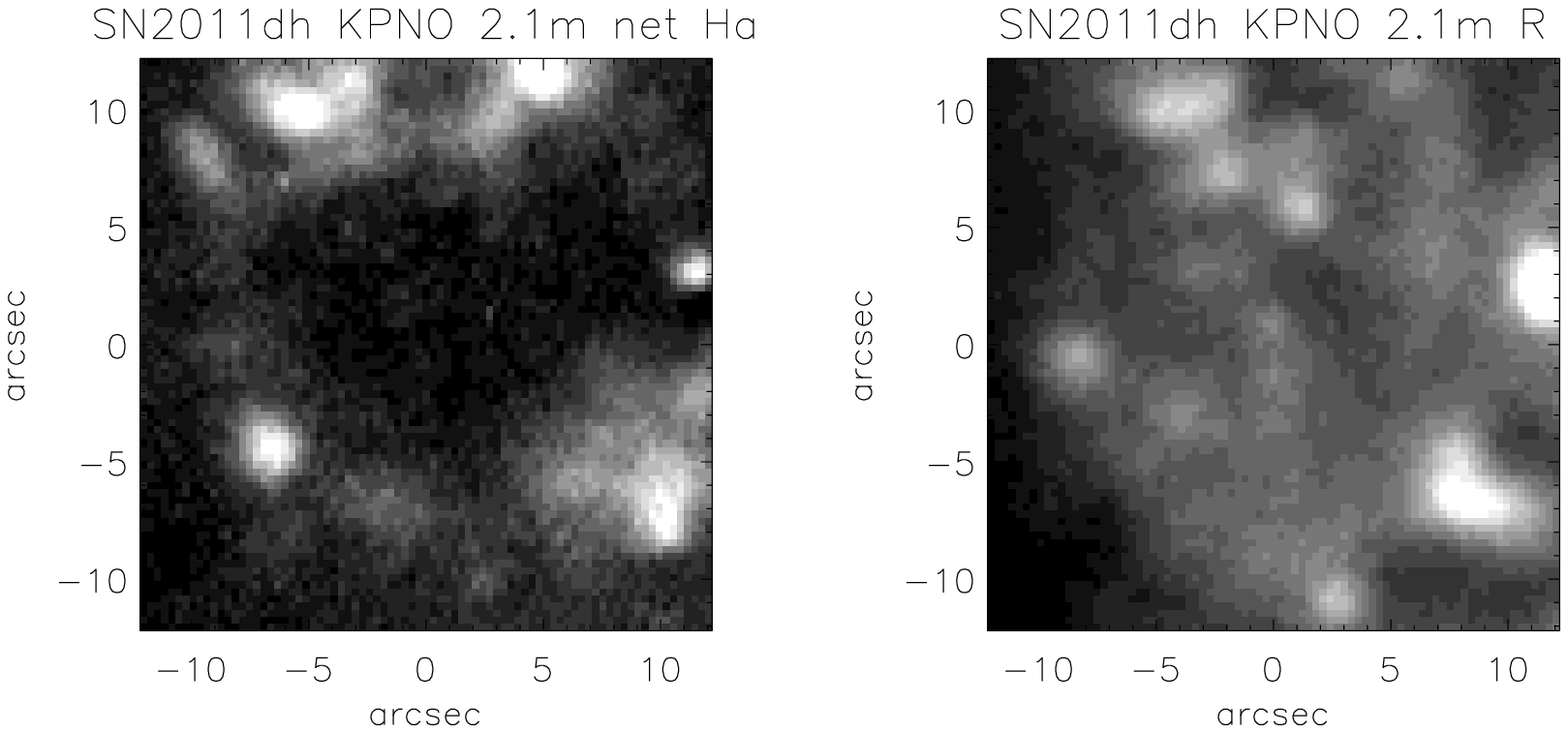}
\caption{(left) KPNO 2.1m net H$\alpha$ image (from Kennicutt et al. 2003) 
showing the nebular environment of SN 2011dh (at centre of image,  
Class 2).
The 25$\times$25 arcsec$^{2}$ field of view projects to 1$\times$1
kpc$^{2}$ at the 8.39 Mpc distance of M~51a; (right) R-band image.}
\label{sn2011dh}
\end{figure*}

This type IIb supernova was discovered in late May 2011,  
2.6$'$ SE of the  nucleus of M~51a (Griga et al. 2011), corresponding
to a radial distance of 2$'$.8 (0.5 $R_{\rm 25}$) or 6.8 kpc for
a  8.39 Mpc distance to M 51a. Fig.~\ref{sn2011dh} shows that
the SN position is close to the centre of a ring of star forming regions, based 
on KNPO 2.1m H$\alpha$ imaging (Kennicutt et al. 2003). Of 
these, the closest to the site of the SN lies 8$''$ (0.35 kpc) SE and
has a similar luminosity to the Rosette nebula. Other, spatially
extended, giant H\,{\sc ii} region complexes lie 11--13$''$ to the NE, NW 
and SW. Anderson et al. (2012) report a NCR pixel value of 0.00 for SN 
2011dh from INT imaging, while inspection of archival HST ACS/WFC images 
confirms the absence of nebular emission at the position of SN 2011dh.
The (point source) SN progenitor is detected in ACS/WFC F555W and F658N 
images (e.g.  van Dyk et al. 2011; Maund et al.  2011), which also 
resolve the star forming region to the SE into two main  components, 
separated by $1''$.0 (45 pc). H\,{\sc ii} regions to the NE and  NE are 
largely dominated by one main component, 
while that to the SW is highly complex, comprising multiple compact sources.


\subsection{SN 2012A in NGC 3239}

\begin{figure*}
\includegraphics[bb=20 10 500 235, width=0.75\textwidth]{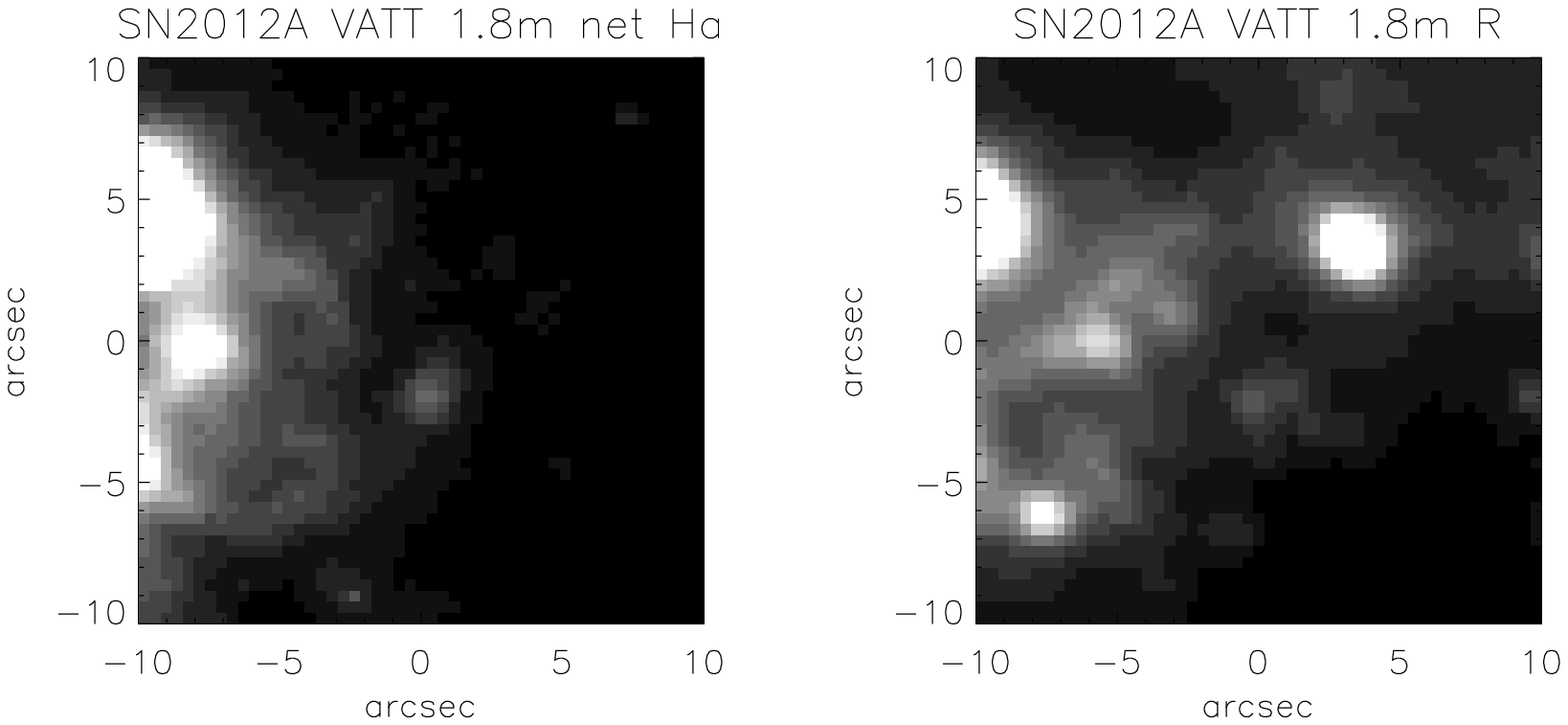}
\caption{left) VATT 1.8m net H$\alpha$ image (from Kennicutt et al. 2008) 
showing the nebular environment of SN 2012A (at centre of image, 
Class 4).
The 20$\times$20 arcsec$^{2}$ field of view projects to 1$\times$1
kpc$^{2}$ at the 10.0 Mpc distance of NGC~3239; (right) R-band image.}
\label{sn2012a}
\end{figure*}

This type II-P supernova was discovered in Jan 2012, is 
50$''$  SE of the nucleus of NGC 3239 (Moore et al. 2012). Based on the
HyperLeda inclination of NGC 3239, this corresponds to a radial distance 
of 64$''$ (0.42 $R_{\rm 25}$) for this galaxy, corresponding to 3.1 kpc 
at 10 Mpc. Fig.~\ref{sn2012a} presents continuum-subtracted 
H$\alpha$ and R-band imaging from VATT 1.8m (Kennicutt et 
al. 2008). SN 2012A is located close to a very bright star forming 
complex, although it is neither spatially coincident with nebular 
emission nor a continuum source. An extended nebula with a luminosity 
comparable to the Rosette nebula lies 2$''$ (120 pc) to the S, with the 
main complex extending over several  hundred pc due E of the SN. Within 
this region, the closest knot of star 
formation lies 7$''$ (0.4 kpc) to  the E of the SN and has an integrated 
H$\alpha$+[N\,{\sc  ii}] flux of 
2$\times  10^{-14}$ erg\,s$^{-1}$\,cm$^{-2}$ (2$''$ aperture radius), 
comparable to  N66 in the SMC. Supergiant H\,{\sc ii} regions are 
found 10$''$ (0.6 kpc) to the NE and 12$''$ (0.75 kpc) to the SE, each 
similar to NGC 604 in luminosity.


\subsection{SN 2012aw in M 95}

\begin{figure*}
\includegraphics[bb=20 10 500 235, width=0.75\textwidth]{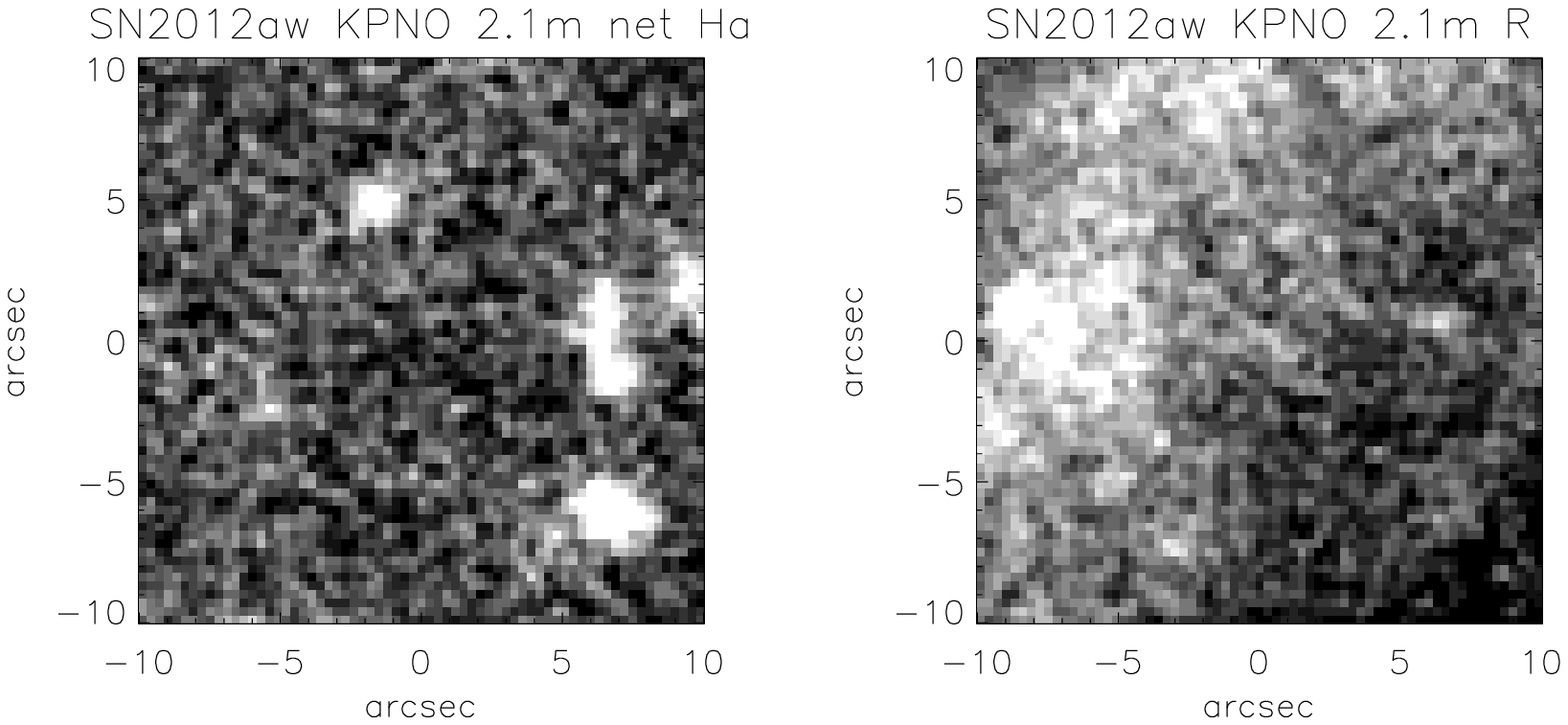}
\caption{(left) KPNO 2.1m net H$\alpha$ image (from Kennicutt et al. 2003) 
showing the nebular environment of SN 2012aw (at centre of image,  
Class 2).
The 20$\times$20 arcsec$^{2}$ field of view projects to 1$\times$1
kpc$^{2}$ at the 10.0 Mpc distance of M~95; (right) R-band image.}
\label{sn2012aw}
\end{figure*}

This type II-P supernova was discovered in Mar 2012,  
2$'$.1  SW of the nucleus of M 95 (NGC 3351). This corresponds to
a radial distance of 2$'$.3 (0.62 $R_{\rm 25}$) using an 
inclination and PA of its major axis from HyperLeda, i.e. a galactocentric 
distance of 6.7 kpc at 10 Mpc. Continuum-subtracted H$\alpha$ imaging from
KPNO 2.1m (Kennicutt et al. 2003) is presented in Fig.~\ref{sn2012aw}.
Nebular emission is not observed at the site of the SN, 
with a H\,{\sc ii} region 5$''$ (260 pc)
to the NNE, somewhat less luminous than the Orion nebula. A further pair
of faint H\,{\sc ii} regions are located 
7$''$ (370 pc) to the W, plus an extended
H\,{\sc ii} region 10$''$ (525 pc) to the SW that is comparable to the
Orion nebula. The closest giant H\,{\sc ii} regions lie 19$''$ (1.0 kpc)
SSE and 22$''$ (1.2 kpc) ENE, with the integrated H$\alpha$+[N\,{\sc  
ii}] flux of the former 4$\times  10^{-14}$ erg\,s$^{-1}$\,cm$^{-2}$ 
(4$''$ aperture radius), implying a luminosity comparable to N66 (SMC).
HST WFPC2 imaging of the site of SN 2012aw was obtained in Nov 1994
using the F555W filter (GO 5397, PI J. Mould).



\clearpage

\clearpage

\clearpage

\clearpage

\clearpage

\clearpage

\clearpage

\clearpage

\clearpage

\clearpage

\clearpage

\clearpage

\clearpage

\section{Core-collapse SNe located at distances of 15--20 Mpc}

Basic properties of ccSNe host galaxies located at distances of
15--20 Mpc (from Tully et al. 2009). Separate Tables are presented 
for low  inclination  ($\leq$65$^{\circ}$) hosts for which accurate ccSNe 
positions  are known (Table B1) and high inclination hosts and ccSNe 
whose coordinates are imprecisely known (Table B2, online-only).

\begin{table*}
  \begin{center}
  \caption{
Basic properties of host galaxies of ccSNe, drawn from RC3 or HyperLeda, 
for which EDD distances lie in the range  15--20 Mpc, restricted to
low inclination ($\leq$65$^{\circ}$) hosts for which accurate ccSNe 
positions are known. SN imposter hosts are 
also omitted (e.g. SN 2003gm in NGC~5334, Smartt et al. 2009)}
%

%
  \label{more_hosts1}
  \begin{tabular}{
r@{\hspace{2mm}}r@{\hspace{2mm}}r@{\hspace{2mm}}r@{\hspace{3mm}}
l@{\hspace{3mm}}r@{\hspace{3mm}}r@{\hspace{3mm}}r@{\hspace{3mm}}
c@{\hspace{3mm}}r@{\hspace{3mm}}l@{\hspace{3mm}}l@{\hspace{3mm}}
}
\hline 
PGC & M & NGC & UGC & Type & $cz$ (km\,s$^{-1}$)  & $i$ & \multicolumn{3}{c}{d (Mpc)} & Ref & ccSNe\\
  \hline
02081 &     &  157 &      &SAB(rs)bc& 1652 & 61.8 & 20.0 &        &     & 
1 & 2009em\\
03572 &     &  337 &      &SB(s)d   & 1648 & 50.6 & 19.5 & $\pm$ & 1.6 & 
1 & 2011dq\\
06826 &     &  701 &      &SB(rs)c  & 1831 & 62.4 & 19.3 & $\pm$ & 3.8 & 1 & 2004fc\\ 
09236 &     & 918 & 01888 & SAB(rs)c? & 1507 & 57.6 & 16.1 & $\pm$ & 3.2 & 1 & 2009js\\
09846 &     &  991 &      & SAB(rc)s & 1532& 28.1 & 17.3 & $\pm$ & 1.1 & 1 & 1984L\\
10464 &     & 1084 &      & SA(s)c &  1407& 49.9 & 17.3 & $\pm$ & 1.1 & 1 & 
1996an, 1998dl, 2009H \\
11479 &     & 1187 &      & SB(r)c  & 1390 & 44.3 & 18.9 & $\pm$ & 2.6 & 1 & 1982R, 2007Y\\
13179 &     & 1365 &      & SB(s)b  & 1636 & 62.7 & 18.0 & $\pm$ & 1.8 & 1 & 1983V, 2001du\\
14617 & \multicolumn{3}{c}{-- ESO G420-G009 --}     & SB(s)c & 1367 & 41.7 & 17.7 & $\pm$ & 1.2 
& 2 & 2003bg \\
14620 &     & 1536 &    & SB(s)c pec? & 1217 & 44.8 &18.0 & $\pm$ & 1.0 & 1 & 1997D \\ 
15850 &     & 1640 &    & SB(r)b      & 1604 & 17.2 &16.8 & $\pm$ & 3.5 & 1 & 1990aj \\ 
29469 &    &    & 05460 & SB(rs)d & 1093 & 39.7 & 20.0 &   &   & 1 & 2011ht \\
31650 &      & 3310 & 05786 & SAB(r)bc pec& 993 & 16.1 & 20.0 &  &   & 1 & 1991N \\
32529 &     & 3423 & 05962 &SA(s)cd & 1011 & 32.1 & 17.0 &$\pm$  &2.5  & 1 & 2009ls\\
34767 &   & 3631 & 06360 & SA(s)c & 1156 & 34.7 & 18.0 &   &   & 1 & {\it 1964A, 1965L}, 1996bu \\
36243 &     & 3810 & 06644 &SA(rs)c &  992 & 48.2 & 16.3 & $\pm$ & 1.7 & 1 & 1997dq, 
2000ew\\
37229 &     & 3938 & 06856 &SA(s)c   & 809 & 14.1 & 17.1 & $\pm$ & 0.8 & 1 &{\it 1961U}, 
1964L, 2005ay\\
37290 &     & 3949 & 06869 &SA(s)bc? &800& 56.5 & 17.1 & $\pm$ & 0.8 & 1 & 2000db\\
37306 &     & 3953 & 06870 &SB(r)bc &1052& 62.1 & 17.1 & $\pm$ & 0.8 & 1 & 2006bp\\
37735 &     &      & 06983 &SB(rs)cd & 1082 & 37.4 & 17.1 & $\pm$ & 0.8 & 1 & 1994P \\
37845 &     & 4030 & 06993 & SA(s)bc & 1465 & 47.1 & 19.5 & $\pm$ & 1.5 & 1 & 2007aa\\
38068 &     & 4051 & 07030 &SAB(rs)bc&700 &30.2 & 17.1 & $\pm$ & 0.8 & 1 & {\it 1983I}, 2003ie, 2010br\\
39578 & 99  & 4254 & 07345 & SA(s)c   & 2407& 20.1  &  1.8 & $\pm$  & 0.8 & 1 & {\it 1967H, 1972Q}, 1986I \\ 
40001 & 61  & 4303 & 07420 & SAB(rs)bc& 1566& 18.1   & 17.6 & $\pm$  & 0.9 & 
1 &  {\it 1926A}, 1961I, {\it 1964F}, 1999gn, 2006ov, 2008in\\
40153 &100 & 4321 & 07450 & SAB(s)bc & 1571 & 23.4 & 15.2 & $\pm$ & 1.5 & 1 & 1979C \\
40745 &     & 4411B & 07546&SAB(s)cd & 1272 & 26.6 & 16.8 & $\pm$ & 0.8 & 1 & 1992ad \\
41050 &     & 4451 & 07600 & Sbc?    & 864 & 53.6    & 16.8 & $\pm$ & 0.8 & 1  & 1985G\\
41746 &     & 4523 & 07713 & SAB(s)m  &262 &25.1   & 16.8 & $\pm$ & 0.8 & 1  & 1999gq\\
42833 &     & 4651 & 07901 & SA(rs)c  &788 & 49.5    & 16.8 & $\pm$ & 0.8 & 1  & 1987K, 2006my\\
43321  &    & 4699 &       & SAB(rs)b & 1394 & 42.6 & 15.3 & $\pm$ & 1.0 & 1 & 1983K \\
43972  &     & 4790 &       & SB(rs)c? & 1344 & 58.8 & 15.3 & $\pm$ & 1.0 & 1 & 2012au \\
44797 &     & 4900 & 08116 & SB(rs)c & 960 & 19.0 & 15.6 & $\pm$ & 1.0 & 1 & 1999br\\
45948 &     & 5033 & 08307 & SA(s)c   & 875 & 64.6  &  18.5 & $\pm$ & 1.1 & 1 &1985L, 2001gd\\
52935 & \multicolumn{3}{c}{-- Arp 261 --} & IB(s)m pec& 1856 & 58.8 & 20 &   &    & 1 & 1995N \\
58827 &    & 6207 & 10521 & SA(s)c   &852 & 64.7  & 18.1 & $\pm$ & 2.1 & 1 & 2004A\\ 
59175 &    & 6221 &       & SB(s)c   &1499 & 50.9 & 15.6 & $\pm$ & 1.7 & 1 & 1990W \\
70094 & \multicolumn{3}{c}{--- IC 5267 --- } & SA0/a(s) & 1712 & 48.4 & 18.7 & $\pm$ & 1.6 & 1 & 2011hs \\
\hline
  \end{tabular}   
  \end{center}
\begin{small}
\begin{flushleft}
1: Tully et al. (2009), 2: NED (Virgo + GA + Shapley)
\end{flushleft}
\end{small}
\end{table*}


\begin{table*}
  \begin{center}
  \caption{
Basic properties of host galaxies of ccSNe, drawn from RC3 or HyperLeda, 
for which EDD distances lie in the range  15--20 Mpc,  restricted to
high inclination ($\geq$65$^{\circ}$) hosts and ccSNe whose coordinates 
are imprecisely known (in italics). SN imposters are omitted.}
%

%
  \label{more_hosts2}
  \begin{tabular}{
r@{\hspace{2mm}}r@{\hspace{2mm}}r@{\hspace{2mm}}r@{\hspace{3mm}}
l@{\hspace{3mm}}r@{\hspace{3mm}}r@{\hspace{3mm}}r@{\hspace{3mm}}
c@{\hspace{3mm}}r@{\hspace{3mm}}l@{\hspace{3mm}}l@{\hspace{3mm}}
}
\hline 
PGC & M & NGC & UGC & Type & $cz$ (km\,s$^{-1}$)  & $i$ & \multicolumn{3}{c}{d (Mpc)} & Ref & ccSNe\\
  \hline
09057 &     & 908  &      & SA(s)c & 1509 & {\bf 65.1} & 19.0 & $\pm$ & 1.6 & 1 & 1994ai \\
09354 & \multicolumn{3}{c}{----- Mk 1039 -----} & Sc? & 2111 & {\bf 75.6} & 19.2 
& $\pm$ & 4.0 & 1 & {\it 1985S} \\
10065 &     & 1035 &      & SA(s)c? & 1241 & {\bf 74.5} & 17.3 & $\pm$ & 1.1 & 1 & {\it 1990E} \\
12007 &     & 1255 &      & SAB(rs)bc& 1686& 58.8 & 20.0 & $\pm$ & 1.2 & 1 & {\it 1980O}\\
13586 &     & 1433 &       & (R)SB(r)ab & 1075 & {\bf 67.4} & 16.8 & $\pm$ & 1.0 & 1 & 1985P \\
13633 &     &      & 02813 & Im?    & 1392 & {\bf 76.6} & 16.1 & $\pm$ & 1.5 & 1 & 2008fb\\
13727 &     & 1448 &      & SAcd?   & 1168 & {\bf 86.1} & 16.8 & $\pm$ & 1.0 & 1 & {\it 1983S}, 2003hn\\
13985 & \multicolumn{3}{c}{-- ESO G302-G014 --}     & IB(s)m & 872 & {\bf 73.6} & 16.8 & $\pm$ & 1.0 &1 & 2008jb \\
14123 &     &      & 02890 & Sdm pec? & 1155 & {\bf 90} & 16.1 & $\pm$ & 1.5 & 1 & 2009bw\\
19531 &     & 2280 &      & SA(s)cd  & 1899 & {\bf 66.2} & 20.0 & $\pm$ & 1.4 & l  & 2001fz\\
19579 &     & 2273B& 03530 & SB(rs)cd: & 2101 & {\bf 67.9} & 17.9 & $\pm$ & 2.2 & 1 & 2011fd\\
36699 &     & 3877 & 06745 &SA(s)c? &  895 &{\bf 83.2} & 17.1 &$\pm$ & 0.8 & 1  & 1998S\\
37912 & IC 755 & 4019 & 07001 & SBb? & 1524 & {\bf 90} & 16.8 & $\pm$ & 0.8 & 1 & 1999an \\
38302 &     & 4088 & 07081 &SAB(rs)bc &757 & {\bf 71.2} &  17.1 & $\pm$ & 0.8 & 1  & 1991G, 2009dd\\
38580 &     & 4129 &       &SB(s)ab? & 1177 & {\bf 90} & 18.0 & $\pm$ & 3.8 & 
1 & 2002E\\
38618 &     & 4136 & 07134 &SAB(r)c  & 609 & 22 & 16.3 & $\pm$ & 0.9 & 1 & {\it 1941C}\\ 
38795 &     & 4157 & 07183 &SAB(s)b? & 774 & {\bf 90} & 17.1 & $\pm$ & 0.8 & 1 & {\it 1937A}, 2003J\\
39724 &     & 4274 & 07377 &(R)SB(r)ab & 930 & {\bf 68} &16.3  & $\pm$  & 0.9 & 1  & 1999ev\\
39974 &     & 4302 & 07418 & Sc? & 1149 & {\bf 90} & 16.8 & $\pm$ & 0.8 & 1 & 1986E \\
40530 & \multicolumn{2}{c}{-- IC 3311 --} & 
                      07510 & Sdm?  & --122     &  {\bf 90} & 20.0  & 
$\pm$  & 1.0  & 2  & 2004gk \\
41608 & \multicolumn{2}{c}{-- IC 3476 --} &
                      07695 & IB(s)m? & --169 & 51.2 & 16.8 & $\pm$ & 0.8 
& 1& {\it 1970A} \\
41789 &     & 4527 & 07721 & SAB(s)bc & 1736 & {\bf 81.2} & 17.6 &$\pm$ & 0.9 & 1 & 2004gn\\
42069 &     & 4568 & 07776 & SA(rs)bc & 2255 & {\bf 67.5} & 16.8 &$\pm$ & 0.8 & 1 & 1990B, 2004cc \\
42975 &     & 4666 & 07926 & SABc:    & 1529 & {\bf 69.6} & 15.7 &$\pm$ & 2.9 & 1 & {\it 1965H} \\
43189 &     & 4688 & 07961 & SB(s)cd & 986  & 23.7    &15.6 & $\pm$ & 1.0 & 1 & {\it 1966B}\\
43969  &    & 4809 & 08034 & Im pec& 915 & {\bf 90} & 15.6 & $\pm$ & 1.0 & 1 & 2011jm\\
53247 &     & 5775 & 09579 & SBc? & 1681 & {\bf 83.2} & 19.8 & $\pm$ & 1.0 & 1 & 1996ae \\
54117 &     & 5879 & 09753 &SA(rs)bc? & 772 & {\bf 72.7} & 15.5 & $\pm$ & 0.9 & 1 &{\it 1954C}\\
54470 &     & 5907 & 09801 & SA(s)c? & 667 & {\bf 90} & 17.2 & $\pm$ & 0.9 & 1 & {\it 1940A}\\
\hline
  \end{tabular}   
  \end{center}
\begin{small}
\begin{flushleft}
1: Tully et al. (2009), 2: Solanes et al. (2002)
\end{flushleft}
\end{small}
\end{table*}


\label{lastpage}

\end{document}